\documentclass[12pt]{article}
\usepackage{jheppub}
\pdfoutput=1

\usepackage{amsmath,bbm,array,amsfonts,graphicx,wrapfig,lscape,float,slashbox,multirow,longtable,rotating,subfigure}

\newcommand{\be}{\begin{equation}}
\newcommand{\ee}{\end{equation}}
\newcommand{\beq}{\begin{equation}}
\newcommand{\beql}[1]{\begin{equation}\label{#1}}
\newcommand{\eeq}{\end{equation}}
\newcommand{\ba}{\begin{array}}
\newcommand{\ea}{\end{array}}
\newcommand{\bea}{\begin{eqnarray}}
\newcommand{\beal}[1]{\begin{eqnarray}\label{#1}}
\newcommand{\eea}{\end{eqnarray}}
\newcommand{\ben}{\begin{enumerate}}
\newcommand{\een}{\end{enumerate}}
\newcommand{\bean}{\begin{eqnarray*}}
\newcommand{\eean}{\end{eqnarray*}}
\newcommand{\eref}[1]{(\ref{#1})}
\newcommand{\sref}[1]{\S\ref{#1}}
\newcommand{\tref}[1]{Table~\ref{#1}}
\newcommand{\nn}{\nonumber}

\newcommand{\fref}[1]{Figure \ref{#1}}
\newcommand{\btab}[1]{\begin{tabular}{#1}}
\newcommand{\etab}{\end{tabular}}

\newcommand{\mesonic}{\mathcal{M}^{mes}}

\newcommand{\comment}[1]{}

\newcommand{\qed}{\nobreak \ifvmode \relax \else
      \ifdim\lastskip<1.5em \hskip-\lastskip
      \hskip1.5em plus0em minus0.5em \fi \nobreak
      \vrule height0.75em width0.5em depth0.25em\fi}

\newcolumntype{C}[1]{>{\centering\arraybackslash}m{#1}}

\title{New Directions in Bipartite Field Theories}

\author[a]{Sebastian Franco,}
\author[a]{Daniele Galloni,}
\author[b]{Rak-Kyeong Seong}

\affiliation[a]{
Institute for Particle Physics Phenomenology, Department of Physics\\
Durham University, Durham DH1 3LE, United Kingdom
}

\affiliation[b]{
Theoretical Physics Group, The Blackett Laboratory,
Imperial College London, \\
Prince Consort Road, London SW7 2AZ, United Kingdom
}

\emailAdd{sebastian.franco@durham.ac.uk, daniele.galloni@durham.ac.uk, rak-kyeong.seong@imperial.ac.uk}

\abstract{
We perform a detailed investigation of Bipartite Field Theories (BFTs), a general class of 4d $\mathcal{N}=1$ gauge theories which are defined by bipartite graphs. This class of theories is considerably expanded by identifying a new way of assigning gauge symmetries to graphs. A new procedure is introduced in order to determine the toric Calabi-Yau moduli spaces of BFTs. For graphs on a disk, we show that the matroid polytope for the corresponding cell in the Grassmannian coincides with the toric diagram of the BFT moduli space. A systematic BFT prescription for determining graph reductions is presented. We illustrate our ideas in infinite classes of BFTs and introduce various operations for generating new theories from existing ones. Particular emphasis is given to theories associated to non-planar graphs.
}

\preprint{
\begin{flushright}IPPP/12/89\end{flushright} \vspace{-0.9cm}
\begin{flushright}DCPT/12/178\end{flushright} \vspace{-0.9cm}
\begin{flushright}Imperial/TP/12/RS/04\end{flushright}
}

\begin{document}

\maketitle

\section{Introduction \label{sintro}}

A modern trend in the study of SUSY gauge theories -- with or without Lagrangian description and in various dimensions -- is to define them in terms of geometric or combinatorial objects. These are for example bipartite graphs on 2-tori \cite{Hanany:2005ve,Franco:2005rj}, Riemann surfaces \cite{Gaiotto:2009we,Benini:2009mz,Bah:2011je} and 3-manifolds \cite{Dimofte:2011ju}. In such constructions, complicated theories can typically be engineered by gluing elementary building blocks. Furthermore, field theory equivalences -- such as Seiberg duality \cite{Seiberg:1994pq}, S-duality \cite{Argyres:2007cn} and mirror symmetry \cite{Intriligator:1996ex} -- are mapped to rearrangements of the underlying geometric object.
 
Along the lines of this general paradigm, a new class of gauge theories, whose UV Lagrangian is defined in terms of a bipartite graph on a bordered Riemann surface, was introduced in \cite{Franco:2012mm}. Such theories are called Bipartite Field Theories (BFTs). A similar class of theories was simultaneously introduced in \cite{Xie:2012mr}. There are subtle differences between such theories and BFTs. In \sref{section_BFTs}, we present some comments clarifying the relation between them, emphasizing that the theories in \cite{Xie:2012mr} need not be regarded as a distinct class but can be included in the more general BFT family. Certain subclasses of BFTs have already appeared in the context of interesting physical systems, including D3-branes over toric Calabi-Yau (CY) 3-folds \cite{Franco:2005rj}, cluster integrable systems \cite{Franco:2011sz,Eager:2011dp,Amariti:2012dd,Franco:2012hv} and, more recently, leading singularities in scattering amplitudes \cite{Nima}. Moreover, similar gauge theories and graphs on Riemann surfaces continue to arise in other areas, most notably in relation to the BPS spectrum of 4d $\mathcal{N}=2$ gauge theories \cite{Cecotti:2011rv,Alim:2011ae,Alim:2011kw,Xie:2012dw,Xie:2012hs, Xie:2012jd, Gaiotto:2012db}. This suggests we are only scratching the surface in terms of possible applications of BFTs. In addition, BFTs may provide a more profound understanding of the physical connections between some of these systems.

Bipartite graphs on a disk classify cells in the Grassmannian \cite{Postnikov_plabic}. As explained in \cite{Franco:2012mm}, several concepts in this area -- such as the boundary measurement, matching polytopes, cells and their boundaries, and equivalence and reduction moves -- have beautiful realizations in terms of BFTs. This list is extended in the current paper by explaining the emergence of matroid polytopes for BFTs. Given the connection between scattering amplitudes in $\mathcal{N}=4$ SYM and the Grassmannian \cite{ArkaniHamed:2009dn}, these objects play an important role in the calculation of leading singularities \cite{Nima}. In this context, bipartite graphs are interpreted as on-shell diagrams. BFTs provide an intuitive perspective on these mathematical structures and a natural platform for extending them in new directions, such as the non-planar case.\footnote{Here we use the usual scattering notion of non-planar graph, namely a graph that cannot be embedded in a disk without crossings.} Starting the investigation of BFTs associated to non-planar graphs is indeed one of the central goals of this paper.

An important conclusion of this article is that the universe of BFT theories is indeed much richer -- in fact twice as large -- than originally envisioned in \cite{Franco:2012mm}. This follows from a careful consideration of anomaly-free symmetries associated to bipartite graphs, which leads to two natural alternatives for gauging. These choices give rise to two independent classes of gauge theories. One of them requires the specification of an embedding of the underlying bipartite graph into a Riemann surface, while the other one is, for Abelian theories, independent of any embedding.

This article is organized as follows. \sref{section_BFTs} reviews the general concept of a BFT and discusses the different classes of theories that arise from two possible ways of gauging symmetries. The computation of master and moduli spaces for BFTs, which are toric CY manifolds, is explained in \sref{section_moduli_spaces}. \sref{section_equivalence_and_reduction} discusses graph equivalence and reduction from a BFT viewpoint, explaining how it is possible to reduce graphs by higgsing. \sref{section_geometry_from_BFT} introduces an alternative way for constructing the CY manifolds corresponding to the master and moduli spaces of BFTs, based on the map between perfect matchings and paths on the bipartite graph. Using this approach it is shown that, in the case of bipartite graphs on a disk, the matroid polytope for the associated cell in the Grassmannian coincides with the toric diagram of the moduli space of the corresponding BFT. Infinite classes of BFTs and several operations for generating new theories are presented in \sref{section_infinite_families} to \sref{section_edge_splitting}. \sref{section_Seiberg} is devoted to Seiberg and Toric duality for BFTs on higher genus Riemann surfaces. \sref{section_gauging_2} discusses some of the most distinctive features of the BFTs associated to the gauging choice that is independent of any embedding of the underlying bipartite graph into a Riemann surface. Conclusions and directions for future research are collected in \sref{conclusions}.

\bigskip

\noindent {\bf Note added:} While this paper was ready for submission, \cite{Vafa_et_al} appeared, with interesting additional results on BFTs.

\bigskip

\section{Bipartite Field Theories}

\label{section_BFTs}

A BFT is a 4d $\mathcal{N}=1$ quiver gauge theory whose Lagrangian is defined in terms of a bipartite graph living on a Riemann surface, possibly containing boundaries. The presence of boundaries is determined by the existence of external nodes. We further restrict to graphs in which external nodes are attached to a single edge in the bipartite graph. 

The next section discusses in detail two possible ways of gauging the symmetries in these theories. These alternatives give rise to two independent classes of BFTs. As we will explain, for one of the possible gaugings the resulting theories are independent of any embedding of the bipartite graphs into a Riemann surface, in other words they can be defined without appealing to any Riemann surface at all. Keeping the two possibilities in mind, it is still useful to invoke an underlying Riemann surface in order to provide a unified presentation of the two classes of BFTs. 

The basic elements of the graph have the following translation into the gauge theory are:

\medskip

\begin{itemize}
\item {\bf\underline{Faces}:} $U(N)$ symmetry groups.
\item {\bf \underline{Edges}:} chiral multiplets $X_{ij}$ transforming in the bifundamental representation of the two groups, $U(N)_i \times U(N)_j$, associated to the two faces adjacent to the edge. The orientation of bifundamental fields is determined by the convention that they go clockwise around white nodes and counterclockwise around black nodes. 
\item {\bf \underline{Nodes}:} a white/black internal node corresponds to a positive/negative monomial in the superpotential involving the chiral fields corresponding to all the edges terminating on it. The clockwise or counterclockwise orientation associated to nodes determines the cyclic ordering of fields in each superpotential term. External nodes are connected to a single edge and are not mapped to superpotential terms.
\end{itemize}

\medskip

\noindent Below additional properties of bipartite graphs, which relate to gauging, are discussed. These need to be taken into account when defining a BFT.

\bigskip

\subsection{Two Alternative Gaugings}

\label{section_two_gaugings}

One possible way of gauging the $U(N)$ symmetries of BFTs was considered in \cite{Franco:2012mm}. More careful thought reveals that there exists yet another natural way of gauging them. We refer to the two possibilities as gaugings 1 and 2 and review them below. Each gauging leads to a different class of consistent theories, expanding the realm of BFTs by effectively doubling it with respect to what was originally considered in \cite{Franco:2012mm}.

\bigskip

\subsubsection{Gauging 1}

The faces sliced by the bipartite graph on the Riemann surface can be divided into two classes. We call them \textbf{internal} or \textbf{external}, depending on whether their perimeter consists entirely of edges or contains parts of the boundaries, respectively. 

Since the graph is bipartite, the number of edges around an internal face is even, and there is an equal number of black and white nodes on its perimeter. This fact, together with a convention for orientation of bifundamental fields, implies that the corresponding node on the BFT quiver has an equal number of incoming and outgoing arrows and is hence anomaly free. This fact is not generically true for external faces. 

\begin{figure}[h]
\begin{center}
\resizebox{0.8\hsize}{!}{
\includegraphics[trim=0cm 0cm 0cm 0cm,totalheight=10 cm]{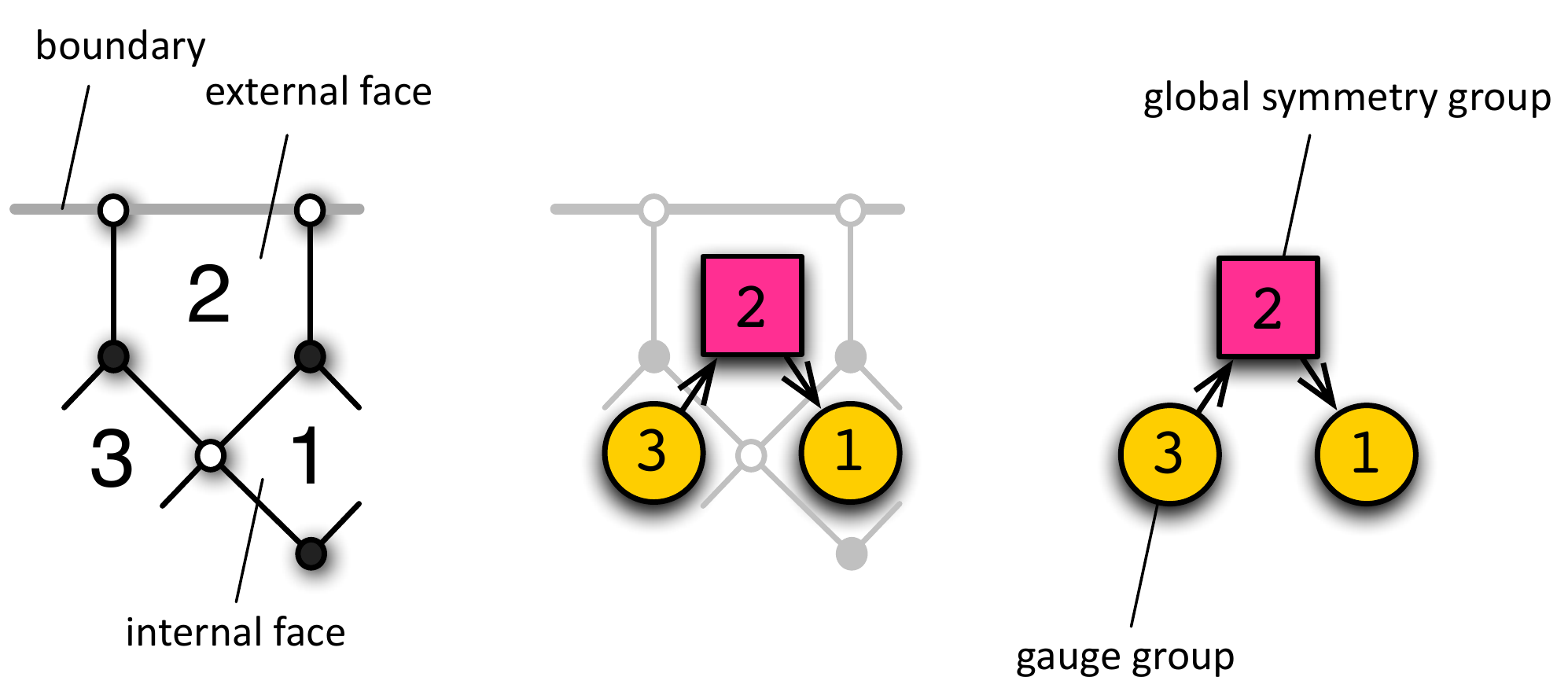}
} \\
  \caption{\textit{A section of a bipartite graph and its dual BFT quiver for gauging 1.} On the gauge theory side, internal and external faces correspond to global and gauge symmetry groups, respectively.
  \label{fbipgraph}}
 \end{center}
 \end{figure}

These observations motivate the definition of {\bf gauging 1}, in which the $U(N)$ groups associated to internal faces are gauged while the ones for external faces remain global symmetries. This is the gauging considered when BFTs were introduced in \cite{Franco:2012mm}. We say that any BFT associated to this choice is of BFT$_1$ type. BFT$_1$ theories are quiver theories, i.e. chiral fields transform in bifundamental or adjoint representations of the gauge and global symmetry groups. \fref{fbipgraph} shows a section of a bipartite graph and its connection to a BFT$_1$. 

Gauging 1 arises naturally when thinking about theories with a D-brane interpretation. In this case, the worldvolume of D-branes spans the two graph directions, which are internal, and has an infinite extension along some transverse dimensions on which the low energy gauge theory lives. Internal faces correspond to D-branes with a finite extension in the internal directions and hence give rise to gauge symmetries in the transverse dimensions. On the other hand, external faces can be interpreted as D-branes that are infinite along some of the internal dimensions, frequently denoted as flavor branes, which give rise to gauge theories with a higher dimensional support, i.e. to global symmetries from the perspective of the transverse dimensions.

Indeed, a subclass of BFT$_1$'s has already appeared in this context in the literature, playing a prominent role. It corresponds to the 4d, $\mathcal{N}=1$ worldvolume theories on D3-branes probing toric Calabi-Yau 3-fold singularities. For this subclass of theories, the Riemann surface is a 2-torus \cite{Franco:2005rj}. In this context, the corresponding bipartite graphs are called \textbf{brane tilings} and have been the subject of extensive investigations \cite{Hanany:2005ve,Franco:2005rj,Franco:2005sm,Kennaway:2007tq,2007arXiv0710.1898I}. The correspondence between these gauge theories and bipartite graphs has indeed been instrumental in several important developments such as the determination of the superconformal field theories that are dual, via the AdS/CFT correspondence, to infinite families of Sasaki-Einstein manifolds \cite{Benvenuti:2004dy,Franco:2005sm,Butti:2005sw}

\tref{tdic} summarizes the dictionary between bipartite graphs on Riemann surfaces and BFT$_1$'s. Let us conclude this section with a few comments on the connection between BFT$_1$'s and the theories introduced in \cite{Xie:2012mr}. In our language, these models are obtained from ours by omitting the chiral fields associated to external legs terminating on black external nodes and the superpotential terms they participate in. In physical terms, the theories in \cite{Xie:2012mr} can be regarded as a sub-class of BFT$_1$'s. Tuning some of the superpotential couplings to zero, more precisely those associated to the white nodes connected to black external nodes, our theories reduce to them plus decoupled singlets, which correspond to the graph legs connected to black external nodes. The additional fields contained in BFT$_1$'s play a nice role in making detailed contact with objects such as matching and matroid polytopes associated to cells in the Grassmannian.

\medskip

\begin{table}[htt!!]
\begin{center}
\begin{tabular}{|l|l|}
\hline
{\bf Graph} & {\bf BFT} \\ \hline \hline
Internal face ($2n$-sided) & Gauge group with $n$ flavors \\ \hline
External face & Global symmetry group \\ \hline \hline
Edge between two faces $i$ and $j$ & Chiral multiplet in the bifundamental \\ & representation of the groups $i$ and $j$. The \\ & orientation of the corresponding arrow is such \\ & that it goes clockwise around white nodes and \\ & counterclockwise around black nodes. \\ \hline \hline
$k$-valent node & Monomial in the superpotential involving $k$ \\ &  multiplets. The signs of the terms are \\ & (+/-) for (white/black) nodes. \\ \hline
\end{tabular}
\caption{The dictionary connecting bipartite graphs on Riemann surfaces and BFTs for gauging 1. \label{tdic}
}
\end{center}
\end{table}

\smallskip

\subsubsection{Gauging 2}

Gauging 1 was motivated by both anomaly considerations and the analogy with theories with a known D-brane realization. However, our previous discussion makes it clear that the symmetries associated to internal faces are not the only ones that are automatically anomaly free. In fact, every closed path in the graph can be associated to an anomaly free symmetry. Those associated to linear combinations of faces are $U(N)$ symmetries. Other types of closed paths, such as the ones along the fundamental directions appearing when the underlying Riemann surface has genus greater than zero, correspond to $U(1)$ symmetries.\footnote{Whether some of these symmetries can be consistently promoted to be non-Abelian is an interesting question that deserves further study. Moreover, it is natural to address this question in the context of a more general study in which arbitrary ranks for all symmetries are considered.} In general, only a minimal set of independent closed paths has to be gauged. Considering this gauging gives rise to a new class of theories which we call BFT$_2$.

Gauging 2 extends gauging 1 by gauging some additional symmetries. While the quiver associated to gauging 1 still provides useful guidance, BFT$_2$'s are not standard quiver theories since chiral fields can be charged under more than two gauge symmetries.

Gauging 2 is the appropriate one for making full contact with the literature on leading singularities, in particular in the non-planar case. For this application it is natural to restrict to the Abelian case, i.e. when all symmetries are $U(1)$, as the following section is going to explain.

It is straightforward to see that the definition of Abelian BFT$_2$'s is actually independent of any embedding of the bipartite graph into a Riemann surface. In fact, an underlying Riemann surface becomes unnecessary for defining this type of theories. This is one of the main reasons why this class of BFTs is more naturally connected to scattering amplitudes, since both types of objects only care about the connectivity of the graph. However, removing the Riemann surface from the discussion needs to be taken with care, since it was not only used for identifying some of the gauge symmetries, but it was also necessary for providing nodes with an orientation that determines the chirality of fields. It is possible to define chirality without the need of a Riemann surface: one simply declares that for any gauge symmetry, the fields associated to edges alternate between being in the fundamental and antifundamental representations as one moves along the corresponding closed path.\footnote{An analogous chirality assignment is also possible for global symmetries associated to open paths in the graph.}

\bigskip
\bigskip

\begin{figure}[h]
\begin{center}
\includegraphics[width=9.5cm]{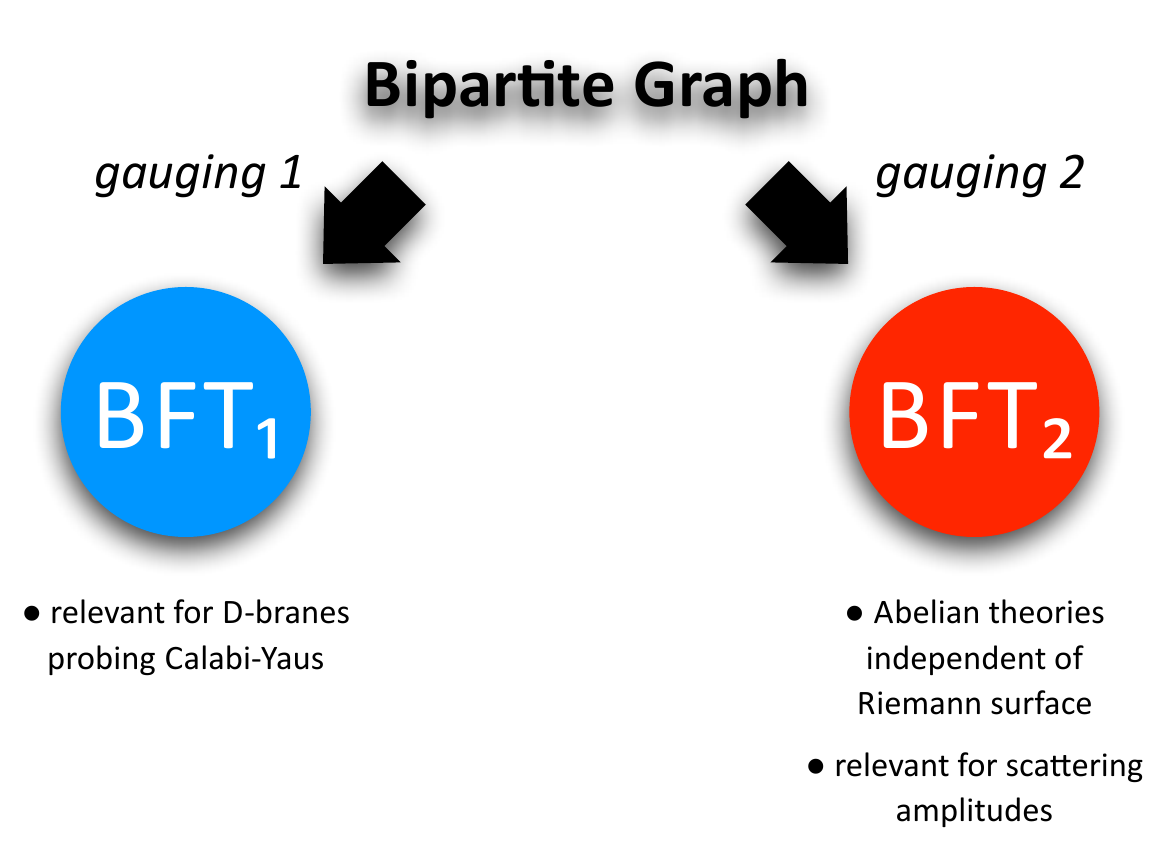}
\caption{\textit{Two types of gaugings in BFTs.} Different gaugings lead to two classes of gauge theories associated to bipartite graphs. Abelian BFT$_2$ theories do not require an embedding of the bipartite graph into a Riemann surface.}
\label{BFTs_1_and_2}
\end{center}
\end{figure}

Gaugings 1 and 2 coincide for planar graphs but they give rise to different gauge theories in the non-planar case. Each alternative has natural applications and certainly deserves independent investigation. \fref{BFTs_1_and_2} summarizes the two classes of BFTs and highlights some areas of applicability for each of them. 

Both BFT$_1$'s and BFT$_2$'s can be analyzed with exactly the same tools. For concreteness, our discussion in \sref{section_infinite_families} to \sref{section_Seiberg} focuses on gauging 1. We hope the reader keeps this choice in mind, since we are not going to constantly refer to it. \sref{section_gauging_2} collects various results illustrating the main changes that arise when considering gauging 2. Whenever we do not refer to any specific gauging in other sections, it means our discussion applies to both sets of theories with the corresponding changes.

\bigskip

\section{Moduli Spaces}

\label{section_moduli_spaces}

BFTs as $4d$ $\mathcal{N}=1$ theories have two classical moduli spaces known as the {\bf master space} $\mathcal{F}^\flat$ \cite{Forcella:2008bb,Forcella:2008eh,Hanany:2010zz} and the {\bf mesonic moduli space} $\mesonic$ \cite{Feng:2002zw,Hanany:2005ve,Franco:2005rj,Benvenuti:2006qr}, which we refer to just as the {\bf moduli space} for brevity. As a first step towards a full investigation of BFTs, this paper focuses on the case where $N=1$, i.e. all symmetries are $U(1)$. This simplification has various motivations. First, Abelian BFTs are relevant for the study of scattering amplitudes, which do not contain any parameter related to a non-trivial $N$. In fact, the scattering problem can be mapped to a $U(1)$ gauge theory living on the graph \cite{Nima} which, in turn, is directly related to Abelian BFTs. This correspondence was studied for graphs on $T^2$ in \cite{2003math.ph..11005K,Franco:2006gc}. Furthermore, although the confining dynamics and Seiberg duality that are discussed in \sref{section_moves} and \sref{section_Seiberg} respectively only occur for $N>1$ , the corresponding graph moves imply the invariance of the Abelian moduli space. Turning this around, the coincidence of the Abelian moduli space of two BFTs is a necessary condition for the corresponding non-Abelian theories to be related by confinement and duality. 

While in some cases, such as BFTs arising on stacks of D-branes, the moduli space of the non-Abelian theory is a symmetrized product of $N$ copies of the Abelian one, a simple connection of this type need not hold for generic BFTs. Elucidating the structure of the moduli space of non-Abelian BFTs is a very interesting question that certainly deserves to be studied in the future. We envision powerful tools such as those based on Hilbert series are going to be useful for this endeavor \cite{Hanany:2007zz,Butti:2007jv,Feng:2007ur,Benvenuti:2006qr}.

Following the arguments above, all our discussions of moduli spaces in the following sections are going to refer to the Abelian theories. Due to the restricted structure of BFTs arising from their definition in terms of bipartite graphs on Riemann surfaces, both the master and moduli spaces are toric Calabi-Yau manifolds \cite{Franco:2012mm}. This section reviews the definition of the moduli spaces in terms of F and D-term constraints and discusses how they can be expressed as symplectic quotients, with a parameterization in terms of gauged linear sigma model (GLSM) fields. These GLSM fields can be identified with perfect matchings of the bipartite graph, which are first reviewed.

\bigskip

\paragraph{Perfect Matchings \cite{Hanany:2005ve,Hanany:2005ss,Kennaway:2007tq,Forcella:2008bb}.}

Given a bipartite graph, an {\bf almost perfect matching} $p$ is a subset of the edges such that:

\begin{itemize}
\item Every internal node is the endpoint of exactly one edge in $p$.
\item Every external node belongs to either one or zero edges in $p$.
\end{itemize}

For brevity, they are going to be referred to as {\bf perfect matchings} in the following discussion. Perfect matchings are in one-to-one correspondence with GLSM fields \cite{Franco:2012mm}.\footnote{This is a generalization of what happens for the BFTs on $T^2$ associated to D3-branes over toric CY 3-folds \cite{Franco:2005rj,Franco:2006gc}.} Remarkably, a very efficient systematic procedure for their determination was introduced in \cite{Franco:2012mm}, which in turns makes the computation of moduli spaces for BFTs straightforward. This method is reviewed below.

\bigskip

\paragraph{Kasteleyn Matrix Technology \cite{Franco:2012mm}.}

The {\bf master Kasteleyn matrix} $K_0$, is an adjacency matrix of the graph in which rows are indexed by white nodes and columns are indexed by black nodes, i.e. for every edge in the bipartite graph between nodes ${\bf w}_\mu$ and ${\bf b}_\nu$, a contribution to the $K_{0,\mu \nu}$ entry is introduced. When more than one edge extends between the same pair of nodes, their contributions are added. $K_0$ has the general form

\beq
K_0 = \left(\begin{array}{c|c|c} & \ \ B_i \ \ & \ \ B_e \ \ \\ \hline
W_i & * & * \\ \hline
W_e & * & 0 \\
\end{array}
\right),
\eeq
where internal and external white nodes are denoted respectively by $W_i$ and $W_e$, and internal and external black nodes are denoted by $B_i$ and $B_e$. 

Notice that since in the presence of boundaries the number of white and black nodes might not be equal, $K_0$ generically needs not to be a square matrix. When it is square, its permanent is a polynomial in which every term corresponds to a perfect matching of the bipartite graph containing all the external nodes.\footnote{The permanent of a square matrix is the determinant with only positive signs.}

Given subsets $W_{e,del}\subseteq W_e$ and $B_{e,del}\subseteq B_e$ of the white and black external nodes, let us define the \textbf{reduced Kasteleyn matrix} as follows:
\beal{es1_11}
K_{(W_{e,del},B_{e,del})} &\equiv& \text{matrix obtained by deleting the rows in $W_{e,del}$}
\nn\\
&& \text{and the columns in $B_{e,del}$ from $K_0$.}
\eea
In analogy with $K_0$, if $K_{(W_e,B_e)}$ is a square matrix, its permanent is a polynomial encoding the perfect matchings containing all external legs except from those in the $W_{e,del}$ and $B_{e,del}$ deleted sets. 

We now have everything in order to determine all perfect matchings in the graph, which are encoded in the {\bf characteristic polynomial}
\beq
\mathcal{P}=\sum_{W_{e,del},B_{e,del}} \det K_{\left( W_{e,del},B_{e,del} \right)},
\label{generalized_Kasteleyn}
\eeq
where the sum runs over all possible subsets $W_{e,del}$ and $B_{e,del}$ of the external nodes (including the cases in which they are empty sets) such that the resulting reduced Kasteleyn matrices are square. Every term in the characteristic polynomial is interpreted as the product of edges in a perfect matching. 

The characteristic polynomial contains all the information relating edges, i.e. bifundamental fields, and perfect matchings. This information can be equivalently recast in terms of a $(e\times c)$-dimensional {\bf perfect matching matrix} $P$, where $e$ is the number of edges $X_{i}$ and $c$ is the number of perfect matchings $p_\alpha$. The components of the matrix are defined as follows 
\beal{es1_13}
P_{i\alpha} = \left \{ \ba{ll}
1 &~~\text{if $X_i\in p_\alpha$} \nn\\
0 &~~\text{if $X_i\notin p_\alpha$} 
 \ea \right.
\label{perfect_matching_matrix}
\eea
where $i=1,\dots,e$ and $\alpha=1,\dots,c$.

For BFTs, perfect matchings are in one-to-one correspondence with GLSM fields which were originally used by Witten in order to study $\mathcal{N}=(2,2)$ supersymmetric field theories \cite{Witten:1993yc}. The correspondence between GLSM fields and perfect matchings in the BFT context can be used to study the moduli spaces. Fayet-Iliopoulos (FI) terms are not going to play a crucial role for our discussion. 

\bigskip

\paragraph{Symplectic Quotient Description of the Moduli Spaces.}

As mentioned above, perfect matchings are used as GLSM fields in order to parameterize the moduli and master spaces of BFTs. By doing so, they can be described as symplectic quotients.

\bigskip

\begin{itemize}
\item \textbf{Master Space $\mathcal{F}^\flat$ \cite{Forcella:2008bb,Forcella:2008eh,Hanany:2010zz}.} F-term relations of the form $\partial_{X_i} W=0$ are encoded in the perfect matching matrix $P_{e\times c}$. Here, $X_i$ relate to internal edges in the bipartite graph. The F-term relations can be implemented by assigning the following charges to perfect matchings

\beal{es1_14}
Q_F = \ker(P_{c\times e}) \ .
\eea
As in \cite{Franco:2012mm}, we give a special treatment to the chiral fields associated to external legs, not imposing the vanishing of the corresponding F-terms. This choice is motivated by the connection with the Grassmannian for planar graphs. Furthermore, since these fields appear in a single superpotential term, imposing the vanishing of their F-terms would set to zero the product of fields they are coupled to. Finally, we expect this assumption can be dynamically explained in explicit D-brane realizations of BFTs. It is natural to envision that in such setups, these fields would arise at the intersection of flavor branes. Their higher dimensional support would then justify considering their expectation values to be non-dynamical parameters from the viewpoint of the lower dimensional BFT. The conclusions present further thoughts about possible D-brane realizations of BFTs for graphs with external legs.

The master space\footnote{Note: This is the equivalent to the \textbf{coherent component} of the master space, and not the full master space. The full master space usually decomposes into smaller irreducible spaces, most of them being $\mathbb{C}^l$. The coherent component is the largest irreducible subspace of the full master space.} is defined by the symplectic quotient
\beal{es1_15}
\mathcal{F}^\flat = \mathbb{C}^c // Q_F \ .
\eea

\medskip

\item \textbf{Mesonic Moduli Space $\mesonic$ \cite{Feng:2000mi,Feng:2002zw,Hanany:2005ve,Franco:2005rj}.} In order to construct the mesonic moduli space, the master space has to be projected onto gauge invariants. It is then useful to introduce the gauge charge matrix $d_{G\times E}$ of the BFT, where $G$ is the number of gauge groups and $E$ is the number of fields.\footnote{$G$ clearly depends on whether one considers gauging 1 or 2.} The elements of the gauge charge matrix are
\medskip
\beal{es1_20}
d_{aj} = 
\left\{ \ba{cl}
- 1 & ~~\text{if $X_j$ is fundamental to $U(N)_a$} \nn\\
+ 1 & ~~\text{if $X_j$ is anti-fundamental to $U(N)_a$} \nn\\
0 & ~~\text{if $X_j$ is adjoint or neutral under $U(N)_a$}
\ea\right.
\eea
\medskip

\noindent where $a=1,\dots,G$ and $j=1,\dots, E$. Note that the number of fundamental and antifundamental fields for every gauge group is the same due to anomaly cancellation.

Each gauge group contributes a D-term. D-terms can be encoded in a charge matrix $Q_{D}$, which is defined through the relation
\beal{es1_21}
d_{G \times E} = Q_{D,G \times c}.P^{t}_{c\times E} .
\eea

The mesonic moduli space $\mesonic$ is then defined as the following symplectic quotient
\beal{es1_22}
\mesonic = \mathbb{C}^c // Q_F // Q_D = \mathcal{F}^\flat // Q_D .
\eea
\end{itemize}
 
Both the mesonic and master spaces of BFTs are toric Calabi-Yau. The toric diagram of the mesonic moduli space is given by
\beal{es1_23}
G = \ker \left(\ba{c} Q_F \\ Q_D \ea \right) .
\eea
Each perfect matching is a point in the toric diagram of $\mesonic$. Columns in the $G$ matrix correspond to perfect matchings and contain the coordinates of the associated point in the toric diagram.

\bigskip

\section{BFT Perspective on Graph Equivalence and Reduction \label{section_equivalence_and_reduction}}

Based on the BFT interpretation of graphs it is possible to introduce a natural notion of {\bf graph equivalence}.\footnote{Following our general discussion in \sref{section_moduli_spaces}, throughout this paper we focus on the moduli space for the Abelian theory. For brevity, it is simply referred to as the moduli space.} We say that:

\begin{center}
\begin{tabular}{|c|}
\hline 
 Two graphs are equivalent if \\ the corresponding BFTs have the same moduli space. \\
 \hline
 \end{tabular}
\end{center}

\noindent Of course the equivalence classes resulting from this definition depend on the specific gauging under consideration. This idea was already advocated in \cite{Franco:2012mm}, after noting that the moduli space is a natural geometric object that remains invariant under certain class of moves and reductions that are reviewed in \sref{section_moves}. As explained in\sref{section_reduction_by_higgsing}, graph equivalence is more subtle and can include transformations beyond those of \sref{section_moves}.

Leading singularities are one of the main areas which play an important role when thinking about applications of BFTs. In this context, one needs to consider gauging 2 and it is possible to see that two graphs are equivalent if the corresponding leading singularities coincide. A general proof of the equality of the BFT moduli space and scattering approaches will be given elsewhere \cite{Nima}. 

\sref{section_reduction_by_higgsing_examples} presents examples illustrating how the moduli space is useful for even determining non-planar/non-planar and non-planar/planar equivalences. Furthermore, this notion of equivalence also applies without changes to graphs without external legs, i.e. those not associated to scattering.

It is also useful to introduce a notion of ordering among equivalent graphs. A natural prescription is to order them according to the number of closed paths along edges.\footnote{Here we mean the number of internal faces in gauging 1 or the actual number of independent closed paths along edges in gauging 2.} A graph is called {\bf reduced} if it has the minimum number of loops within a given equivalence class. From a BFT point of view, a reduced graph corresponds to a quiver with the minimal gauge symmetry. Clearly, reduced graphs in a given equivalence class are not unique, since they are defined up to equivalence moves. 

Reduced graphs are of particular interest. For example, they play a central role in the context of scattering, giving the simplest expressions for leading singularities \cite{Nima}.

There are two natural questions that arise in connection with graph equivalence and reducibility:

\medskip
\begin{itemize}
\item How can one identify efficiently whether two graphs are equivalent?
\item How can one determine whether the graph is reduced?
\end{itemize}
\medskip

\noindent These two questions have elegant answers in the case of planar graphs. In this case, graphs associated to the same permutation are equivalent \cite{Postnikov_plabic}.\footnote{In a graph with boundaries, a permutation of the external nodes is defined as follows. Given two external nodes $b_i$ and $b_j$, we say that $b_i$ is permuted to $b_j$ if they are the starting and ending points of a zig-zag path, respectively.} In addition, a graph is reducible if it contains self-intersecting zig-zag paths \cite{2003math.ph...5057K,Hanany:2005ss} or multiple intersections between different zig-zag paths. Whether some of these ideas can be generalized to non-planar graphs is an interesting question worth pursuing. In any case, it is extremely interesting to explore whether alternative approaches, which do not rely on zig-zag paths, exist. 

From the discussion above, it is clear that the moduli space of the associated BFT provides an ideal diagnostic for graph equivalence, which is physically intuitive and extends without modifications to non-planar graphs. We are going to see later that BFTs also provide efficient methods for determining graph reducibility.

\subsection{Moves and Bubble Reduction}

\label{section_moves}

\begin{figure}[h]
\begin{center}
\includegraphics[width=12cm]{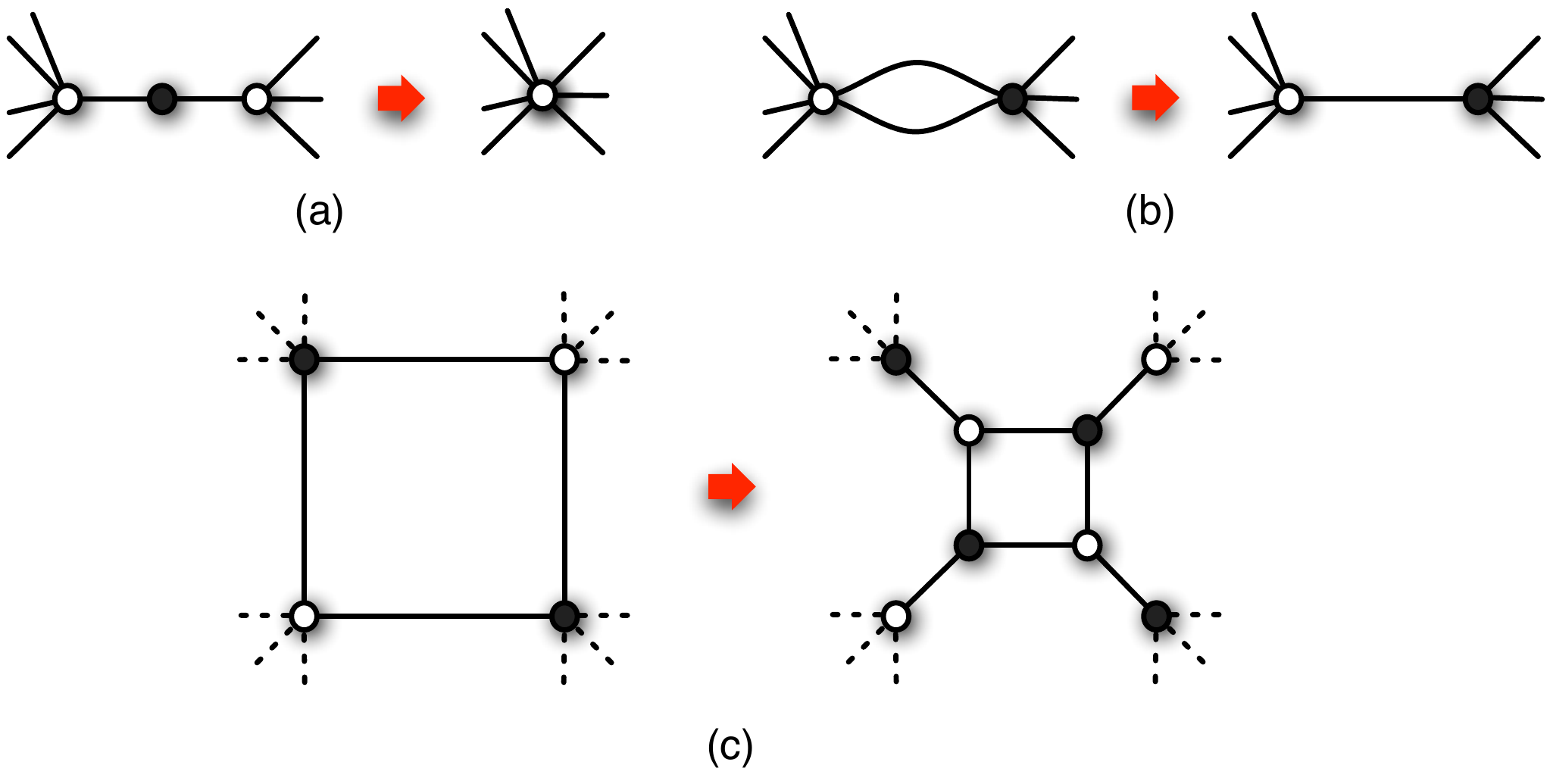}
\caption{\textit{Three basic transformations of a bipartite graph.} They correspond to: (a) integrating out massive fields, (b) confinement of an $N_f=N_c$ gauge group (bubble reduction) and (c) Seiberg duality on an $N_f=2N_c$ gauge group.}
\label{reduction_moves}
\end{center}
\end{figure}

\fref{reduction_moves} shows three basic transformations that can be applied to bipartite graphs. Their field theoretic interpretation has been discussed in \cite{Franco:2012mm,Xie:2012mr}, where the reader can find a detailed discussion. In summary, they correspond to:

\begin{itemize}
\item[{\bf (a)}] Integrating out massive fields. In some cases 2-valent nodes, i.e. mass terms, can appear on external legs of the graph. If this happens, we only integrate them out whenever this operation does not take us outside of the realm of graphs that define our theories as mentioned at the beginning of \sref{section_BFTs}, i.e. graphs in which external nodes are connected to a single edge. This poses no limitation on the theories that can be considered, since it is totally valid to consistently keep massive fields in their analysis.
\item[{\bf (b)}] Confinement of an $N_f=N_c$ gauge group, staying on a branch of moduli space on which mesons do not get expectation values.
\item[{\bf (c)}] Seiberg duality \cite{Feng:2000mi,Feng:2001xr,Feng:2002zw,Seiberg:1994pq,Feng:2001bn,2001JHEP...12..001B,Franco:2003ea} on an $N_f=2N_c$ gauge group.\footnote{The brane tiling transformation has many names: square move, urban renewal, and spider move.} Let us emphasize that this rule correctly describes Seiberg duality even for faces adjacent to external ones. There is no limitation of any sort in the type of $N_f=2N_c$ gauge groups that can be dualized. For general situations, it is only necessary to appropriately take into account the comments in point {\bf (a)}.
\end{itemize}

The {\bf (b)} and {\bf (c)} interpretation of moves require the theory to be non-Abelian, i.e. to have $N>1$. In any case, all these operations preserve the moduli space of the BFT even for $N=1$ and hence lead to equivalent graphs. Bubble reduction in {\bf (b)} decreases the number of loops by one, so it can be used for reducing graphs.

\bigskip

\subsection{Reduction by Higgsing}

\label{section_reduction_by_higgsing}

\begin{figure}[h]
\begin{center}
\includegraphics[width=8.5cm]{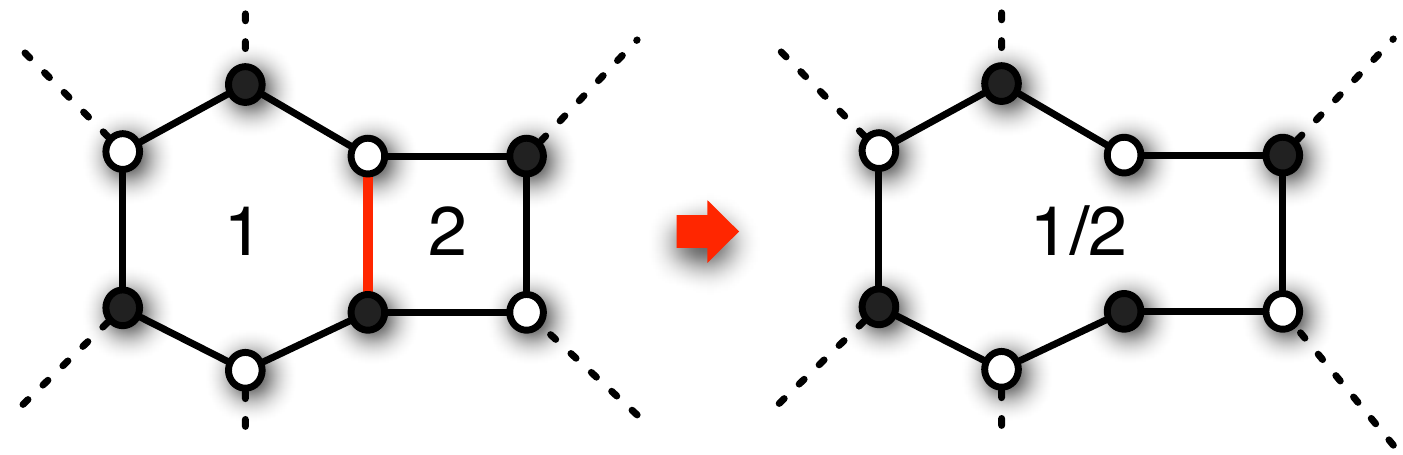}
\caption{\textit{Higgsing.} Removing an edge in the graph corresponds to turning on a vev for a bifundamental scalar, resulting in the merging of two faces.}
\label{tiling_higgsing}
\end{center}
\end{figure}

Edge removal, which has been discussed in detail in \cite{Franco:2012mm}, is another natural operation on graphs. \fref{tiling_higgsing} shows an example of this operation, after which the two original faces at both sides of the removed edge get combined into a single one. 

In the BFT, the deletion of an edge corresponds to giving a non-zero vev to the chiral field associated to the edge. When the removed edge is internal, the transformation corresponds to higgsing \cite{Feng:2002fv} in the BFT. Removing edges decreases the number of loops. In order for it to give rise to a reduced graph, it should also lead to a theory in the same equivalence class of the original one. Naively, this might seem counterintuitive since, in BFT language, it would correspond to a higgsing that preserves the moduli space. Elicit examples are going to be provided in order to show that this is indeed possible. The possibility of reducing graphs by removing edges was first discovered and investigated in the language of leading singularities in scattering amplitudes \cite{Nima}.

The practical implementation of this algorithm is straightforward. The first step is to determine the moduli space of the original theory, making a list of the perfect matchings associated to each point in its toric diagram. It is important to emphasize that one only needs to keep track of which perfect matchings belong to the same point in the toric diagram, while remembering the actual coordinates of these points is not necessary.

Following the map between perfect matchings and chiral fields given in \eref{perfect_matching_matrix}, deleting an edge associated to the field $X_i$, implies the elimination of all perfect matchings $p_\alpha$ with $P_{i\alpha}=1$; in other words all perfect matchings containing the edge under consideration. If after this process the moduli space remains invariant, i.e. if there still is at least one perfect matching for every point in the original toric diagram, it can be concluded that the higgsed theory is equivalent to the original one.

One can scan over all edges of the graph and determine whether they are individually removable. Iterating this process, it is possible to determine all combinations of edges that can be simultaneously removed. Reduced graphs are reached when deleting edges without eliminating points in the toric diagram is no longer feasible.

The procedure outlined above makes it possible to identify all combinations of vevs that produce reduced graphs. Some of these sets of vevs can lead to different reduced graphs. Whenever this happens, the original graph has multiple reductions. This phenomenon is a manifestation of having multiple leading singularities. An attractive feature of the BFT approach is that multiple reductions can be systematically identified. 
\\

\begin{figure}[h]
\begin{center}
\includegraphics[width=5cm]{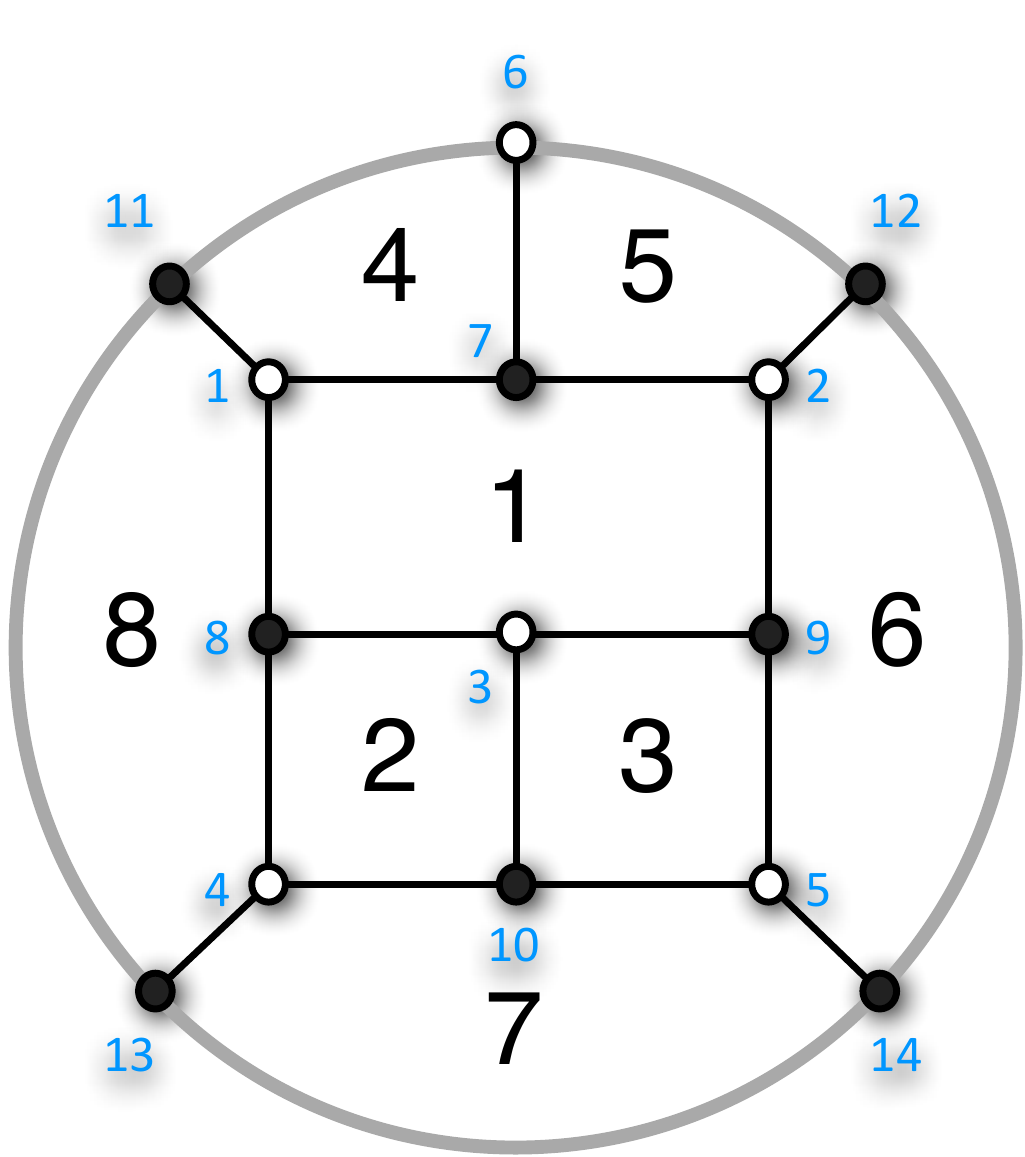}
\caption{\textit{An example of a planar reducible graph.}}
\label{tiling_reduction_higgsing_example}
\end{center}
\end{figure}

\noindent\underline{\textit{Example.}}
Let us illustrate these ideas with an explicit example. Consider the graph shown in \fref{tiling_reduction_higgsing_example}. It can be analyzed using the techniques discussed in \sref{section_reduction_by_higgsing}.

For future reference, we quote some of the intermediate details of the calculation. The Master Kasteleyn matrix is

{\footnotesize
\beq
K_0 =
\left(
\begin{array}{c|cccc|cccc}
 & \ \ 7 \ \ & \ \ 8 \ \ & \ \ 9 \ \ & \ \ 10 \ \ & \ \ 11 \ \ & \ \ 12 \ \ & \ \ 13 \ \ & \ \ 14 \ \ \\ \hline
\ 1 \ & \ X_{41} \ & \ X_{18} \ & 0 & 0 & \ X_{84} \ & 0 & 0 & 0 \\
2 & X_{15} & 0 & \ X_{61} \ & 0 & 0 & X_{56} & 0 & 0 \\
3 & 0 & X_{21} & X_{13} & \ X_{32} \ & 0 & 0 & 0 & 0 \\
4 & 0 & X_{82} & 0 & X_{27} & 0 & 0 & \ X_{78} \ & 0 \\
5 & 0 & 0 & X_{36} & X_{73} & 0 & 0 & 0 & \ X_{67} \ \\ \hline
6 & X_{54} & 0 & 0 & 0 & 0 & 0 & 0 & 0 
\end{array}
\right).
\eeq}
One can then determine the perfect matching matrix giving the translation between chiral fields and perfect matchings, which becomes:

{\scriptsize
\beq
P=\left(
\begin{array}{c|cccccccccccccccccccccc}
& \ p_1 \ & \ p_2 \ & \ p_3 \ & \ p_4 \ & \ p_5 \ & \ p_6 \ & \ p_7 \ & \ p_8 \ & \ p_9 \ & p_{10} & p_{11} & p_{12} & p_{13} & p_{14} & p_{15} & p_{16} & p_{17} & p_{18} & p_{19} & p_{20} & p_{21} & p_{22} \\ \hline
X_{21} & 1 & 1 & 1 & 1 & 1 & 1 & 1 & 0 & 0 & 0 & 0 & 0 & 0 & 0 & 0 & 0 & 0 & 0 & 0 & 0 & 0 & 0 \\
X_{27} & 1 & 1 & 1 & 1 & 1 & 0 & 0 & 1 & 1 & 0 & 0 & 0 & 0 & 0 & 0 & 0 & 0 & 0 & 0 & 0 & 0 & 0 \\
X_{36} & 1 & 1 & 1 & 0 & 0 & 0 & 0 & 0 & 0 & 1 & 1 & 1 & 1 & 1 & 0 & 0 & 0 & 0 & 0 & 0 & 0 & 0 \\
X_{41} & 1 & 0 & 0 & 1 & 0 & 1 & 0 & 0 & 0 & 1 & 0 & 0 & 0 & 0 & 1 & 1 & 0 & 0 & 0 & 0 & 0 & 0 \\
X_{56} & 1 & 1 & 0 & 0 & 0 & 0 & 0 & 1 & 0 & 1 & 1 & 1 & 0 & 0 & 1 & 0 & 1 & 1 & 0 & 0 & 0 & 0 \\
X_{13} & 0 & 0 & 0 & 0 & 0 & 0 & 0 & 1 & 1 & 0 & 0 & 0 & 0 & 0 & 1 & 0 & 1 & 1 & 1 & 1 & 0 & 0 \\
X_{15} & 0 & 0 & 1 & 0 & 0 & 0 & 0 & 0 & 1 & 0 & 0 & 0 & 1 & 1 & 0 & 0 & 0 & 0 & 1 & 1 & 0 & 0 \\
X_{18} & 0 & 0 & 0 & 0 & 0 & 0 & 0 & 1 & 1 & 0 & 1 & 0 & 1 & 0 & 0 & 0 & 1 & 0 & 1 & 0 & 1 & 0 \\
X_{67} & 0 & 0 & 0 & 1 & 1 & 0 & 0 & 1 & 1 & 0 & 0 & 0 & 0 & 0 & 0 & 1 & 0 & 0 & 0 & 0 & 1 & 1 \\
X_{54} & 0 & 1 & 0 & 0 & 1 & 0 & 1 & 1 & 0 & 0 & 1 & 1 & 0 & 0 & 0 & 0 & 1 & 1 & 0 & 0 & 1 & 1 \\
X_{61} & 0 & 0 & 0 & 1 & 1 & 1 & 1 & 0 & 0 & 0 & 0 & 0 & 0 & 0 & 0 & 1 & 0 & 0 & 0 & 0 & 1 & 1 \\
X_{32} & 0 & 0 & 0 & 0 & 0 & 0 & 0 & 0 & 0 & 1 & 1 & 1 & 1 & 1 & 0 & 1 & 0 & 0 & 0 & 0 & 1 & 1 \\
X_{78} & 0 & 0 & 0 & 0 & 0 & 1 & 1 & 0 & 0 & 0 & 1 & 0 & 1 & 0 & 0 & 0 & 1 & 0 & 1 & 0 & 1 & 0 \\
X_{73} & 0 & 0 & 0 & 0 & 0 & 1 & 1 & 0 & 0 & 0 & 0 & 0 & 0 & 0 & 1 & 0 & 1 & 1 & 1 & 1 & 0 & 0 \\
X_{82} & 0 & 0 & 0 & 0 & 0 & 0 & 0 & 0 & 0 & 1 & 0 & 1 & 0 & 1 & 1 & 1 & 0 & 1 & 0 & 1 & 0 & 1 \\
X_{84} & 0 & 1 & 1 & 0 & 1 & 0 & 1 & 0 & 0 & 0 & 0 & 1 & 0 & 1 & 0 & 0 & 0 & 1 & 0 & 1 & 0 & 1 \\
\end{array}
\right) .
\label{P_matrix_multiplicities_3_loops_5_legs}
\eeq}

Under vanishing D-terms for gauge groups 1, 2 and 3, the moduli space is a 5d toric CY. The 22 perfect matchings form a toric diagram consisting of 10 points given by the following matrix:

{\footnotesize
\beq
G=\left(
\begin{array}{C{0.35cm}C{0.35cm}C{0.35cm}C{0.35cm}C{0.35cm}C{0.35cm}C{0.35cm}C{0.35cm}C{0.35cm}C{0.35cm}}
 -1 & 0 & 0 & 0 & -1 & 1 & -1 & 0 & 0 & 0 \\
 0 & 0 & 0 & 0 & 1 & 0 & 1 & 1 & 0 & 1 \\
 0 & 0 & 1 & 0 & 1 & 0 & 0 & 1 & -1 & 0 \\
 1 & 1 & 0 & 0 & 0 & 0 & 1 & 0 & 1 & 0 \\
 1 & 0 & 0 & 1 & 0 & 0 & 0 & -1 & 1 & 0 \\
\hline
\bf{3} & \bf{3} & \bf{3} & \bf{3} & \bf{3} & \bf{2} & \bf{2} & \bf{1} & \bf{1} & \bf{1} \\
\hline
\end{array}\right),
\label{G_matrix_multiplicities_3_loops_5_legs}
\eeq}
where the last row summarizes the perfect matching multiplicity for each point in the toric diagram. The 10 points in the toric diagram correspond to the following sets of perfect matchings
\beq
\begin{array}{c}
\{ p_{1}, p_{10}, p_{15} \} \ , \ \{ p_{2}, p_{12}, p_{18} \} \ , \ \{ p_{3}, p_{14}, p_{20} \} \ , \ \{ p_{4}, p_{9}, p_{16} \} \ , \ \{ p_{6}, p_{13}, p_{19} \} \\
\{ p_{5}, p_{22} \} \ , \ \{ p_{11}, p_{17} \} \\
\{ p_{7} \} \ , \ \{ p_{8} \} \ , \ \{ p_{21} \}
\end{array}
\eeq
where the numbering in \eref{P_matrix_multiplicities_3_loops_5_legs} is used.

\begin{figure}[h]
 \centering
 \begin{tabular}[c]{ccc}
\epsfig{file=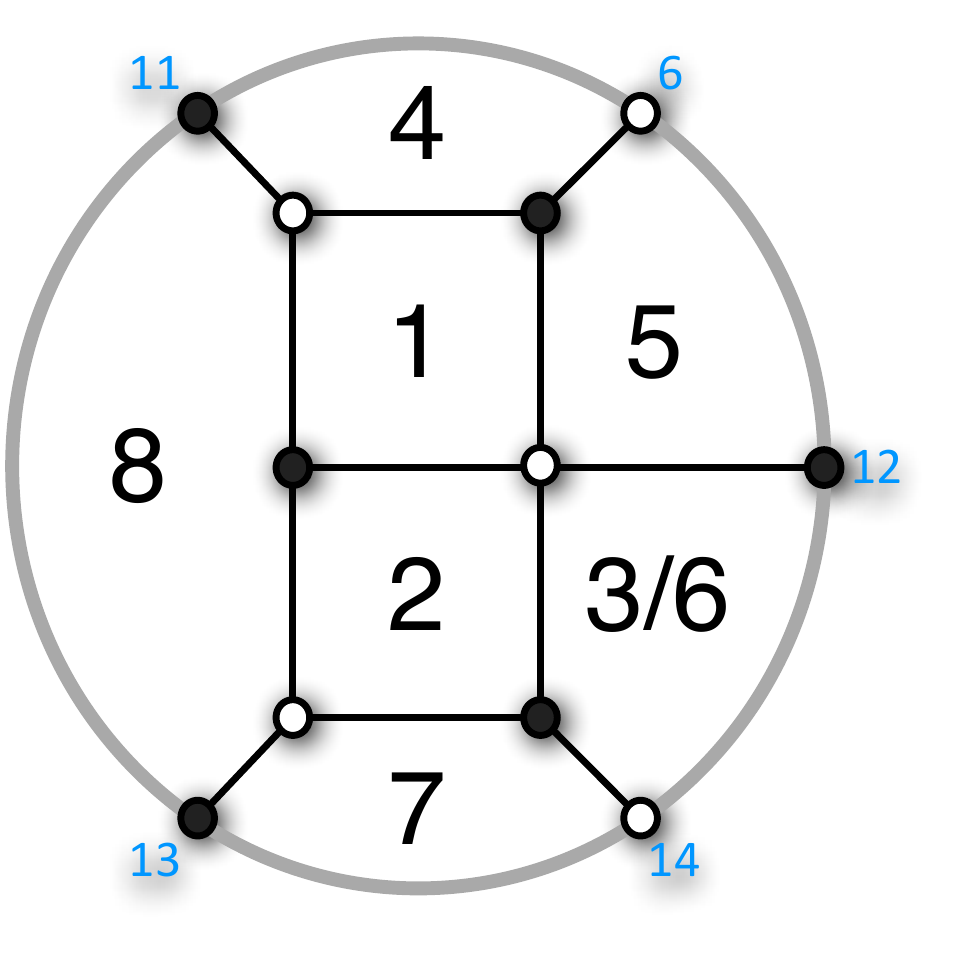,width=0.25\linewidth,clip=} & \ \ \ \ \ &
\epsfig{file=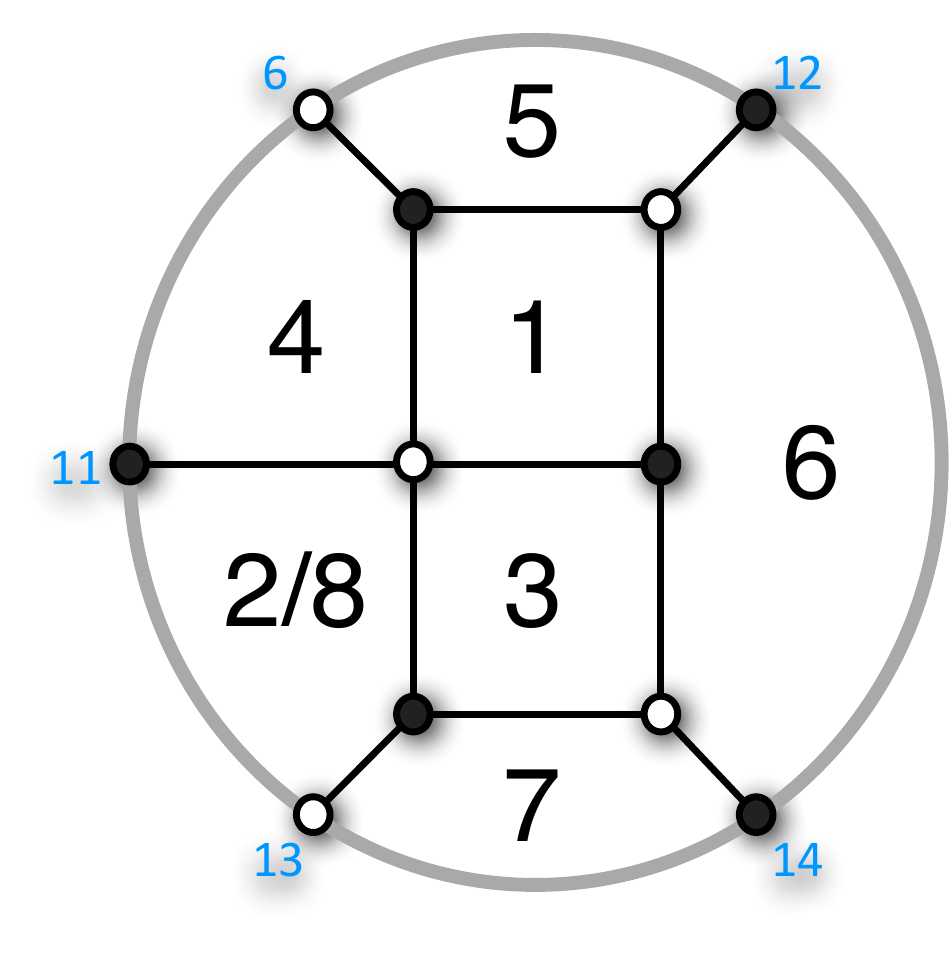,width=0.25\linewidth,clip=} \\ 
\hspace{-0.45cm}\mbox{(a)} &  \ \ \ \ \ & \hspace{0.45cm}\mbox{(b)}
 \end{tabular}
\caption{\textit{Reductions of the theory in \fref{P_matrix_multiplicities_3_loops_5_legs} obtained by higgsing.} The fields acquiring a non-zero vev are: $X_{36}$ for (a) and $X_{82}$ for (b).}
\label{tiling_reduction_higgsing_example_results} 
\end{figure}

Only two fields can independently get vevs without deleting any point in the toric diagram. They are $X_{36}$ (which removes $p_1$, $p_2$, $p_3$, $p_{10}$, $p_{11}$, $p_{12}$, $p_{13}$ and $p_{14}$) and $X_{82}$ (which removes $p_{10}$, $p_{12}$, $p_{14}$, $p_{15}$, $p_{16}$, $p_{18}$, $p_{20}$ and $p_{21}$). The resulting graphs are shown in \fref{tiling_reduction_higgsing_example_results}. It is straightforward to verify that these graphs are reduced, since it is impossible to turn on a second vev without eliminating some of the points in the toric diagram.

In this example, the reduced graphs can alternatively be reached by the moves and bubble reduction discussed in \sref{section_moves}, starting from Seiberg dualizing either face 2 or 3. The full scope of reductions by higgsing is going to be investigated in \sref{section_reduction_by_higgsing_examples}. In some cases higgsing produces reductions that cannot be achieved by any move or bubble reduction and that it can also reduce non-planar graphs to planar ones.

\bigskip

\section{Geometry from Gauge Theory: an Alternative Approach}

\label{section_geometry_from_BFT}

\sref{section_moduli_spaces} discussed how the master and moduli spaces of a BFT are parametrized in terms of perfect matchings, and explained how to determine the resulting geometry; in other words how to find the positions of perfect matchings in the corresponding toric diagram. The constraints following from F and D-term equations can be implemented by assigning charges to perfect matchings. While the methods in \sref{section_moduli_spaces} are used in explicit examples throughout the paper, this section introduces an alternative procedure for finding toric diagrams of moduli spaces, which provides additional intuition.

\bigskip

\subsection{Master Space \label{ssmaster}}

To every perfect matching one can associate an oriented path in the graph, which is given by its difference with a reference perfect matching $p_0$. The choice of $p_0$ is not important, since different choices correspond to overall modular transformations of the toric diagram.  The resulting paths can be expressed in terms of a basis, for which a convenient choice is given by:

\medskip

\begin{itemize}

\item {\bf Faces:} A variable $w_i$, $i=1,\ldots, F$, is considered for each path going clockwise around a face, either internal or external. Face variables are subject to the constraint
\beq
\prod_{i=1}^F w_i= 1. \nonumber
\eeq
As a result, one of the face variables can always be regarded as redundant such that it is expressed in terms of other $w$'s. For concreteness, in cases with boundaries the variable associated to one of the external faces is chosen to be discarded.

\item {\bf Fundamental cycles:} There are $\alpha_i$ and $\beta_i$ pairs of variables, $i=1,\ldots g$, that are associated to the fundamental cycles in the genus $g$ Riemann surface $\Sigma$.

\item {\bf Boundaries:} For a number of boundaries $B\geq 1$, one needs to include paths connecting the different boundary components. This can be achieved with $B-1$ paths, which are called $b_i$, $i=1,\ldots, B-1$. The specific choice of these $B-1$ representative paths is unimportant. 

\end{itemize}

\medskip

The coordinates in this basis for the path associated to each perfect matching give the position of the corresponding point in the toric diagram of the master space. This is after projection to one lower dimension by using the Calabi-Yau condition, which forces all points to lie on a hyperplane at unit distance from the origin. The dimension of the master space is then given by the number of paths in the basis plus one, which becomes

\beq
\begin{array}{ccccccl}
B & \neq & 0: & \ \ \ \ \ & d_{master} & = & F+B+2g-1 \\ 
B & = & 0: & \ \ \ \ \ & d_{master} & = & F+2g 
\end{array}
\eeq
where we have distinguished the cases with and without boundaries. The coordinates defined above fully distinguish different perfect matchings where every point in the toric diagram of the master space corresponds to a single perfect matching. 
\\

\begin{figure}[h]
\begin{center}
\includegraphics[width=3.7cm]{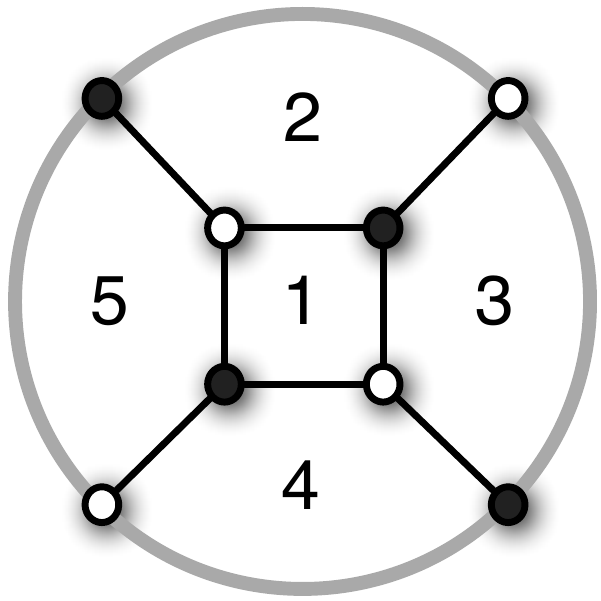}
\caption{\textit{A planar graph defining a BFT.}}
\label{tiling_square_4legs}
\end{center}
\end{figure}

\begin{figure}[h]
\begin{center}
\includegraphics[width=11cm]{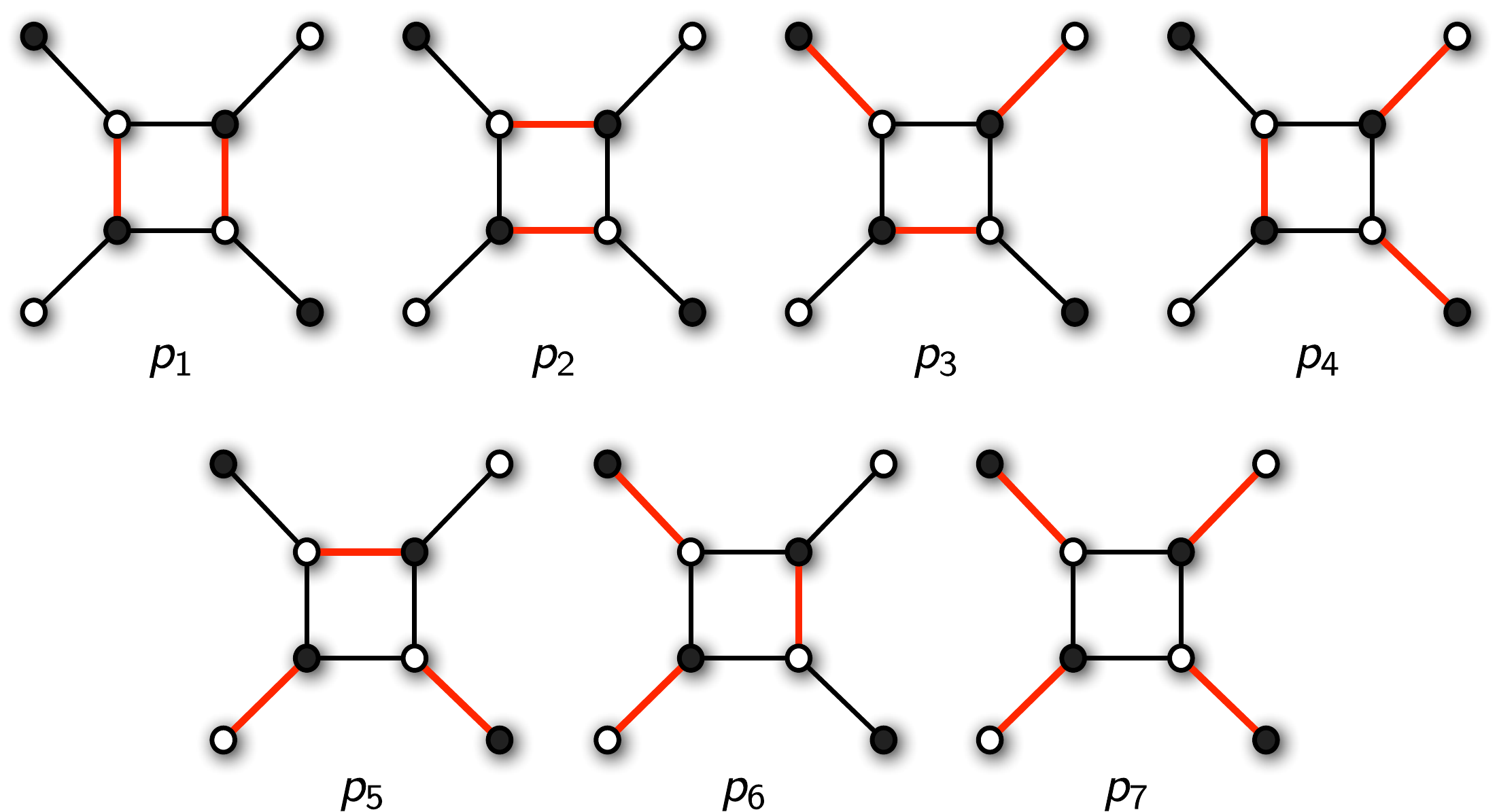}
\caption{\textit{The seven perfect matchings for the BFT in \fref{tiling_square_4legs}.} Edges in the perfect matchings are indicated in red.}
\label{pms_square_4legs}
\end{center}
\end{figure}

\noindent\underline{\textit{Example.}}
Let us illustrate this procedure with an example. Consider the bipartite graph shown in \fref{tiling_square_4legs}. This model has one internal face $w_1$ and four external faces $w_2$, $w_3$, $w_4$ and $w_5$. One can use $\prod_{i=1}^5 w_i = 1$ to eliminate $w_2$ from all expressions, by setting $w_2=w_1^{-1} w_3^{-1} w_4^{-1} w_5^{-1}$. The corresponding perfect matchings are shown in \fref{pms_square_4legs}.

\begin{figure}[h]
\begin{center}
\includegraphics[width=11cm]{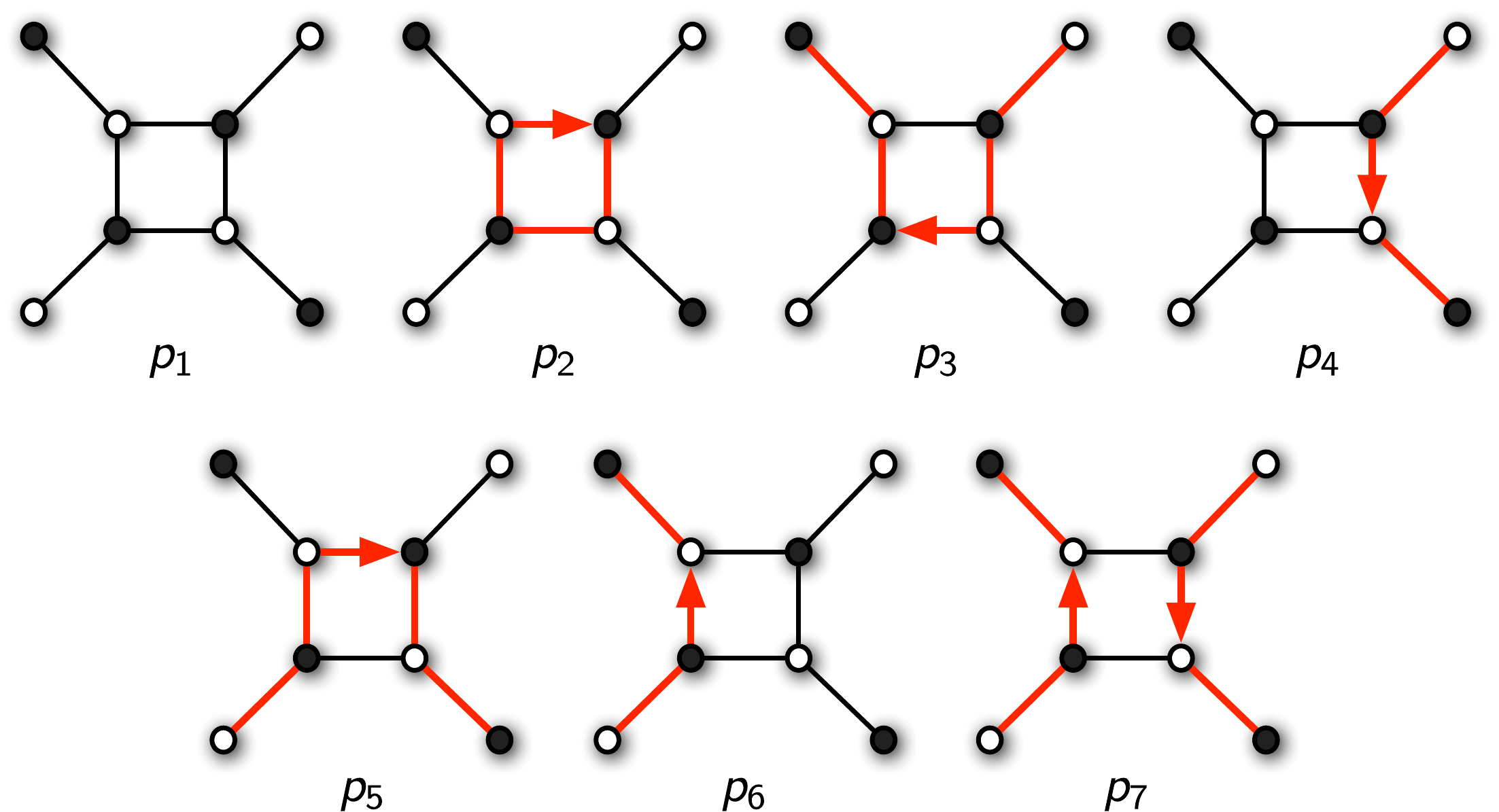}
\caption{\textit{Paths and Perfect Matchings.} These are the paths in the graph obtained by subtracting the reference perfect matching $p_1$ from the perfect matchings in \fref{pms_square_4legs}.}
\label{paths_square_4legs}
\end{center}
\end{figure}

Taking $p_1$ as the reference, the paths shown in \fref{paths_square_4legs} are obtained.

\beq
\begin{array}{|c|c|c|}
\hline
& {\rm Path} & \ {\rm Coordinates:~}(w_1,w_3,w_4,w_5) \ \\ \hline \hline
\ \ \ p_1 \ \ \ & 1 & (0,0,0,0) \\ 
p_2 & w_1 & (1,0,0,0) \\
p_3 & \ \ w_3^{-1} w_4^{-1} w_5^{-1} \ \ & (0,-1,-1,-1) \\
p_4 & w_3^{-1} & (0,-1,0,0) \\
p_5 & w_1 w_4 & (1,0,1,0) \\
p_6 & w_5^{-1} & (0,0,0,-1) \\
p_7 & w_3^{-1} w_5^{-1} & (0,-1,0,-1) \\
\hline
\end{array}
\eeq

\medskip

\noindent The table above gives the coordinates of points in the toric diagram of the 5-dimensional master space.

\bigskip

\subsection{Moduli Space}

\label{section_moduli_space_loops}

Going from the master to the moduli space corresponds to demanding invariance under all gauge symmetries. The discussion in this section is specialized for gauging 1, in which gauge symmetries correspond to internal faces of the graph. Extending it to gauging 2 is straightforward and simply amounts to requiring further invariance under additional gauge symmetries. In terms of the procedure introduced in this section, this projection simply amounts to dropping the coordinates associated to independent internal $w_i$'s. Once again, the resulting coordinates correspond to the projection of the toric diagram of the moduli space to one dimension less using the CY condition. The dimension of the moduli space is then

\beq
\begin{array}{ccccccl}
B & \neq & 0: & \ \ \ \ \ & d_{moduli} & = & F_e+B+2g-1 \\ 
B & = & 0: & \ \ \ \ \ & d_{moduli} & = & 2g+1 
\end{array}
\eeq
Notice that, while all internal faces are independent for $B \neq 0$, only $F-1$ of them are independent for $B=0$. 
\\

\noindent\underline{\textit{Example.}}
Returning to the example in \sref{ssmaster}, the toric diagram of the moduli space is obtained by dropping the $w_1$ coordinate, after which one obtains

\beq
\begin{array}{|c|c|c|}
\hline
& {\rm Path} & \ {\rm Coordinates:~}(w_3,w_4,w_5) \ \\ \hline \hline
\ \ \ p_1 \ \ \ & 1 & (0,0,0) \\ 
p_2 & w_1 & (0,0,0) \\
p_3 & \ \ w_3^{-1} w_4^{-1} w_5^{-1} \ \ & (-1,-1,-1) \\
p_4 & w_3^{-1} & (-1,0,0) \\
p_5 & w_1 w_4 & (0,1,0) \\
p_6 & w_5^{-1} & (0,0,-1) \\
p_7 & w_3^{-1} w_5^{-1} & (-1,0,-1) \\
\hline
\end{array}
\eeq

\bigskip

This example exhibits a well-known phenomenon \cite{Hanany:2005ve,Franco:2005rj,Franco:2005sm,Kennaway:2007tq,2007arXiv0710.1898I} which has been discussed in the context of general BFTs in \cite{Franco:2012mm}. Single points in the toric diagram of the moduli space can correspond to multiple perfect matchings. The discussion in this section provides an intuitive understanding of the origin of such multiplicities. If the loops $(p-p_0)$ and $(p'-p_0)$ associated to two different perfect matchings $p$ and $p'$ differ only by internal faces of the graph, then they map to the same point in the toric diagram of the moduli space. Equivalently, this happens when $(p-p')$ can be expressed solely in terms of internal face variables. In the example, $p_2-p_1 = w_1$, and this difference disappears when projecting down to the moduli space.

\bigskip

\subsection{On the Relation between the Moduli Space and the Matroid Polytope}

The toric diagram of the BFT master space precisely coincides with a polytope which is known as the \textbf{matching polytope} introduced in \cite{Postnikov_toric}. This does not restrict to the planar case, but generalizes also to the non-planar case. The approach for computing master and moduli spaces discussed in the previous section is particularly suitable for elucidating the relation between the moduli space of the BFT and the {\bf matroid polytope}. This is another construction appearing in the mathematical literature for the study of planar graphs \cite{Postnikov_toric}, which is discussed in the remainder of this section. 

Let us begin by briefly reviewing the definition of matroid polytopes in the specific context of planar graphs. In combinatorics, a matroid generalizes the concept of linear independence in vector spaces. In the following discussion we restrict to its explicit incarnation for bipartite graphs. Perfect matchings are in one-to-one correspondence with {\bf perfect orientations}, which are flows in the graph such that there are two outgoing and one incoming arrows at each internal white node and two incoming and one outgoing arrows at each internal black node. The reader is referred to \cite{Postnikov_plabic} for details on the map between perfect matchings and perfect orientations (see also \cite{Franco:2012mm} for a review). Given a perfect orientation, its {\bf source set} is defined as the set of external nodes that source arrows coming into the graph. Two perfect matchings give rise to perfect orientations with identical source sets if their difference is an internal closed loop. Furthermore, in order for two perfect matchings to differ by an internal loop, their external leg content must be the same. 
We conclude that the source sets $s_\mu$ are in one-to-one correspondence with perfect matchings, considered modulo internal edges. This identification with perfect matchings is very useful for practical applications.

One can now define the matroid polytope, which is encoded in an $n_{legs}\times n_{source}$ matrix $Q$, whose definition is similar to the perfect matching matrix $P$ for the matching polytope. Denoting the external edges by $X^{e}_i$ and the source sets by $s_\mu$, we have
\beq
Q_{i\mu}=\left\{ \begin{array}{ccccc} 1 & \rm{ if } & X^{(e)}_i  & \in & s_\mu \\
0 & \rm{ if } & X^{(e)}_i  & \notin & s_\mu
\end{array}\right.
\label{definition_matroid_polytope}
\eeq
\\

\noindent\underline{\textit{Example.}}
Let us apply this definition to the example from the sections above. The perfect matchings $p_1$ and $p_2$ coincide over external legs, so they correspond to the same source set $s_1$. Source sets are labelled according to
\beq
\begin{array}{rclcrcl}
p_1, p_2 & \to & s_1 & \ \ \ \ & p_5 & \to & s_4 \\
p_3 & \to & s_2 & \ \ \ \ & p_6 & \to & s_5 \\
p_4 & \to & s_3 & \ \ \ \ & p_7 & \to & s_6 
\end{array}
\eeq
Applying \eref{definition_matroid_polytope}, one obtains
{\small
\beq
Q=\left(
\begin{array}{c|cccccc}
& \ s_1 \ & \ s_2 \ & \ s_3 \ & \ s_4 \ & \ s_5 \ & \ s_6 \ \\ \hline
X_{52} \ & 0& 1& 0& 0& 1& 1 \\
X_{32} \ & 0& 1& 1& 0& 0& 1  \\
X_{34} \ & 0& 0& 1& 1& 0& 1 \\
X_{54} \ & 0& 0& 0& 1& 1& 1  
\end{array}
\right).
\eeq}

In general, the basic structure of the matching polytope and matroid polytope matrices $P$ and $Q$ can be summarized as follows:

\beq
\begin{array}{ccc}
\mbox{\underline{{\bf Matching Polytope P}}} & & \mbox{\underline{{\bf Matroid Polytope Q}}} \\ \\
\begin{array}{cc}
\begin{array}{c} \\ \mbox{edges} \end{array} &

\begin{array}{c}\ \ \ \mbox{perfect matchings} \\ \left\updownarrow\begin{array}{cc} \ & \overleftrightarrow{\left(
\begin{array}{cccccc}
* & * & * & * & * & * \\
* & * & * & * & * & * \\
* & * & * & * & * & * \\
* & * & * & * & * & * 
\end{array}\right)}\end{array}\right.\end{array}\end{array}
& \ \ \
\underrightarrow{\Psi}
\ \ \ &
\begin{array}{cc}
\begin{array}{c} \\ \mbox{external}\\ \mbox{legs} \end{array} &

\begin{array}{c}\ \ \ \mbox{source sets} \\ \left\updownarrow\begin{array}{cc} \ & \overleftrightarrow{\left(
\begin{array}{cccc}
* & * & * & * \\
* & * & * & * \\
* & * & * & * 
\end{array}\right)}\end{array}\right.\end{array}\end{array}
\end{array}
\eeq

\bigskip
\noindent The projection $\Psi$ taking from $P$ to $Q$ acts on rows by keeping only those associated to external legs and on columns by identifying perfect matchings that differ by closed loops. The alert reader might notice that this projection is very similar to the one discussed in \sref{section_moduli_space_loops}, connecting the master and moduli spaces. 

At this point one can conclude that, for BFTs associated to planar graphs, the toric diagram of the moduli space and the matroid polytope are, at the very least, extremely similar constructions that can alternatively be used for addressing the same questions. The following subsection explains that the two objects indeed coincide. The concept of moduli space is more physically motivated and, in the case of theories with a microscopic realization in string theory, is directly linked to the geometry probed by stacks of D-branes. Furthermore, the range of applicability of the moduli space is far more general. The moduli space is defined for generic BFTs, including theories that are non-planar or even without boundaries. 

\bigskip

\subsubsection{The Equivalence}

Below it is shown how the toric diagram of the moduli space of a BFT associated to a planar graph is bijectively related to the matroid polytope. Provided the color of external nodes is given, it is possible to explicitly construct the map between the two objects. The following discussion assumes that the graph has internal lines. The case where all lines are external is going to be discussed at the end of this section. 

\begin{figure}[h]
\begin{center}
\includegraphics[width=6cm]{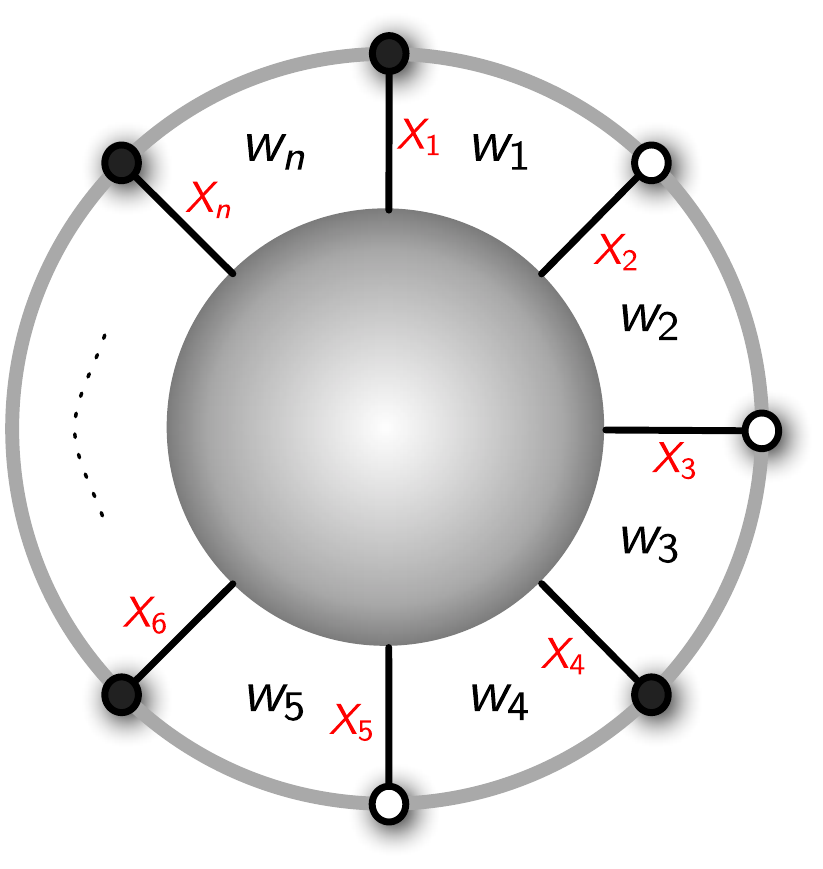}
\caption{\textit{General graph on a disk.} The striped blob at the center of the graph is an arbitrary connection between the external edges. The external edges and faces are labeled in blue.}
\label{matroidDisk}
\end{center}
\end{figure}

External edges of a disk can be numbered in a clockwise fashion, and external faces can be analogously cyclically numbered: the external face between edge $ X_i $ and $ X_{i+1} $ is labeled $w_i$. This is schematically drawn in \fref{matroidDisk}, from which it is clear that there is a one-to-one correspondence between external edges and faces.

It is convenient to order the rows of the matroid polytope using the cyclic numbering prescribed in \fref{matroidDisk}, i.e. assigning $ X_1 $ to the first row and proceeding cyclically. The rows of the toric diagram $ G $ describing the moduli space, obtained using the map presented in \sref{section_geometry_from_BFT}, can analogously be ordered using the numbering in \fref{matroidDisk}.

Since there are internal lines, it is always possible to choose a reference perfect matching that does not contain any external lines. In this way, perfect matchings specifying the columns of the matroid polytope and those specifying the columns of $ G $ can be immediately recognized, since the columns are distinguished only by external legs. For the purposes of the bijection the precise network of internal edges is irrelevant: it does not alter the external edges, thus preserving the matroid polytope as well as the powers of external faces required to specify the perfect matching.

It is now possible to construct the bijection between the matroid polytope and the toric diagram of the moduli space of a BFT on a disk. Each path from perfect matchings contains an even number of external lines, because each path that leaves the boundary must eventually return to the boundary. Since each $ w_i $ consists of two external edges, one leaving the boundary and one entering it, the product of two consecutive external faces $ w_i w_{i+1} $ occupies edges $ i $ and $ i+2 $.

Provided the toric diagram of the moduli space, the matroid polytope is thus obtained by replacing in $ G $ consecutive $ \pm 1 $'s appearing in rows $ i,i+1,\ldots,i+k $ by a 1 in row $ i $ and a 1 in row $ i+k+1 $. All remaining rows are assigned a 0. To obtain the toric diagram from the matroid polytope, it is a simple matter of performing the inverse process, i.e. replacing zeroes separating two 1's by a sequence of $\pm 1 $. The sign is finally determined by the color of the external nodes, which determines the orientation of the path.

Any change in the basis for the toric diagram presented in 
\sref{section_geometry_from_BFT} is going to preserve the bijection. As a final remark, changing the reference perfect matching corresponds to a modular transformation of the toric diagram. Thus, the case of BFTs with only external edges is going to work analogously to the case described above, provided that the reference perfect matching is also given when constructing the explicit map.
\\

\section{Infinite Families of Non-Planar BFTs}

\label{section_infinite_families}

This section illustrates how our techniques apply to the study of BFTs, putting special emphasis on the non-planar case. For this purpose this section introduces an infinite family of BFTs in this section and discusses several approaches for generating new models in the sections that follow.

Let us define a one parameter class of models on a cylinder, which we denote $C_n$, where $n$ measures the length of the graph along the periodic direction. 

\bigskip

\subsection{The $C_n$ Family}

\label{section_Cn}

\begin{figure}[h!]
\begin{center}
\includegraphics[trim=0cm 0cm 0cm 0cm,totalheight=4.5 cm]{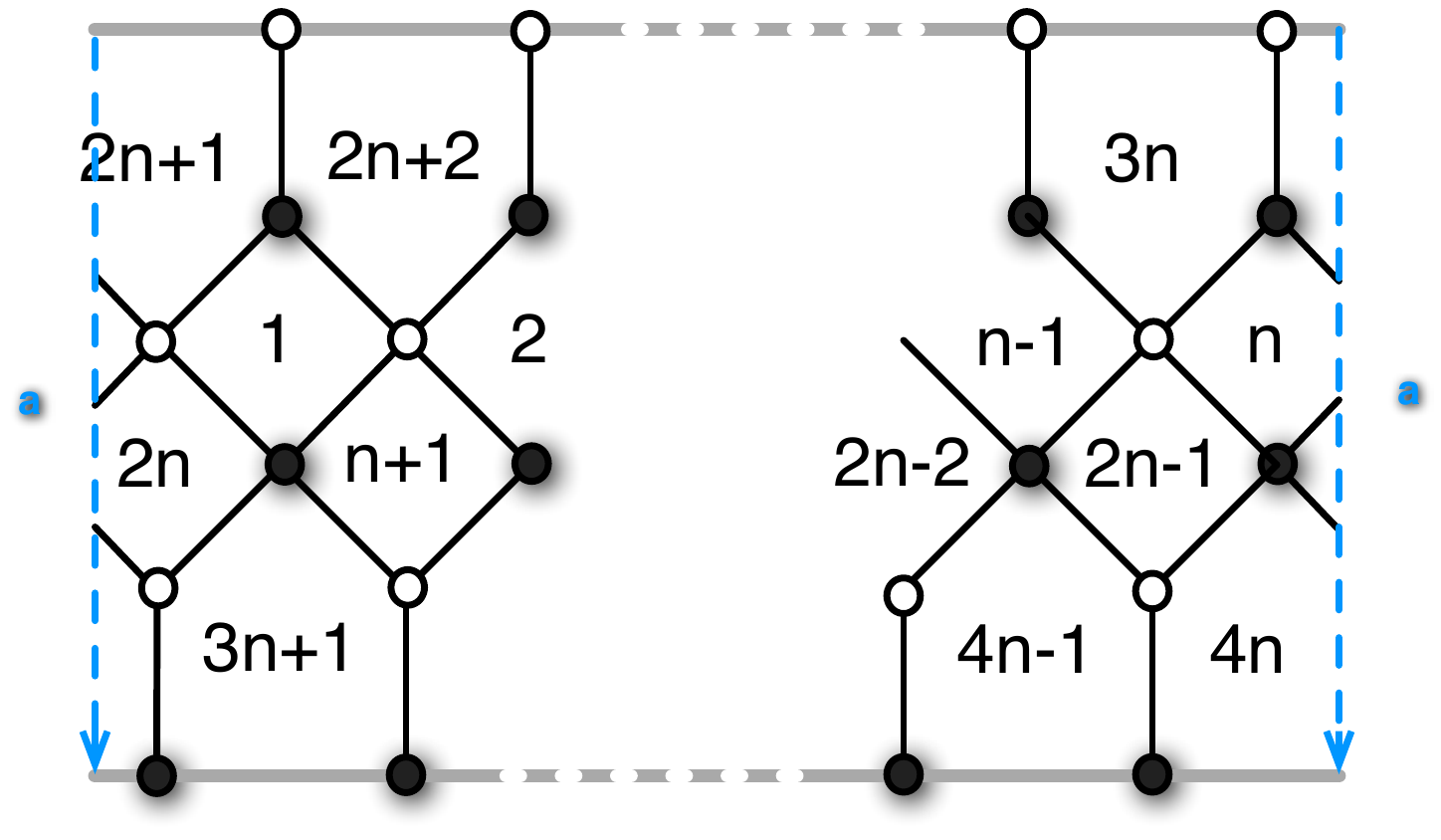}
  \caption{\textit{The general form of the $C_n$ tiling.}
  \label{fm4a}}
 \end{center}
 \end{figure}

The general bipartite graph of the $C_n$ model is shown in \fref{fm4a}.
\\

For $n=1,\dots,4$, the master Kasteleyn matrix takes the form: 
\\
{\tiny
\beq
\begin{array}{ccc}
K_{0}^{C_1}=
\left(\ba{c|cc|c} 
\; & 4 & 5 & 6 \\
\hline
1 & X_{1}+X_{2} & X_{3}+X_{4} & 0 \\
2 & 0 & X_{5}+X_{6} & X_{8} \\
\hline
3 & X_{7} & 0 & 0
\ea \right) & \ \ \ \ & 
K_{0}^{C_3}=
\left(
\begin{array}{c|cccccc|ccc}
\; & 10 & 11 & 12 & 13 & 14 & 15 & 16 & 17 & 18 \\
\hline
1 & X_{1} & 0 & X_{6} & X_{7} & 0 & X_{12} & 0 & 0 & 0 \\
2 & X_{2} & X_{3} & 0 & X_{8} & X_{9} & 0 & 0 & 0 & 0 \\
3 & 0 & X_{4} & X_{5} & 0 & X_{10} & X_{11} & 0 & 0 & 0 \\
4 & 0 & 0 & 0 & X_{13} & 0 & X_{18} & X_{24} & 0 & 0 \\
5 & 0 & 0 & 0 & X_{14} & X_{15} & 0 & 0 & X_{22} & 0 \\
6 & 0 & 0 & 0 & 0 & X_{16} & X_{17} & 0 & 0 & X_{23} \\
\hline
7 & X_{19} & 0 & 0 & 0 & 0 & 0 & 0 & 0 & 0 \\
8 & 0 & X_{20} & 0 & 0 & 0 & 0 & 0 & 0 & 0 \\
9 & 0 & 0 & X_{21} & 0 & 0 & 0 & 0 & 0 & 0
\end{array}
\right) \\ \\ \\
K_{0}^{C_2}=
\left(
\begin{array}{c|cccc|cc}
\; & 7 & 8 & 9 & 10 & 11 & 12 \\
\hline
1& X_{1} & X_{4} & X_{5} & X_{8} & 0 & 0 \\
2& X_{2} & X_{3} & X_{6} & X_{7} & 0 & 0 \\
3& 0 & 0 & X_{9} & X_{12} & X_{16} & 0 \\
4& 0 & 0 & X_{10} & X_{11} & 0 & X_{15} \\
\hline
5& X_{13} & 0 & 0 & 0 & 0 & 0 \\
6& 0 & X_{14} & 0 & 0 & 0 & 0
\end{array}
\right) & \ \ \ \ &
K_{0}^{C^{4,0}} =
\left(
\begin{array}{c|cccccccc|cccc}
\; & 13 & 14 & 15 & 16 & 17 & 18 & 19 & 20 & 21 & 22 & 23 & 24 \\
\hline
1 & X_{1} & 0 & 0 & X_{8} & X_{9} & 0 & 0 & X_{16} & 0 & 0 & 0 & 0 \\
2 & X_{2} & X_{3} & 0 & 0 & X_{10} & X_{11} & 0 & 0 & 0 & 0 & 0 & 0 \\
3 & 0 & X_{4} & X_{5} & 0 & 0 & X_{12} & X_{13} & 0 & 0 & 0 & 0 & 0 \\
4 & 0 & 0 & X_{6} & X_{7} & 0 & 0 & X_{14} & X_{15} & 0 & 0 & 0 & 0 \\
5 & 0 & 0 & 0 & 0 & X_{17} & 0 & 0 & X_{24} & X_{32} & 0 & 0 & 0 \\
6 & 0 & 0 & 0 & 0 & X_{18} & X_{19} & 0 & 0 & 0 & X_{29} & 0 & 0 \\
7 & 0 & 0 & 0 & 0 & 0 & X_{20} & X_{21} & 0 & 0 & 0 & X_{30} & 0 \\
8 & 0 & 0 & 0 & 0 & 0 & 0 & X_{22} & X_{23} & 0 & 0 & 0 & X_{31} \\
\hline
9 & X_{25} & 0 & 0 & 0 & 0 & 0 & 0 & 0 & 0 & 0 & 0 & 0 \\
10 & 0 & X_{26} & 0 & 0 & 0 & 0 & 0 & 0 & 0 & 0 & 0 & 0 \\
11 & 0 & 0 & X_{27} & 0 & 0 & 0 & 0 & 0 & 0 & 0 & 0 & 0 \\
12 & 0 & 0 & 0 & X_{28} & 0 & 0 & 0 & 0 & 0 & 0 & 0 & 0
\end{array}
\right)
\end{array}
\eeq
}

For $n>1$, the master Kasteleyn matrix takes the general form:

\beql{K0_general_Gn}
\scriptsize
K_{0}^{C_n} = 
\left(
\begin{array}{cccccccccc|ccccc}
 X_1 & 0 & \dots & 0 & X_{2n} 		& X_{2n+1} & 0 & \dots & 0 & X_{4n} 		& 0 & \; & \dots & \; & 0 \\
X_2 & X_3 & \; & \; & 0 				& X_{2n+2} & X_{2n+3} & \; & \; & 0 		& \; & \; & \; & \; & \; \\ 
0 & X_4 & \ddots & \; & \vdots		& 0 & X_{2n+4} & \ddots & \; & \vdots 	& \vdots & \;& \ddots& \; & \vdots\\
\vdots & \; & \ddots & X_{2n-3} & 0	& \vdots & \; & \ddots & X_{4n-3} & 0		& \; & \; & \;& \; & \; \\
0 & \dots & 0 & X_{2n-2} & X_{2n-1}	& 0 & \dots & 0 & X_{4n-2} & X_{4n-1}		& 0 & \;& \dots & \; & 0 \\
0 & \; & \dots & \; & 0 				& X_{4n+1} & 0 & \dots & 0 & X_{6n}		& X_{8n} & 0 & \multicolumn{2}{c}{\dots} & 0\\ 
\; & \; & \; & \; & \; 				& X_{4n+2} & X_{4n+3} & \; & \; & 0		& 0 & X_{7n+1} & \; & \; & \multirow{2}{*}{\vdots}\\ 
\vdots & \;& \ddots& \; & \vdots		& 0 & X_{4n+4} & \ddots & \; & \vdots 	& \multirow{2}{*}{\vdots} & \; & X_{7n+2} & \; & \\
\; & \; & \;& \; & \; 				& \vdots & \; & \ddots & X_{6n-3} & 0		& & \; & \; & \ddots &  0 \\
0 & \;& \dots & \; & 0 				& 0 & \dots & 0 & X_{6n-2} & X_{6n-1}		& 0 & \multicolumn{2}{c}{\dots} & 0 & X_{8n-1}\\
\hline
X_{6n+1} & 0 & \multicolumn{2}{c}{\dots} & 0			& 0 & \; & \dots & \; & 0 		& 0 & \; & \dots & \; & 0 \\ 
0 & X_{6n+2} & \; & \; & \multirow{2}{*}{\vdots}		& \; & \; & \; & \; & \; 		& \; & \; & \; & \; & \; \\ 
\multirow{2}{*}{\vdots} & \; & X_{6n+3} & \; &			& \vdots & \;& \ddots& \; & \vdots	& \vdots & \;& \ddots& \; & \vdots\\
& \; & \; & \ddots &  0 								& \; & \; & \;& \; & \; 			& \; & \; & \;& \; & \; \\
0 & \multicolumn{2}{c}{\dots} & 0 & X_{7n}			& 0 & \;& \dots & \; & 0 		& 0 & \;& \dots & \; & 0 \\
\end{array}
\right) \\
\\
\vspace{0.5cm}
\eeq

Let us summarize some general properties of this class of models:

\medskip

\begin{itemize}
\item \textit{Faces/Groups:} The number of internal as well as external faces is $2n$, which gives a total number of $4n$ faces.
\item \textit{Edges/Fields:} The total number of edges is $8n$, with $6n$ internal edges and $2n$ external legs.
\item \textit{Nodes/$W$-Terms:} There are $2n$ internal white nodes and $2n$ internal black nodes, which correspond to positive and negative terms in the superpotential. 
\item \textit{Zig-Zag Paths:} For all $C_n$ models, the number of internal zig-zag paths is $3$. The number of external zig-zag paths is $2n$.
\item \textit{Master Space:} It is a toric Calabi-Yau manifold of dimension $(4n+1)$.
\item \textit{Mesonic Moduli Space:} It is a toric Calabi-Yau manifold of dimension $(2n+1)$.
\end{itemize}

\bigskip

\subsubsection{Examples}

Let us now discuss the first members of the $C_n$ family in further detail. As a reference for the reader, the perfect matching matrices are collected in appendix \ref{appendix_Cn}.

\bigskip

\paragraph{$C_1$ Model.} 

\begin{figure}[H]
\begin{center}
 \begin{tabular}[c]{ccc}
\includegraphics[trim=0cm 0cm 0cm 0cm,totalheight=4 cm]{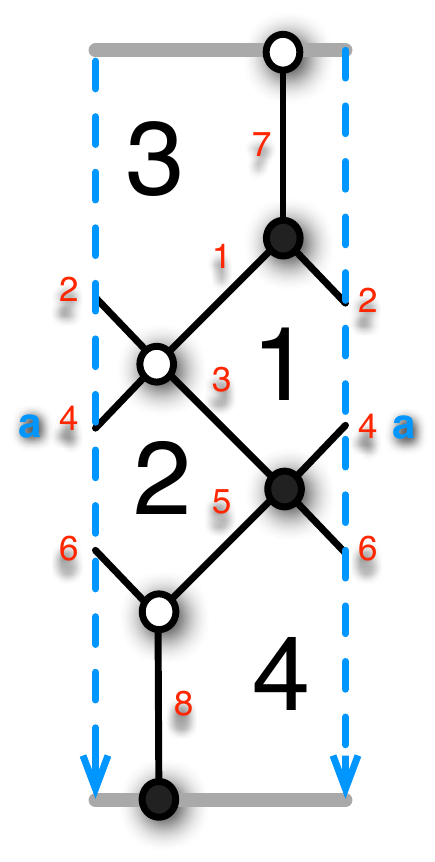} 
& \hspace{2cm} &
\includegraphics[trim=0cm 0cm 0cm 0cm,totalheight=4 cm]{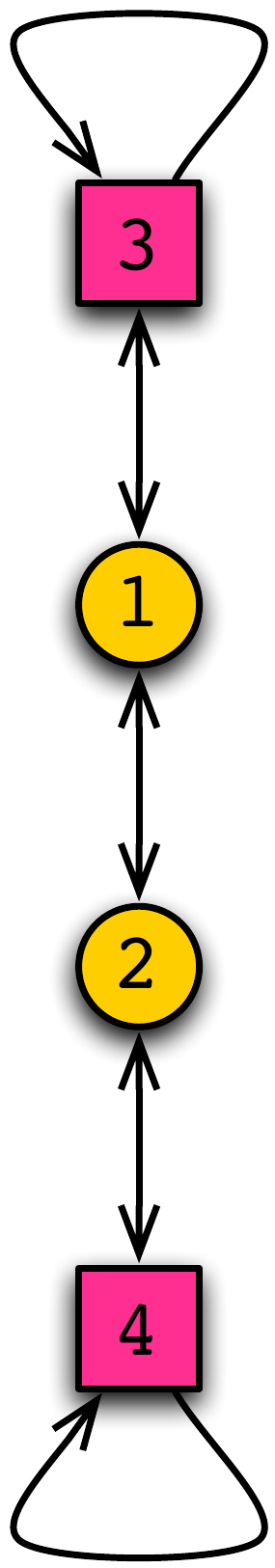} 
\\ 
\hspace{0cm}\mbox{(a)} &  \hspace{2cm} & \hspace{0cm}\mbox{(b)}
 \end{tabular}
  \caption{\textit{The bipartite graph and quiver for $C_1$.}
  \label{fm1a}}
 \end{center}
 \end{figure}

The bipartite graph and quiver diagram for this theory are shown in \fref{fm1a}.

The moduli space is obtained by imposing invariance under $Q_1$ and $Q_2$, where $Q_i$ is a shorthand for the gauge symmetry associated to face $i$. The moduli space then becomes a 3d CY with toric diagram given by

{\footnotesize
\beq
G_{C_1}=\left(
\begin{array}{C{0.4cm}C{0.4cm}C{0.4cm}C{0.4cm}C{0.4cm}}
 -1 & -2 & 0 & 0 & 1 \\
 1 & 2 & 0 & 1 & 0 \\
 1 & 1 & 1 & 0 & 0 \\
\hline
 \bf{2}  &  \bf{1}  &  \bf{1}  &  \bf{1}  &  \bf{1} \\
\hline
\end{array}\right).
\label{G_matrix_multiplicities_C10}
\eeq}

\bigskip

\paragraph{$C_2$ Model.} 

\begin{figure}[H]
\begin{center}
 \begin{tabular}[c]{ccc}
\includegraphics[trim=0cm 0cm 0cm 0cm,totalheight=4 cm]{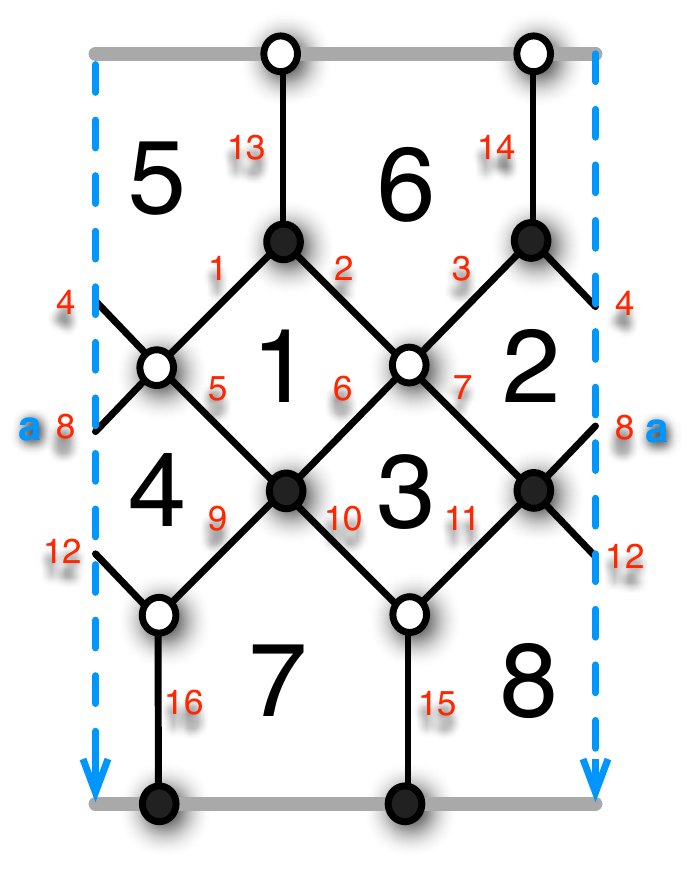}
& \hspace{2cm} &
\includegraphics[trim=0cm 0cm 0cm 0cm,totalheight=4 cm]{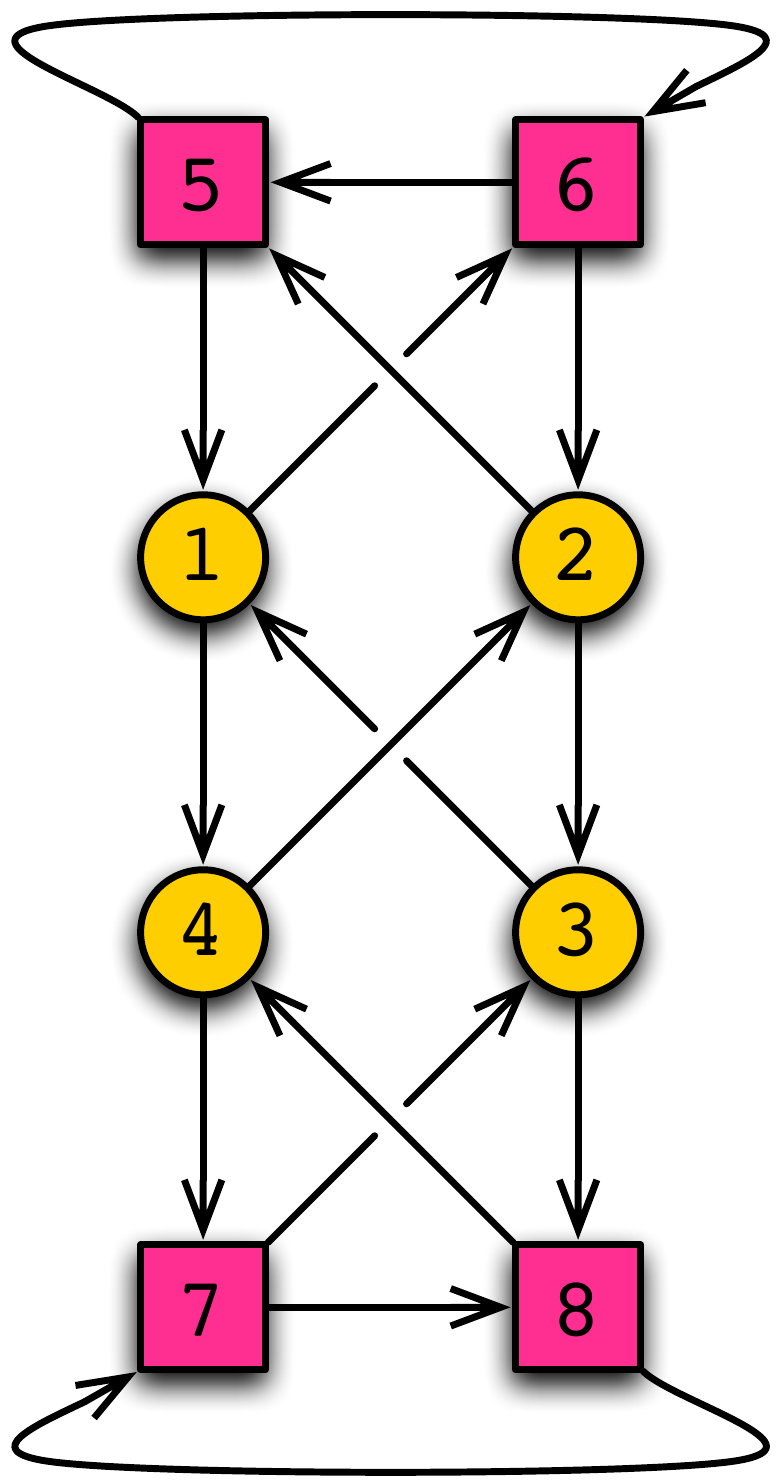}
\\ 
\hspace{0cm}\mbox{(a)} &  \hspace{2cm} & \hspace{0cm}\mbox{(b)}
 \end{tabular}
  \caption{\textit{The bipartite graph and quiver for $C_2$.}
  \label{fm2a}}
 \end{center}
 \end{figure}

\fref{fm2a} shows the corresponding graph and quiver. 

In order to obtain the moduli space, we quotient by $Q_1,\ldots,Q_4$. A 5d CY with toric diagram is obtained. It is given by

{\footnotesize
\beq
G_{C_2}=\left(
\begin{array}{C{0.35cm}C{0.35cm}C{0.35cm}C{0.35cm}C{0.35cm}C{0.35cm}C{0.35cm}C{0.35cm}C{0.35cm}C{0.35cm}C{0.35cm}C{0.35cm}C{0.35cm}}
 0 & -1 & 1 & 0 & 0 & 0 & 0 & 1 & 0 & -1 & 0 & 0 & 0 \\
 1 & -1 & 0 & -1 & 0 & 1 & -2 & 1 & -1 & 0 & 0 & -1 & 0 \\
 0 & 1 & 0 & 0 & 0 & -1 & 1 & -1 & 1 & 0 & 0 & 1 & 1 \\
 0 & 1 & 0 & 1 & 1 & 1 & 1 & 0 & 0 & 1 & 0 & 1 & 0 \\
 0 & 1 & 0 & 1 & 0 & 0 & 1 & 0 & 1 & 1 & 1 & 0 & 0 \\ \hline
 \bf{3}  &  \bf{3}  &  \bf{3}  &  \bf{3}  &  \bf{2}  &  \bf{1}  &  \bf{1}  &  \bf{1}  &  \bf{1}  &  \bf{1}  &  \bf{1}  &  \bf{1}  &  \bf{1} \\ 
\hline
\end{array}
\right) .
\label{G_matrix_multiplicities_C20}
\eeq}

\bigskip

\paragraph{$C_3$ Model.} 

\begin{figure}[ht!!]
\begin{center}
 \begin{tabular}[c]{ccc}
\includegraphics[trim=0cm 0cm 0cm 0cm,totalheight=4 cm]{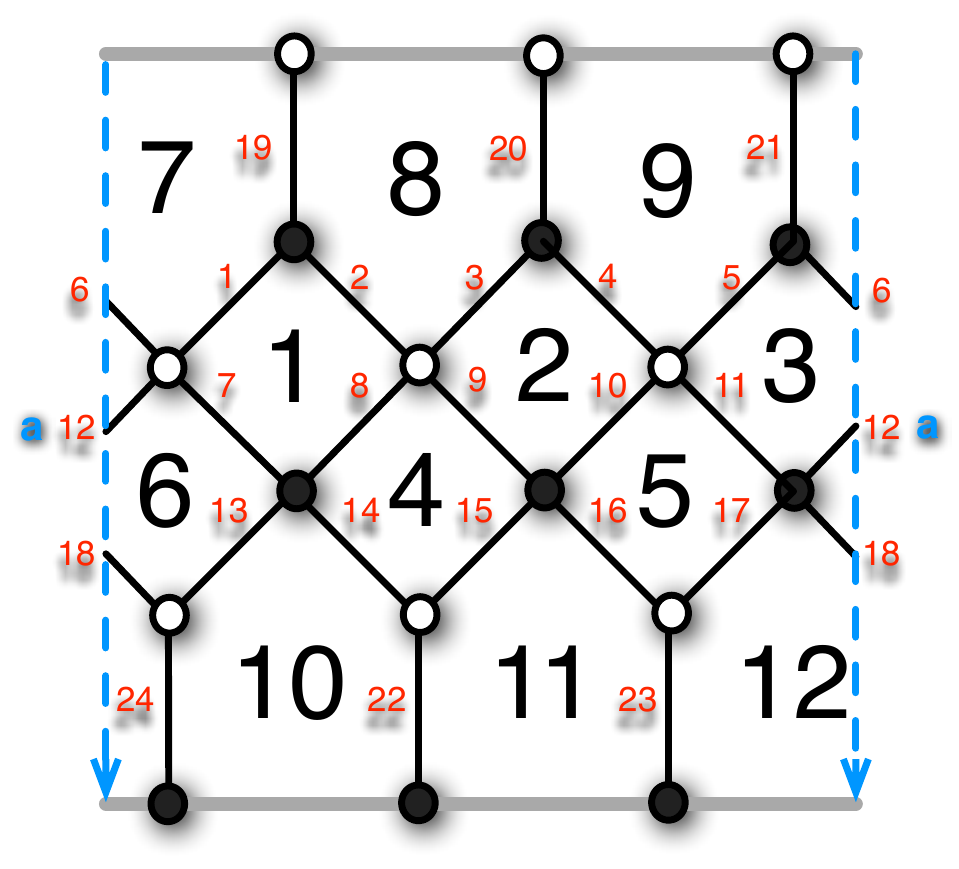}
& \hspace{2cm} &
\includegraphics[trim=0cm 0cm 0cm 0cm,totalheight=4 cm]{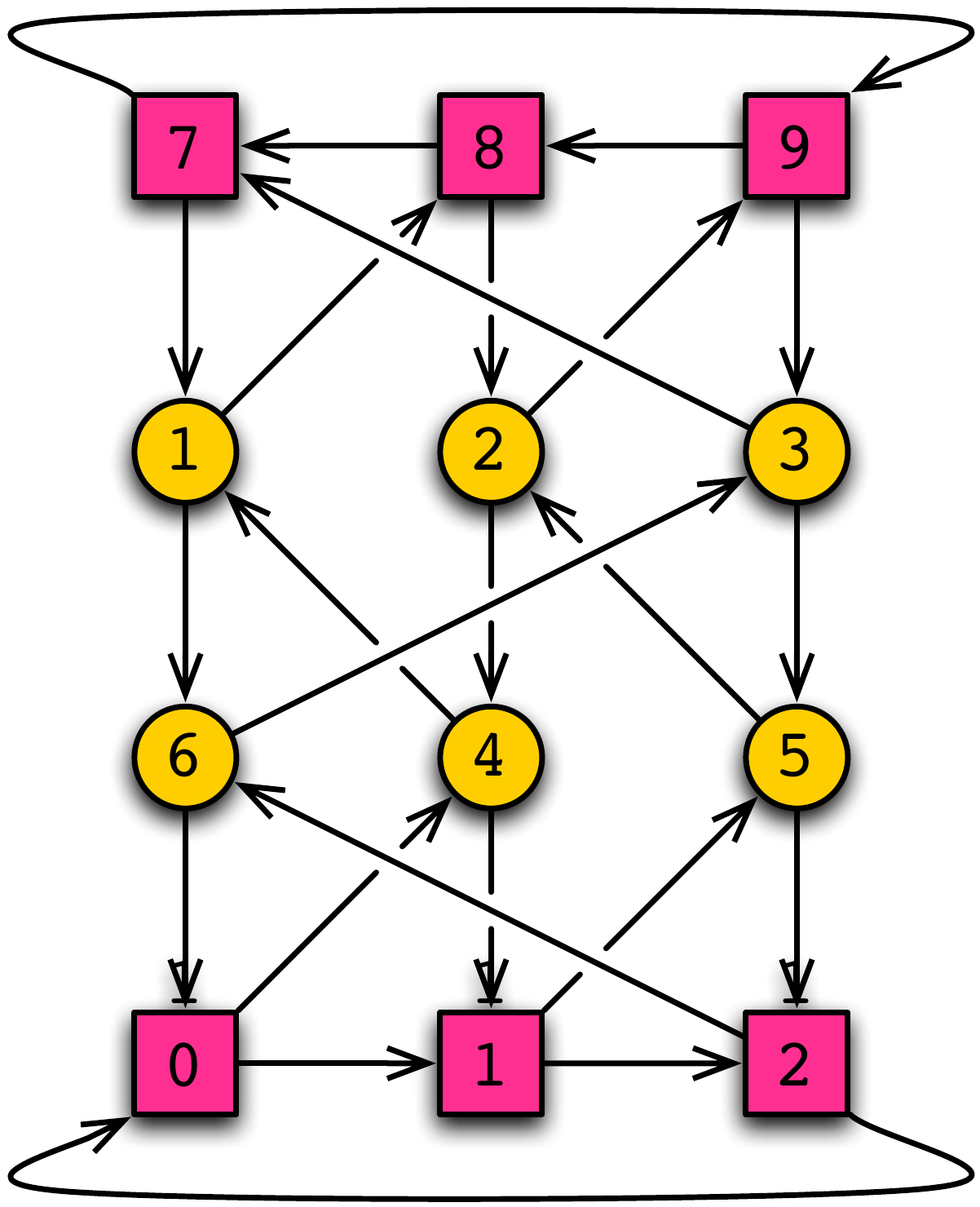}
\\ 
\hspace{0cm}\mbox{(a)} &  \hspace{2cm} & \hspace{0cm}\mbox{(b)}
 \end{tabular}
  \caption{\textit{The bipartite graph and quiver for $C_3$.}
  \label{fm3a}}
 \end{center}
 \end{figure}

The graph and quiver for this theory are given in \fref{fm3a}.

The $C_3$ model has a total 96 perfect matchings. The moduli space is obtained by quotienting with $Q_1,\ldots,Q_6$. It is a 7d CY with toric diagram given by

{\scriptsize
\beq
\begin{array}{rcl}
G_{C_3} &=&
\left(
\begin{array}{C{0.35cm}C{0.35cm}C{0.35cm}C{0.35cm}C{0.35cm}C{0.35cm}C{0.35cm}C{0.35cm}C{0.35cm}C{0.35cm}C{0.35cm}C{0.35cm}C{0.35cm}C{0.35cm}C{0.35cm}C{0.35cm}C{0.35cm}C{0.35cm}C{0.35cm}}
 -2 & -1 & 0 & -1 & 0 & 0 & -1 & 0 & 0 & -1 & -1 & -1 & -1 & 0 & -1 & -1 & -1 & 0 & 0
   \\
 1 & 0 & 0 & 1 & 1 & 0 & 1 & -1 & 0 & 0 & 1 & 0 & 0 & -1 & 2 & 1 & 0 & 0 & 0 \\
 1 & 1 & 0 & 0 & 1 & 0 & 1 & 0 & -1 & 0 & 2 & 0 & 1 & -1 & 1 & 0 & 0 & 1 & 0 \\
 0 & 0 & 0 & 0 & -1 & 0 & 0 & 1 & 1 & 1 & -1 & 1 & 0 & 1 & -1 & 0 & 1 & 0 & 0 \\
 0 & 0 & -1 & 0 & 0 & 1 & 0 & -1 & 1 & -1 & -1 & 1 & -1 & 0 & 1 & 1 & 0 & 0 & 0 \\
 1 & 1 & 1 & 1 & 0 & 0 & 0 & 1 & 0 & 1 & 1 & 0 & 1 & 1 & 0 & 0 & 1 & 0 & 1 \\
 0 & 0 & 1 & 0 & 0 & 0 & 0 & 1 & 0 & 1 & 0 & 0 & 1 & 1 & -1 & 0 & 0 & 0 & 0 \\
\hline
\bf{5} & \bf{5} & \bf{5} & \bf{5} & \bf{5} & \bf{4} & \bf{4} & \bf{4} & \bf{3} & \bf{3} & \bf{3} & \bf{3} & \bf{3} & \bf{3} & \bf{3} & \bf{3} & \bf{3} & \bf{3} & \bf{3} \\
\hline
\end{array}
\right. \cdots \\ \\ 
& \cdots &
\left. \begin{array}{C{0.35cm}C{0.35cm}C{0.35cm}C{0.35cm}C{0.35cm}C{0.35cm}C{0.35cm}C{0.35cm}C{0.35cm}C{0.35cm}C{0.35cm}C{0.35cm}C{0.35cm}C{0.35cm}C{0.35cm}C{0.35cm}C{0.35cm}C{0.35cm}C{0.35cm}}
 1 & 0 & -1 & -1 & -2 & -1 & -1 & -1 & -1 & -1 & 0 & 0 & 0 & 0 & 0 & -1 & 0 & 0 & 1
   \\
 -1 & 1 & 1 & -1 & 2 & -1 & 2 & 0 & 0 & 1 & 1 & -1 & -1 & 0 & 0 & 1 & 1 & 0 & 0 \\
 -1 & 0 & 1 & -1 & 2 & 0 & 2 & -1 & 0 & 1 & 1 & -1 & 0 & -1 & 0 & 1 & 1 & 0 & 0 \\
 1 & 0 & -1 & 2 & -1 & 1 & -2 & 1 & 0 & 0 & -1 & 2 & 1 & 1 & 0 & 0 & -1 & 1 & 0 \\
 0 & 0 & 0 & 0 & 0 & 0 & 0 & 0 & 0 & 1 & 1 & 0 & 0 & 0 & 0 & -1 & -1 & 0 & 0 \\
 0 & 0 & 1 & 1 & 1 & 1 & 1 & 1 & 1 & 0 & 0 & 0 & 0 & 0 & 0 & 1 & 1 & 0 & 0 \\
 1 & 0 & 0 & 1 & -1 & 1 & -1 & 1 & 1 & -1 & -1 & 1 & 1 & 1 & 1 & 0 & 0 & 0 & 0 \\
\hline
\bf{3} & \bf{3} & \bf{2} & \bf{1} & \bf{1} & \bf{1} & \bf{1} & \bf{1} & \bf{1} & \bf{1} & \bf{1} & \bf{1} & \bf{1} & \bf{1} & \bf{1} & \bf{1} & \bf{1} & \bf{1} & \bf{1} \\ 
\hline
\end{array}
\right) .
\end{array}
\label{G_matrix_multiplicities_C30}
\eeq}

\section{Untwisting}

\label{section_untwisting}

This is the first section which discusses a method for generating new BFTs. The section studies untwisting \cite{Feng:2005gw,Franco:2007ii,Franco:2011sz,Hanany:2012vc}, an operation acting on zig-zag paths. 

\bigskip

\subsection{Zig-Zag Paths and Untwisting} 

\begin{figure}[h]
\begin{center}
\includegraphics[width=4.5cm]{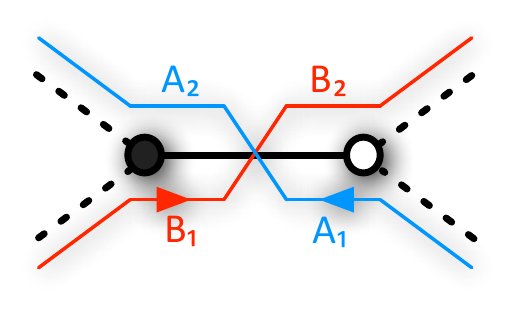}
\caption{\textit{Double line implementation of zig-zag paths.}}
\label{double_line_zig_zags}
\end{center}
\end{figure}

A \textbf{zig-zag path} is an oriented path along the edges of a bipartite graph such that it turns maximally left at white nodes and maximally right at black nodes. They can be elegantly encoded in a {\it double line} notation, also denoted {\it alternating strand} notation, in which every edge has two associated zig-zag paths going in opposite directions and crossing at its middle point, as shown in \fref{double_line_zig_zags}. In this representation, each zig-zag path is translated into a path across the edges  such that white nodes are always to the right and black nodes are always to the left.

Zig-zag paths have several important applications. For graphs on a disk, they can be used to define {\it permutations}, which encode how external nodes are paired by zig-zag paths. In turn, permutations identify whether two different graphs are related by a sequence of moves and bubble reductions. On a related application, reducibility of graphs on a disk translates into the existence of zig-zag paths with multiple crossings or self-intersections.\footnote{The presence of self-intersecting zig-zag paths has been linked to inconsistencies in the BFTs on D3-branes over toric CY 3-folds. These issues do not seem to extend to generic BFTs, although a detailed investigation of potential problems associated to self-intersecting zig-zag paths is certainly desirable.} For both theories on the disk and on $T^2$ without boundaries, zig-zag paths can be used to reconstruct the underlying bipartite graph.\footnote{This is probably true more generally, although such a construction has not been worked out in the literature so far.} Finally, for BFTs on $T^2$ without boundaries, zig-zag paths correspond to external legs in the (p,q)-web diagram \cite{Aharony:1997bh}. This is a graph dual to the 2d toric diagram for the CY 3-fold moduli space. 

\begin{figure}[h]
\begin{center}
\includegraphics[width=10cm]{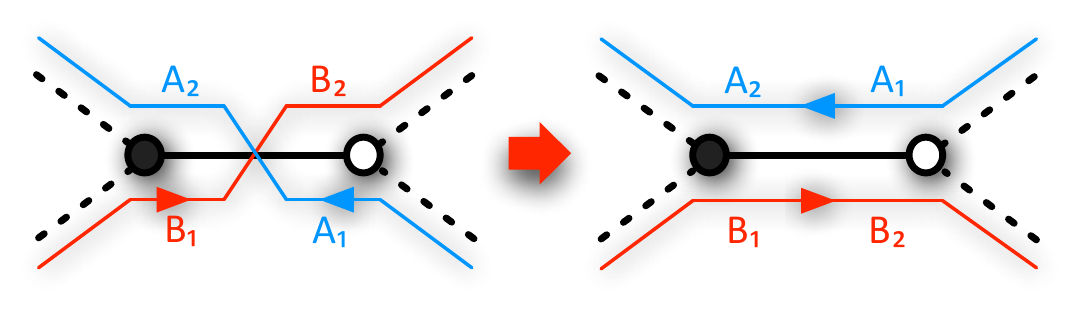}
\caption{\textit{The action of the untwisting map.}}
\label{untwisting_map}
\end{center}
\end{figure}

{\bf Untwisting} is a transformation of a bipartite graph defined by its action on zig-zag paths, which is shown in \fref{untwisting_map}. Equivalently, untwisting can be defined as a transformation that maps zig-zag paths to paths along edges which constitute a face, and vice versa. The underlying Riemann surface in general changes under untwisting.

The untwisting map is very important for certain sub-classes of BFTs. For example, for graphs without boundaries on $T^2$, which correspond to D3-branes probing toric CY 3-folds, untwisting gives rise to the mirror configuration of intersecting D6-branes \cite{Feng:2005gw}. For the same class of graphs, untwisting can also be interpreted as relating a graph defining a cluster integrable system to the corresponding spectral curve \cite{GK,Franco:2011sz,Eager:2011dp}. More recently, it has been used to identify BFTs that share the same master space, in a correspondence which is called \textbf{specular duality} \cite{Hanany:2012hi,Hanany:2012vc}. It is natural to expect the importance of untwisting to be much broader and that it plays a profound role, yet to be unveiled, for general BFTs.\footnote{More concretely, we expect untwisting to be important in the context of BFT$_1$'s. As we are going to explain in \sref{section_gauging_2}, BFT$_2$'s are invariant under untwisting.}

\begin{figure}[h]
\begin{center}
\includegraphics[width=9cm]{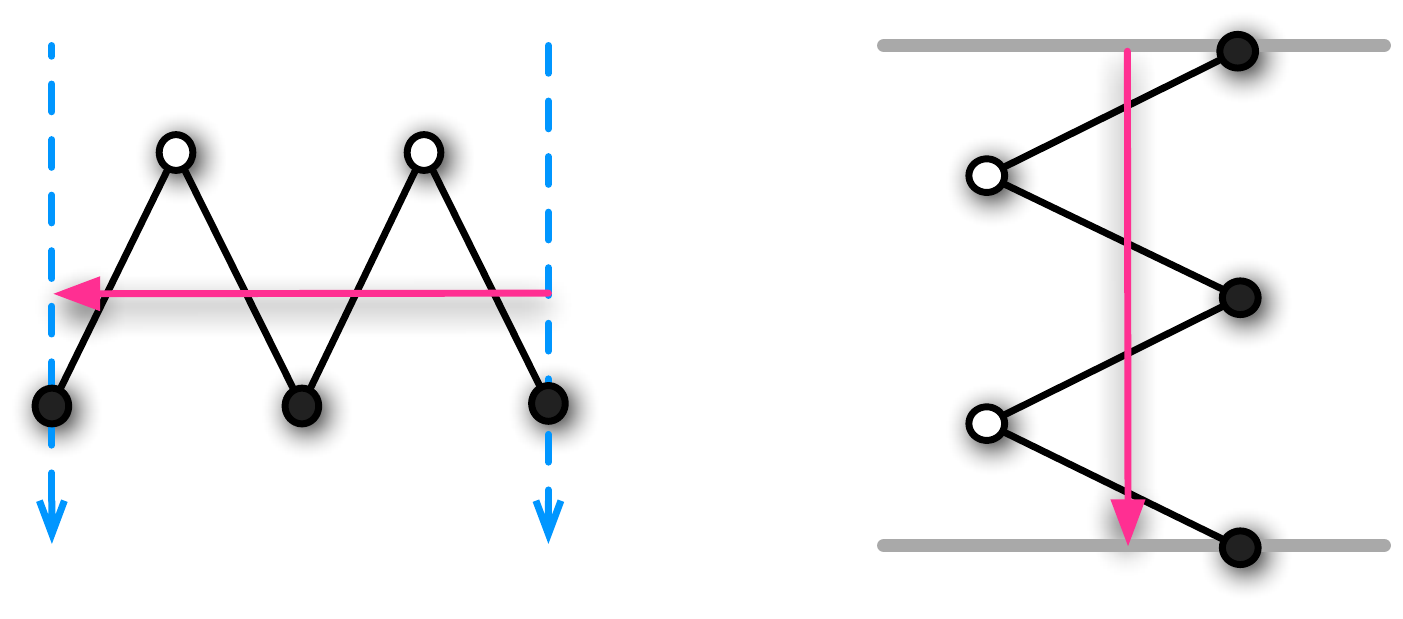}
\caption{\textit{Closed and open zig-zag paths.}}
\label{open_closed_zig_zags}
\end{center}
\end{figure}

Let us identify some general properties of the untwisting map. As mentioned above, it transforms zig-zag paths into faces and vice versa. It is useful to distinguish between open and closed zig-zag paths, which are illustrated in \fref{open_closed_zig_zags}. Untwisting has the following effect:

\medskip

\begin{itemize}
\item Open zig-zag paths become external faces, and vice versa
\item Closed zig-zag paths become internal faces, and vice versa
\end{itemize} 

\medskip

\begin{figure}[h]
\begin{center}
\includegraphics[width=9cm]{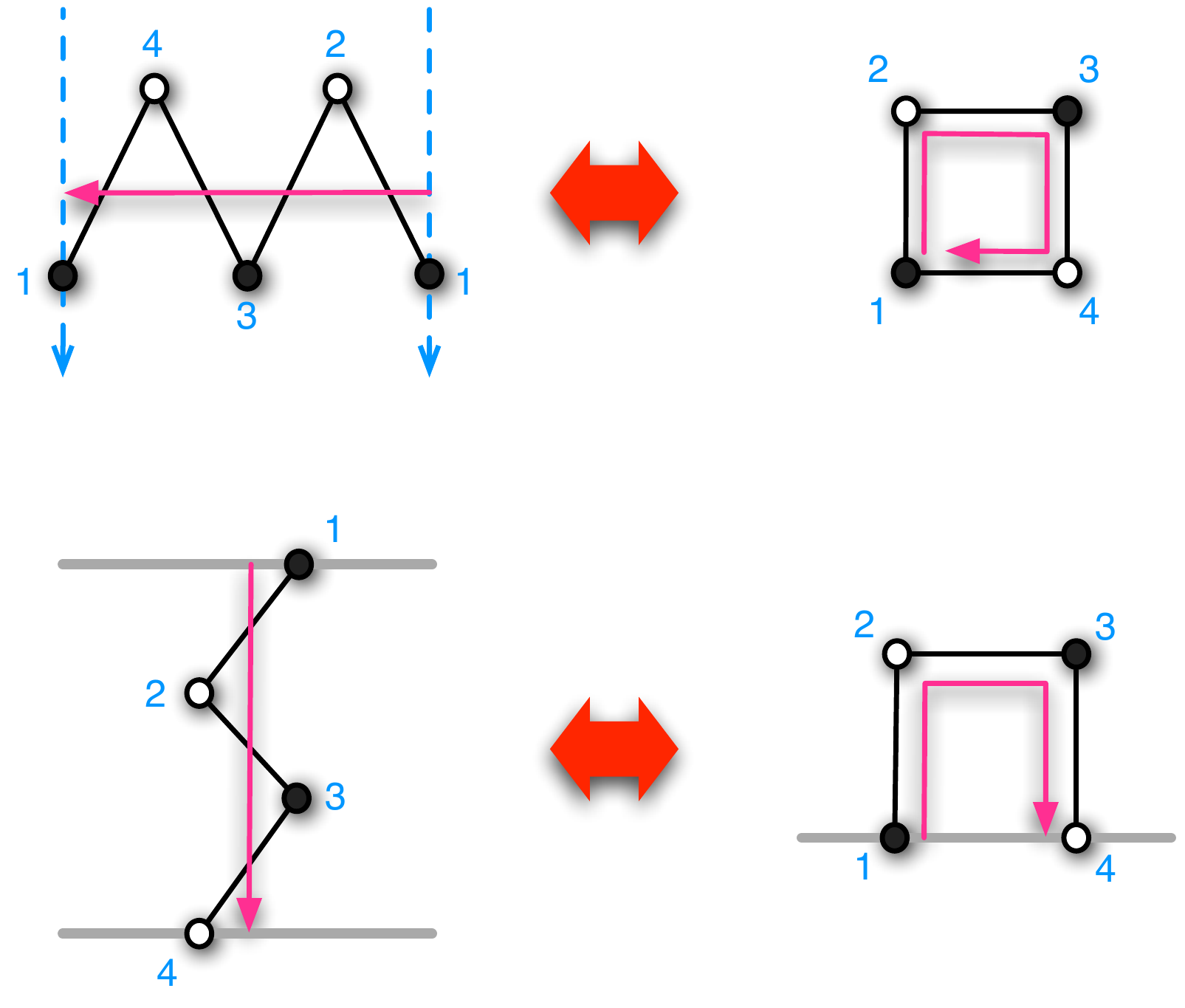}
\caption{\textit{Untwisting on closed and open zig-zag paths.}}
\label{fzzpaths2}
\end{center}
\end{figure}

This behavior is illustrated in \fref{fzzpaths2}. The number of boundaries generically changes under untwisting. Let us define a {\bf boundary cycle} as a subset of the external nodes that is obtained by following open zig-zag paths until returning to the starting point. In other words, one starts from an external node $i$ and follow the zig-zag path emanating from it until reaching a new external node $j$. Next, we consider the zig-zag that leaves from node $j$ and takes us to an external node $k$. This operation is repeated until one comes back to the starting point. A boundary cycle is the set of all external nodes visited during such an excursion.  Each boundary cycle gives rise to a boundary component in the untwisted graph.

The master Kasteleyn matrix and its reductions remain invariant under untwisting. As a consequence, in an extension of specular duality \cite{Hanany:2012vc} to general BFTs, the master spaces of the original and untwisted theories are the same.  This follows from the fact that the master space only cares about how edges are connected into nodes, i.e. how chiral multiplets are coupled by the superpotential. This information is not modified by untwisting. 

The following discussion is going to use untwisting to generate bipartite graphs on bounded Riemann surfaces with arbitrarily large genus starting from relatively simple ones. 

\bigskip

\subsection{A New Class of BFTs: Untwisting $C_n$}

\sref{section_Cn} considered the infinite class of $C_n$ theories, defined by bipartite graphs on a cylinder. The following section introduces a new class of BFTs, denoted by $\widetilde{C_n}$, which is generated by untwisting the $C_n$ theories. Following the general discussion in the section above, the master and reduced Kasteleyn matrices as well as the master space for these models are identical to those for $C_n$ theories. Let us summarize some general properties of the $\widetilde{C_n}$ family:

\bigskip

\begin{itemize}
\item \textit{Riemann Surface:} The underlying Riemann surface has genus $n-1$ and a single boundary.
\item \textit{Faces/Groups:} There are 3 internal faces for any $n$. All internal faces are $2n$-sided. The number of external faces is $2n$. These are respectively the number of closed and open zig-zag paths in the $C_n$ models.
\item \textit{Edges/Fields:} The number of edges is the same as for the $C_n$ models.  The total number of edges is $8n$ out of which $6n$ are internal and $2n$ are external.
\item \textit{Nodes/W-Terms:} The number of nodes is the same as for the $C_n$ models. There are $2n$ white nodes and $2n$ black nodes.
\item \textit{Zig-Zag Paths:} There are $2n$ closed and $2n$ open zig-zag paths, which map via untwisting to the internal and external faces of the untwisted $C_n$.
\item \textit{Master Space:} It is the same as the one for $C_n$, i.e. a toric Calabi-Yau manifold of dimension $(4n+1)$.
\item  \textit{Mesonic Moduli Space:} It is a toric Calabi-Yau manifold of dimension $(4n-2)$.
\end{itemize}

\bigskip

\subsubsection{Examples}

This section discusses in detail the first examples of the $\widetilde{C_n}$ family. The perfect matchings and master space of these theories are identical to those of the $C_n$ models, and are obtained using the master Kasteleyn matrix given in \eref{K0_general_Gn}. Some of the resulting perfect matching matrices are given in appendix \ref{appendix_Cn}. The different face structure of the untwisted theories results in a different gauging, which enters in the computation of the moduli space.

\bigskip

\paragraph{$\widetilde{C_1}$ Model.} 

\begin{figure}[h]
\begin{center}
 \begin{tabular}[c]{ccc}
\includegraphics[trim=0cm 0cm 0cm 0cm,totalheight=4.5 cm]{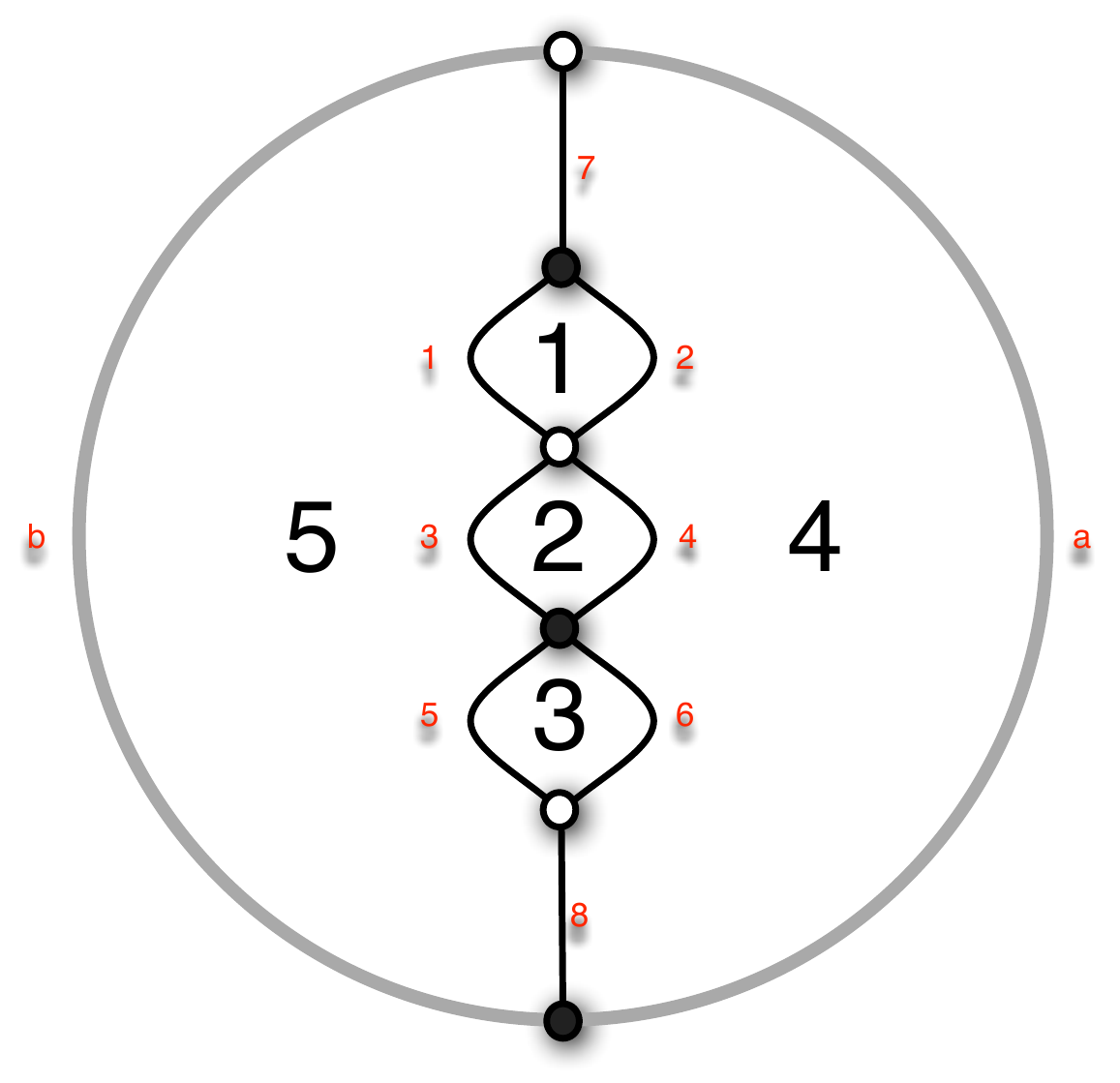}
& \hspace{2cm} &
\includegraphics[trim=0cm 0cm 0cm 0cm,totalheight=4.5 cm]{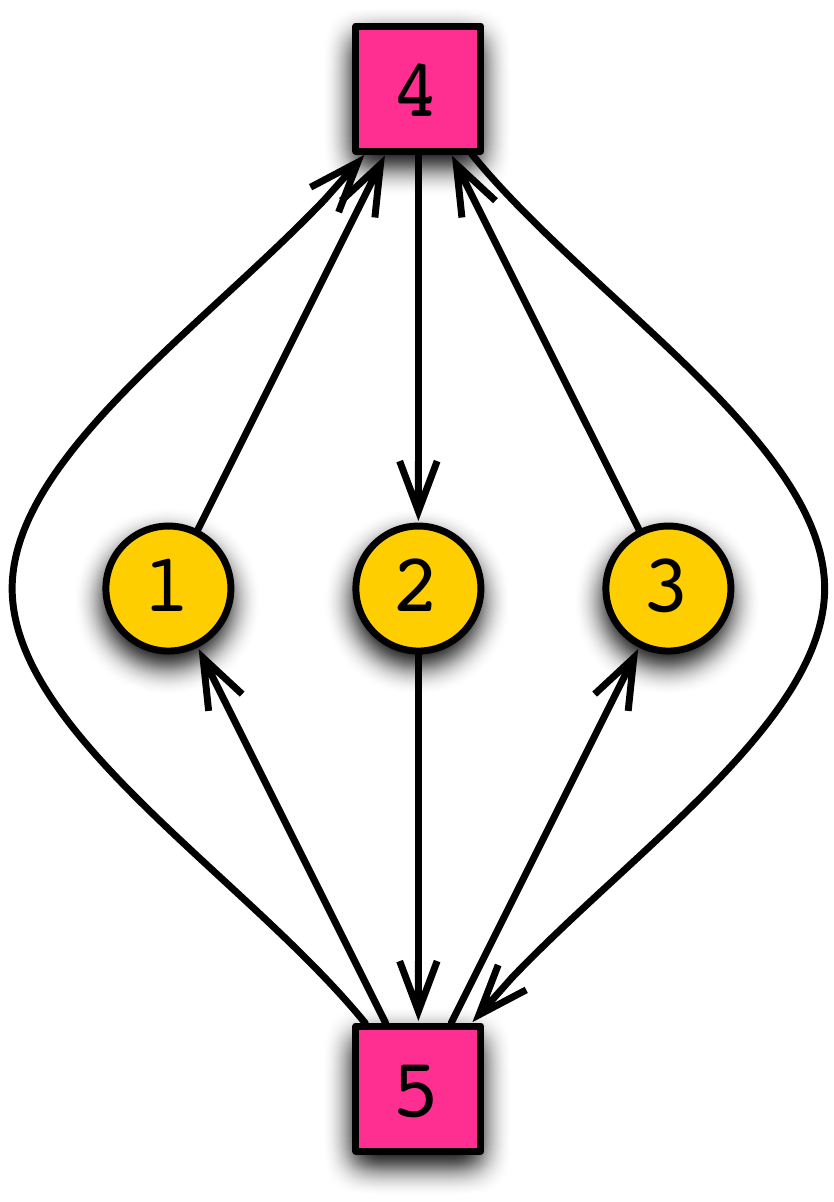}
\\ 
\hspace{0cm}\mbox{(a)} &  \hspace{2cm} & \hspace{0cm}\mbox{(b)}
 \end{tabular}
  \caption{\textit{The bipartite graph and quiver for $\widetilde{C_1}$.} The bipartite graph lives on a disk.
  \label{fm1aa}}
 \end{center}
 \end{figure}

The bipartite graph and quiver for this model are shown in \fref{fm1aa}. The fact that this graph is reducible is particularly obvious due to the presence of bubbles.

The moduli space is obtained by imposing gauge invariance under $Q_1$, $Q_2$ and $Q_3$. It is a 2d CY with toric diagram given by

{\footnotesize
\beq
G_{\widetilde{C_1}}=\left(
\begin{array}{C{0.35cm}C{0.2cm}}
 1 & 0 \\
 0 & 1 \\
\hline
\bf{4} & \bf{2} \\
\hline
\end{array}\right).
\label{G_matrix_multiplicities_tildeC10}
\eeq}
Interestingly it coincides with the one for $C_1$.

\bigskip

\begin{figure}[h]
\begin{center}
 \begin{tabular}[c]{ccc}
\includegraphics[trim=0cm 0cm 0cm 0cm,totalheight=4.5 cm]{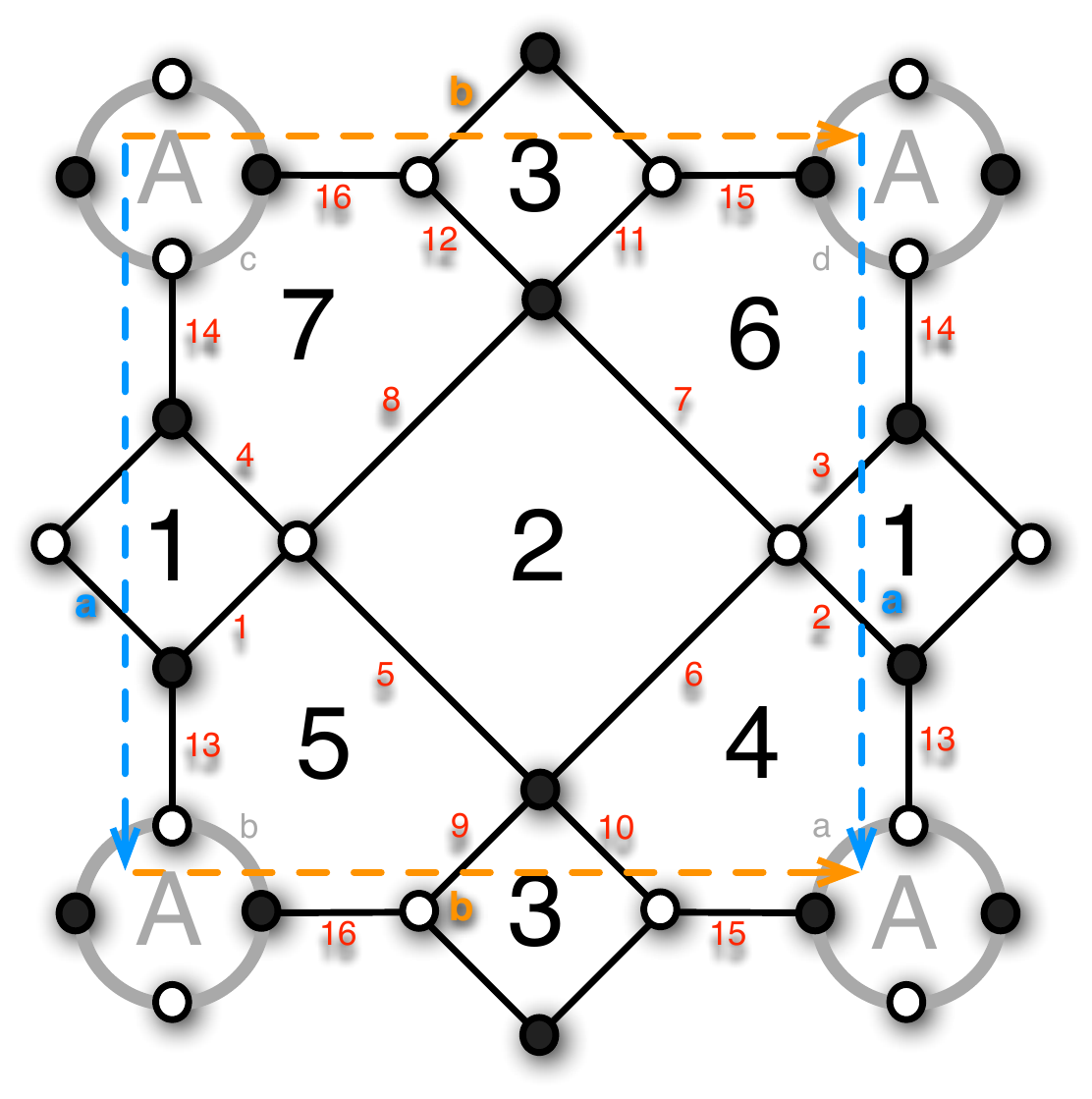}
& \hspace{2cm} &
\includegraphics[trim=0cm 0cm 0cm 0cm,totalheight=4.5 cm]{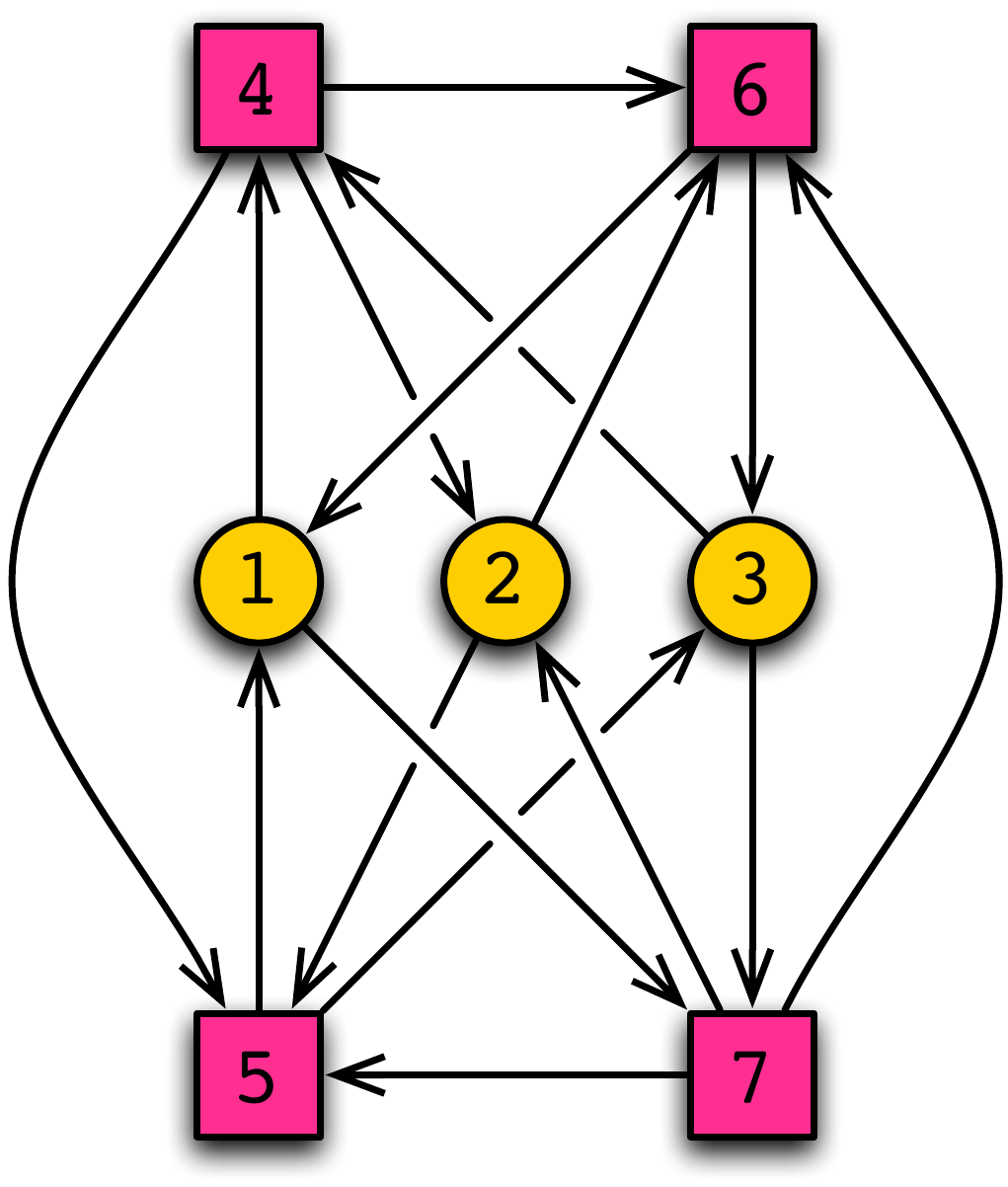}
\\ 
\hspace{0cm}\mbox{(a)} &  \hspace{2cm} & \hspace{0cm}\mbox{(b)}
 \end{tabular}
  \caption{\textit{The bipartite graph and quiver for $\widetilde{C_2}$.} The bipartite graph lives on a torus with a single boundary.
  \label{fm2aa}}
 \end{center}
 \end{figure}

\paragraph{$\widetilde{C_2}$ Model.} 
The bipartite graph and quiver for this model are shown in \fref{fm2aa}. As in all the $C_n$ models, the moduli space is obtained from quotienting by $Q_1$, $Q_2$ and $Q_3$. The moduli space is a 6d CY with toric diagram given by
\\

{\footnotesize
\beq
G_{\widetilde{C_2}}=\left(
\begin{array}{C{0.35cm}C{0.35cm}C{0.35cm}C{0.35cm}C{0.35cm}C{0.35cm}C{0.35cm}C{0.35cm}C{0.35cm}C{0.35cm}C{0.35cm}C{0.35cm}C{0.35cm}C{0.35cm}C{0.35cm}C{0.35cm}C{0.35cm}C{0.2cm}}
 0 & 0 & 1 & 0 & 2 & 1 & 0 & -1 & 1 & 0 & -2 & -1 & -1 & 0 & -1 & 0 & 0 & 1 \\
 0 & 0 & 0 & 0 & -1 & -1 & 0 & 0 & -1 & -1 & 1 & 0 & 1 & 0 & 1 & 0 & 1 & 0 \\
 0 & 0 & -1 & -1 & -1 & -1 & 0 & 0 & 0 & 0 & 1 & 1 & 0 & 0 & 1 & 1 & 0 & 0 \\
 1 & 0 & 1 & 1 & 1 & 1 & 1 & 1 & 1 & 1 & 0 & 0 & 0 & 0 & 0 & 0 & 0 & 0 \\
 0 & 1 & 1 & 1 & 1 & 1 & 1 & 1 & 1 & 1 & 0 & 0 & 0 & 0 & 0 & 0 & 0 & 0 \\
 0 & 0 & -1 & 0 & -1 & 0 & -1 & 0 & -1 & 0 & 1 & 1 & 1 & 1 & 0 & 0 & 0 & 0 \\ \hline
\bf{4} & \bf{2} & \bf{1} & \bf{1} & \bf{1} & \bf{1} & \bf{1} & \bf{1} & \bf{1} & \bf{1} & \bf{1} & \bf{1} & \bf{1} & \bf{1} & \bf{1} & \bf{1} & \bf{1} & \bf{1} \\ \hline
\end{array}
\right).
\eeq}

\bigskip

\begin{figure}[h!!]
\begin{center}
 \begin{tabular}[c]{ccc}
\includegraphics[trim=0cm 0cm 0cm 0cm,totalheight=7 cm]{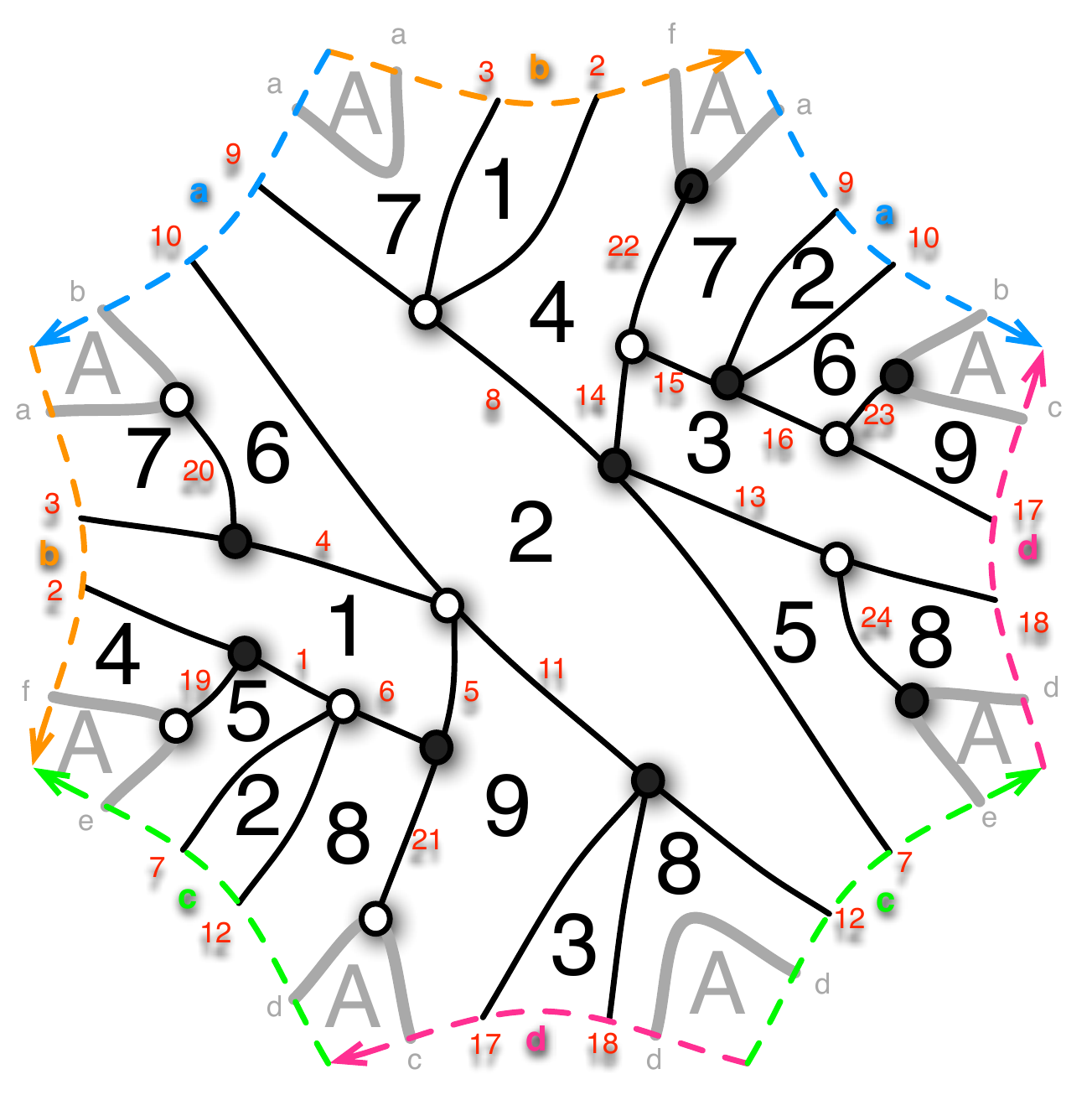}
& \hspace{2cm} &
\raisebox{0.8cm}{\includegraphics[trim=0cm 0cm 0cm 0cm,totalheight=5 cm]{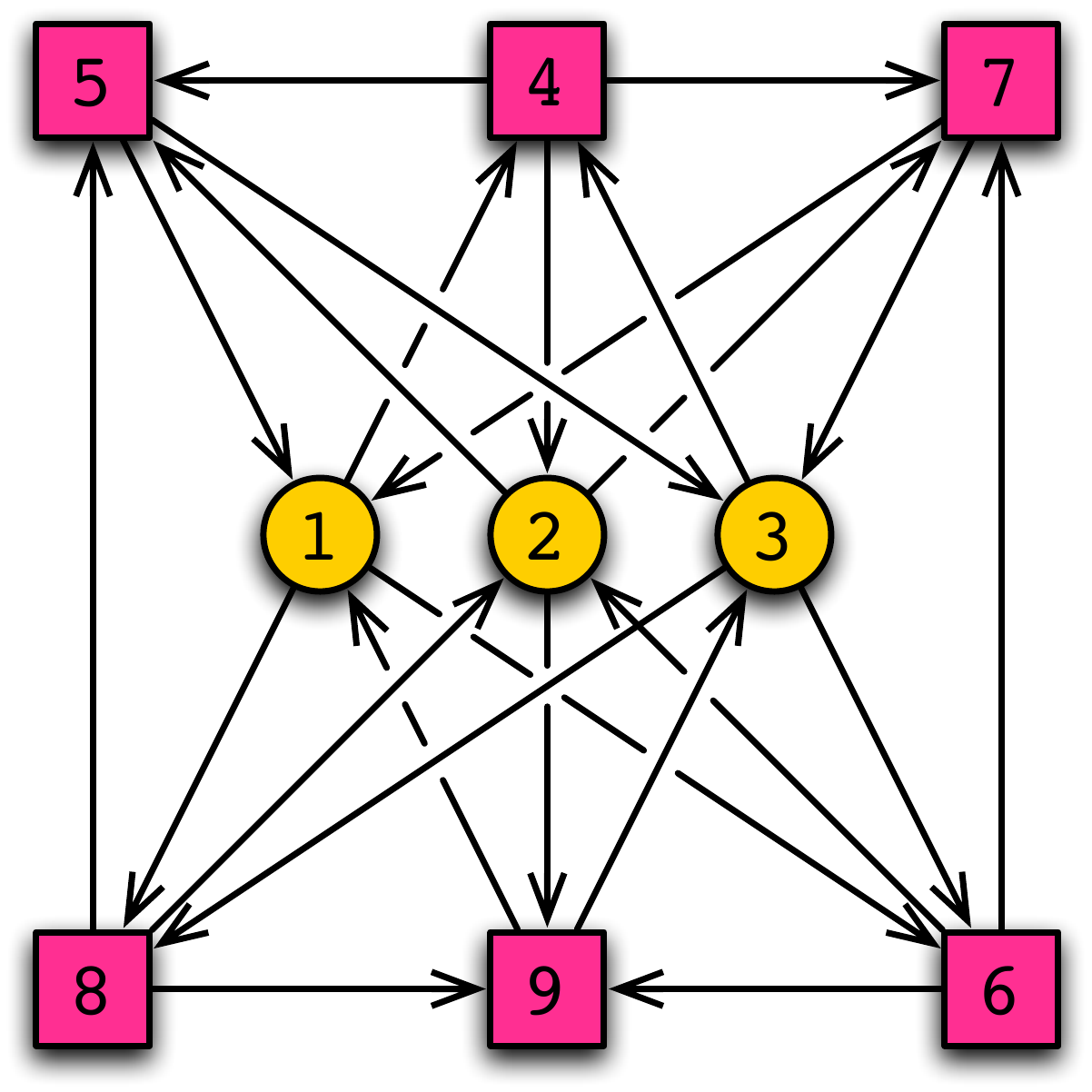}}
\\ 
\hspace{0cm}\mbox{(a)} &  \hspace{2cm} & \hspace{0cm}\mbox{(b)}
 \end{tabular}
  \caption{\textit{The bipartite graph and quiver for $\widetilde{C_3}$.} The bipartite graph lives on a genus 2 Riemann surface with a single boundary.
  \label{fm3aa}}
 \end{center}
 \end{figure}

\paragraph{$\widetilde{C_3}$ Model.} 

The bipartite graph and quiver for this model are shown in \fref{fm3aa}. In order to get the moduli space, we quotient by $Q_1$, $Q_2$ and $Q_3$, obtaining a 10d CY with toric diagram given by

{\tiny
\beq
\begin{array}{rl} G_{\widetilde{C_3}}=\left(
\begin{array}{C{0.35cm}C{0.35cm}C{0.35cm}C{0.35cm}C{0.35cm}C{0.35cm}C{0.35cm}C{0.35cm}C{0.35cm}C{0.35cm}C{0.35cm}C{0.35cm}C{0.35cm}C{0.35cm}C{0.35cm}C{0.35cm}C{0.35cm}C{0.35cm}C{0.35cm}C{0.35cm}C{0.35cm}C{0.35cm}C{0.35cm}}
 -1 & 0 & -1 & -1 & -1 & -1 & -2 & -2 & -1 & -1 & -1 & -1 & -1 & -2 & -1 & 0 & -1 & 0 & 0 & -1 & 0 & 0 & 0 \\
 0 & 0 & 0 & 0 & -1 & 0 & 0 & 1 & 0 & -1 & 0 & 0 & -1 & 0 & -1 & -1 & 0 & -1 & -1 & 0 & -1 & -1 & 0 \\
 0 & 0 & 0 & 0 & 0 & 0 & 0 & 0 & 0 & 0 & 0 & 0 & 1 & 1 & 1 & 0 & 0 & 0 & 0 & 0 & 0 & -1 & -1 \\
 0 & 1 & 1 & 1 & 1 & 1 & 1 & 1 & 1 & 1 & 1 & 1 & 0 & 0 & 0 & 0 & 0 & 0 & 0 & 0 & 0 & 1 & 1 \\
 -1 & 0 & -1 & 0 & -2 & 0 & -1 & 1 & 0 & -1 & 1 & 1 & -3 & -2 & -2 & -3 & -2 & -2 & -2 & -1 & -1 & -2 & 0 \\
 1 & 0 & 1 & 0 & 2 & 1 & 1 & 0 & 1 & 1 & 0 & 0 & 2 & 1 & 1 & 2 & 1 & 1 & 2 & 1 & 1 & 2 & 1 \\
 0 & 0 & 0 & 0 & 0 & -1 & 0 & -1 & 0 & 0 & -1 & 0 & 1 & 1 & 1 & 1 & 1 & 1 & 1 & 1 & 1 & 0 & -1 \\
 1 & 0 & 1 & 1 & 1 & 1 & 1 & 1 & 0 & 1 & 1 & 0 & 1 & 1 & 1 & 1 & 1 & 1 & 0 & 0 & 0 & 1 & 1 \\
 0 & 0 & -1 & -1 & 0 & -1 & 0 & -1 & -1 & 0 & -1 & -1 & 0 & 0 & 0 & 0 & 0 & 0 & 0 & 0 & 0 & 0 & -1 \\
 1 & 0 & 1 & 1 & 1 & 1 & 1 & 1 & 1 & 1 & 1 & 1 & 1 & 1 & 1 & 1 & 1 & 1 & 1 & 1 & 1 & 1 & 1 \\ \hline
\bf{4} & \bf{2} & \bf{1} & \bf{1} & \bf{1} & \bf{1} & \bf{1} & \bf{1} & \bf{1} & \bf{1} & \bf{1} & \bf{1} & \bf{1} & \bf{1} & \bf{1} & \bf{1} & \bf{1} & \bf{1} & \bf{1} & \bf{1} & \bf{1} & \bf{1} & \bf{1} \\ \hline
\end{array}
\right. & \cdots
\end{array} \nonumber
\eeq}

{\tiny
\beq
\ldots 
\left.
\begin{array}{C{0.35cm}C{0.35cm}C{0.35cm}C{0.35cm}C{0.35cm}C{0.35cm}C{0.35cm}C{0.35cm}C{0.35cm}C{0.35cm}C{0.35cm}C{0.35cm}C{0.35cm}C{0.35cm}C{0.35cm}C{0.35cm}C{0.35cm}C{0.35cm}C{0.35cm}C{0.35cm}C{0.35cm}C{0.35cm}C{0.35cm}}
 -1 & -1 & 0 & 0 & 0 & 0 & -1 & -1 & 0 & 0 & -1 & 0 & 0 & -1 & -1 & 0 & 0 & -1 & -1 & -2 & -1 & 0 & -1 \\
 0 & 1 & -1 & 0 & -1 & 0 & 0 & 1 & -1 & 0 & 1 & -1 & 0 & 0 & 1 & -1 & 0 & 0 & 0 & 1 & 0 & 0 & 1 \\
 -1 & -1 & -1 & -1 & -1 & -1 & -1 & -1 & -1 & -1 & 0 & 0 & 0 & 0 & 0 & 0 & 0 & 0 & 1 & 1 & 1 & 0 & 0 \\
 1 & 1 & 1 & 1 & 1 & 1 & 1 & 1 & 1 & 1 & 1 & 1 & 1 & 1 & 1 & 1 & 1 & 1 & 0 & 0 & 0 & 0 & 0 \\
 -1 & 1 & -1 & 1 & -1 & 1 & 0 & 2 & 0 & 2 & 0 & -2 & 0 & -1 & 1 & -1 & 1 & 0 & -2 & -1 & -1 & -2 & -1 \\
 1 & 0 & 1 & 0 & 2 & 1 & 1 & 0 & 1 & 0 & 0 & 1 & 0 & 0 & 0 & 1 & 0 & 0 & 1 & 0 & 0 & 1 & 0 \\
 0 & -1 & 0 & -1 & 0 & -1 & 0 & -1 & 0 & -1 & -1 & 0 & -1 & 0 & -1 & 0 & -1 & 0 & 0 & 0 & 0 & 0 & 0 \\
 1 & 1 & 1 & 1 & 0 & 0 & 0 & 0 & 0 & 0 & 1 & 1 & 1 & 1 & 0 & 0 & 0 & 0 & 1 & 1 & 1 & 1 & 1 \\
 0 & -1 & 0 & -1 & 0 & -1 & 0 & -1 & 0 & -1 & -1 & 0 & -1 & 0 & -1 & 0 & -1 & 0 & 0 & 0 & 0 & 0 & 0 \\
 1 & 1 & 1 & 1 & 1 & 1 & 1 & 1 & 1 & 1 & 1 & 1 & 1 & 1 & 1 & 1 & 1 & 1 & 1 & 1 & 1 & 1 & 1 \\ \hline
\bf{1} & \bf{1} & \bf{1} & \bf{1} & \bf{1} & \bf{1} & \bf{1} & \bf{1} & \bf{1} & \bf{1} & \bf{1} & \bf{1} & \bf{1} & \bf{1} & \bf{1} & \bf{1} & \bf{1} & \bf{1} & \bf{1} & \bf{1} & \bf{1} & \bf{1} & \bf{1} \\ \hline
\end{array}
\right. \ldots
\nonumber
\eeq}

{\tiny
\beq
\ldots 
\left.
\begin{array}{C{0.35cm}C{0.35cm}C{0.35cm}C{0.35cm}C{0.35cm}C{0.35cm}C{0.35cm}C{0.35cm}C{0.35cm}C{0.35cm}C{0.35cm}C{0.35cm}C{0.35cm}C{0.35cm}C{0.35cm}C{0.35cm}C{0.35cm}C{0.35cm}C{0.35cm}C{0.35cm}C{0.35cm}C{0.35cm}C{0.35cm}}
 0 & 0 & -1 & 0 & -1 & 0 & 0 & -1 & -1 & -2 & -1 & 0 & -1 & 0 & -1 & 0 & 1 & 0 & 0 & 1 & 1 & 0 & 0 \\
 0 & 0 & 1 & 0 & 1 & -1 & 0 & 0 & 0 & 1 & 0 & -1 & 0 & 0 & 1 & 0 & -1 & 0 & 1 & -1 & 0 & 0 & 0 \\
 0 & 0 & 0 & 0 & 0 & 0 & 0 & 0 & 1 & 1 & 1 & 1 & 1 & 0 & 0 & 0 & 0 & 0 & -1 & -1 & -1 & -1 & 0 \\
 0 & 0 & 0 & 0 & 1 & 1 & 1 & 1 & 0 & 0 & 0 & 0 & 0 & 0 & 0 & 0 & 0 & 0 & 1 & 1 & 1 & 1 & 0 \\
 -1 & -1 & 0 & 0 & 0 & -2 & 0 & -1 & -2 & -1 & -1 & -2 & -1 & -2 & -1 & -1 & -2 & -1 & 1 & -1 & 1 & 0 & -1 \\
 0 & 1 & 0 & 0 & 0 & 1 & 0 & 0 & 1 & 0 & 0 & 1 & 0 & 1 & 0 & 0 & 1 & 0 & 0 & 1 & 0 & 0 & 1 \\
 0 & 0 & 0 & 0 & -1 & 0 & -1 & 0 & 0 & 0 & 0 & 1 & 1 & 0 & 0 & 0 & 1 & 1 & -1 & 0 & -1 & 0 & 0 \\
 1 & 0 & 0 & 0 & 1 & 1 & 1 & 1 & 1 & 1 & 1 & 0 & 0 & 1 & 1 & 1 & 0 & 0 & 0 & 0 & 0 & 0 & 0 \\
 0 & 0 & 0 & 0 & 0 & 1 & 0 & 1 & 1 & 1 & 1 & 1 & 1 & 1 & 1 & 1 & 1 & 1 & 0 & 1 & 0 & 1 & 1 \\
 1 & 1 & 1 & 1 & 0 & 0 & 0 & 0 & 0 & 0 & 0 & 0 & 0 & 0 & 0 & 0 & 0 & 0 & 0 & 0 & 0 & 0 & 0 \\ \hline
\bf{1} & \bf{1} & \bf{1} & \bf{1} & \bf{1} & \bf{1} & \bf{1} & \bf{1} & \bf{1} & \bf{1} & \bf{1} & \bf{1} & \bf{1} & \bf{1} & \bf{1} & \bf{1} & \bf{1} & \bf{1} & \bf{1} & \bf{1} & \bf{1} & \bf{1} & \bf{1} \\ \hline
\end{array}
\right. \ldots
\nonumber
\eeq}

{\tiny
\beq
\ldots 
\left.
\begin{array}{C{0.35cm}C{0.35cm}C{0.35cm}C{0.35cm}C{0.35cm}C{0.35cm}C{0.35cm}C{0.35cm}C{0.35cm}C{0.35cm}C{0.35cm}C{0.35cm}C{0.35cm}C{0.35cm}C{0.35cm}C{0.35cm}C{0.35cm}C{0.35cm}C{0.35cm}C{0.35cm}C{0.35cm}C{0.35cm}C{0.2cm}}
 -1 & 0 & -1 & -1 & -1 & -2 & -1 & -1 & -1 & 0 & -1 & 0 & 0 & 0 & 0 & 0 & 0 & -1 & 0 & -1 & 0 & 0 & 1 \\
 1 & 0 & 0 & 0 & 0 & 1 & 0 & 0 & 0 & 0 & 1 & 0 & 0 & 0 & 0 & 0 & 0 & 1 & 0 & 1 & 0 & 1 & 0 \\
 0 & 0 & 0 & 0 & 1 & 1 & 1 & 1 & 1 & 0 & 0 & 0 & 0 & 0 & -1 & -1 & 0 & 0 & 0 & 1 & 1 & 0 & 0 \\
 0 & 0 & 1 & 1 & 0 & 0 & 0 & 0 & 0 & 0 & 0 & 0 & 0 & 0 & 1 & 1 & 0 & 0 & 0 & 0 & 0 & 0 & 0 \\
 0 & 0 & -1 & 0 & -1 & 0 & -1 & 0 & 0 & -1 & 0 & -1 & 0 & 0 & 0 & 1 & 0 & 1 & 1 & 0 & 0 & 0 & 0 \\
 0 & 0 & 1 & 0 & 1 & 0 & 1 & 0 & 0 & 1 & 0 & 1 & 0 & 0 & 1 & 0 & 1 & 0 & 0 & 0 & 0 & 0 & 0 \\
 0 & 0 & 0 & 0 & 0 & 0 & 1 & 0 & 1 & 0 & 0 & 1 & 0 & 1 & 0 & 0 & 0 & 0 & 0 & 0 & 0 & 0 & 0 \\
 0 & 0 & 1 & 1 & 1 & 1 & 0 & 1 & 0 & 1 & 1 & 0 & 1 & 0 & 0 & 0 & 0 & 0 & 0 & 0 & 0 & 0 & 0 \\
 1 & 1 & 0 & 0 & 0 & 0 & 0 & 0 & 0 & 0 & 0 & 0 & 0 & 0 & 0 & 0 & 0 & 0 & 0 & 0 & 0 & 0 & 0 \\
 0 & 0 & 0 & 0 & 0 & 0 & 0 & 0 & 0 & 0 & 0 & 0 & 0 & 0 & 0 & 0 & 0 & 0 & 0 & 0 & 0 & 0 & 0 \\ \hline
\bf{1} & \bf{1} & \bf{1} & \bf{1} & \bf{1} & \bf{1} & \bf{1} & \bf{1} & \bf{1} & \bf{1} & \bf{1} & \bf{1} & \bf{1} & \bf{1} & \bf{1} & \bf{1} & \bf{1} & \bf{1} & \bf{1} & \bf{1} & \bf{1} & \bf{1} & \bf{1} \\ \hline
\end{array}
\right).
\eeq}

\bigskip

The explicit form of this large matrix is not particularly illuminating. It is presented in order to show the very small perfect matching multiplicity of the points in the toric diagram. The 96 perfect matchings of this theory project onto 92 distinct points in the toric diagram of the moduli space, out of which only 2 have multiplicity greater than one. The small multiplicities are directly correlated with the irreducibility of the corresponding bipartite graph.

\section{Higgsing and Unhiggsing \label{shiggs}}

\sref{section_reduction_by_higgsing} above discussed in detail higgsing in BFTs, which corresponds to removal of edges in the bipartite graph that are not external legs. There it is explained that, in some instances, higgsing can lead to equivalent theories, i.e. theories with the same moduli space. More generally, higgsing can be used to generate theories for which the moduli space is different.\footnote{As explained in \sref{section_moduli_spaces} and motivated in part by applications to scattering amplitudes, this paper focuses on the study of the Abelian moduli space. It is important to reiterate that, for general non-Abelian BFTs, it is quite possible that the coincidence of Abelian moduli spaces does not imply a duality between theories.  This is an interesting question that certainly deserves future study. This would imply that even higgsings leading to theories with the same Abelian moduli space might produce genuinely new models. Keeping this possibility in mind, this section is devoted to the more dramatic case in which even the Abelian moduli space of the higgsed theory differs from the original one.} The inverse procedure relates to introducing a new field which is associated to an edge in the bipartite graph such that the new theory is still a BFT.\footnote{In parallel with the definition of higgsing, we do not call unhiggsing the addition of an edge that results in two new external faces, i.e. the introduction of a new external leg. This is because the inverse process only amounts to the spontaneous breaking of two global symmetry groups down to the diagonal combination.} This process is referred to as {\bf unhiggsing}. In the case of BFTs on $T^2$, both methods have been successfully exploited for generating new gauge theories on D3-branes probing CY 3-folds. In \cite{Franco:2012mm}, the connection between certain higgsings of BFTs on a disk and the boundary operator in cells of the positive Grassmannian was elucidated.

\bigskip

\subsection{Another New Class of BFTs: Higgsing $C_n$}

\label{section_Cn'}

This section introduces a new class of BFTs, which are denoted $C'_n$. They are obtained from $C_n$ by higgsing with $n$ non-zero vevs. These vevs correspond to removing edges between pairs of square internal faces, turning them into hexagons. The perfect matching matrices for these theories are collected in appendix \ref{appendix_Cn'}. 

\medskip
As for the $C_n$ and $\widetilde{C_n}$ families, some general properties of this class of theories can be summarized as follows:
\begin{itemize}
\item \textit{Faces/Groups:} There are $n$ internal faces and $2n$ external ones, which gives a total number of $3n$ faces.
\item \textit{Edges/Fields:} The total number of edges is $7n$, with $5n$ internal edges and $2n$ external legs.
\item \textit{Nodes/$W$-Terms:} There are $2n$ internal white nodes and $2n$ internal black nodes. 
\item \textit{Master Space:} It is a toric Calabi-Yau manifold of dimension $(3n+1)$.
\item \textit{Mesonic Moduli Space:} It is a toric Calabi-Yau manifold of dimension $(2n+1)$.
\end{itemize}

\bigskip

\subsubsection{Examples}

\begin{figure}[ht!!]
\begin{center}
 \begin{tabular}[c]{ccc}
\includegraphics[trim=0cm 0cm 0cm 0cm,totalheight=4cm]{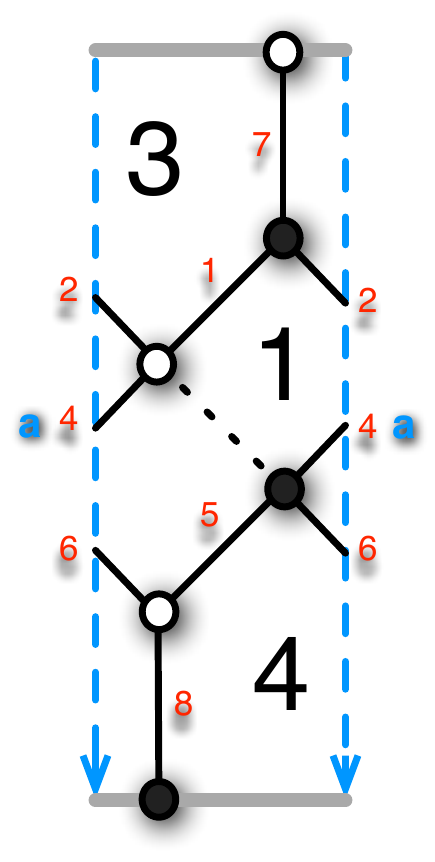}
& \hspace{2cm} &
\includegraphics[trim=0cm 0cm 0cm 0cm,totalheight=4cm]{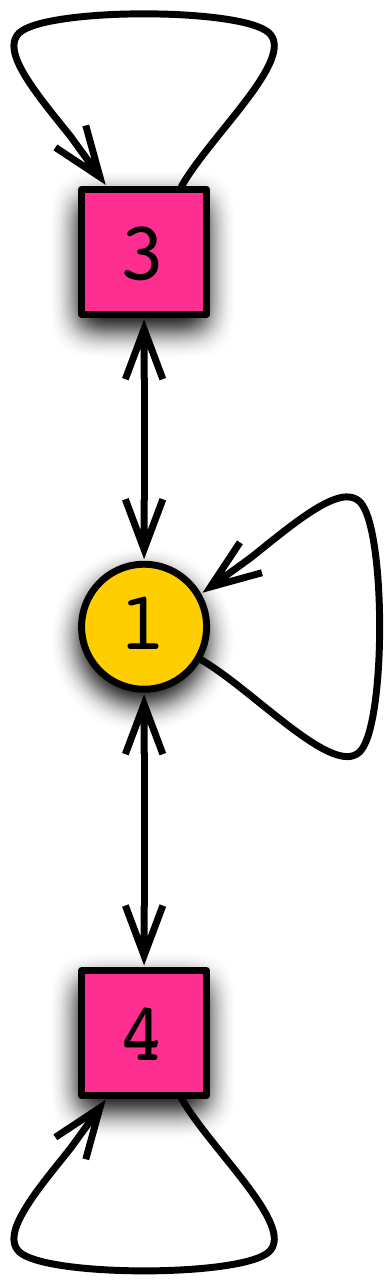}
\\ 
\hspace{0cm}\mbox{(a)} &  \hspace{2cm} & \hspace{-0.3cm}\mbox{(b)}
 \end{tabular}
  \caption{\textit{The bipartite graph and quiver for $C'_1$.}
  \label{fm1b}}
 \end{center}
 \end{figure}

\paragraph{$C'_1$ Model.} 
This model is obtained from $C_1$ by giving a non-zero vev to $X_3$. The resulting bipartite graph and quiver diagram are shown in \fref{fm1b}. 

Using the perfect matching matrix in appendix \ref{appendix_Cn'}, one sees that higgsing eliminates a single perfect matching, removing the corresponding point from the toric diagram of the moduli space. The toric diagram is given by

{\footnotesize
\beq
G_{C'_1}=\left(
\begin{array}{C{0.35cm}C{0.35cm}C{0.35cm}C{0.2cm}}
 -1 & -2 & 0 & 1 \\
 1 & 2 & 0 & 0 \\
 1 & 1 & 1 & 0 \\
\hline
\bf{2} & \bf{1} & \bf{1} & \bf{1} \\
\hline
\end{array}\right).
\label{G_matrix_multiplicities_C1'}
\eeq}

\bigskip

\paragraph{$C'_2$ Model.} 

\begin{figure}[ht!!]
\begin{center}
 \begin{tabular}[c]{ccc}
\includegraphics[trim=0cm 0cm 0cm 0cm,totalheight=4 cm]{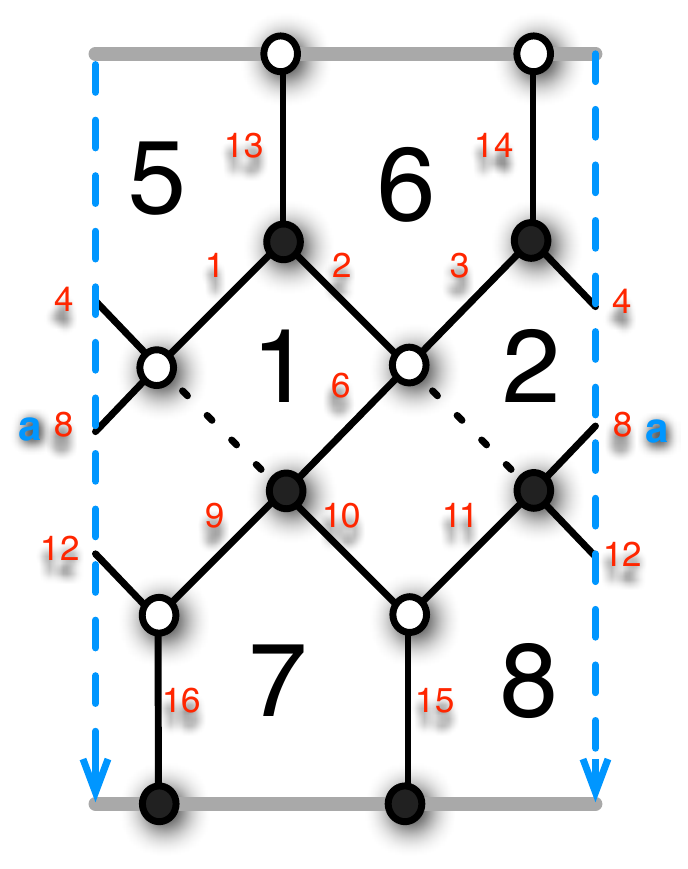}
& \hspace{2cm} &
\includegraphics[trim=0cm 0cm 0cm 0cm,totalheight=4 cm]{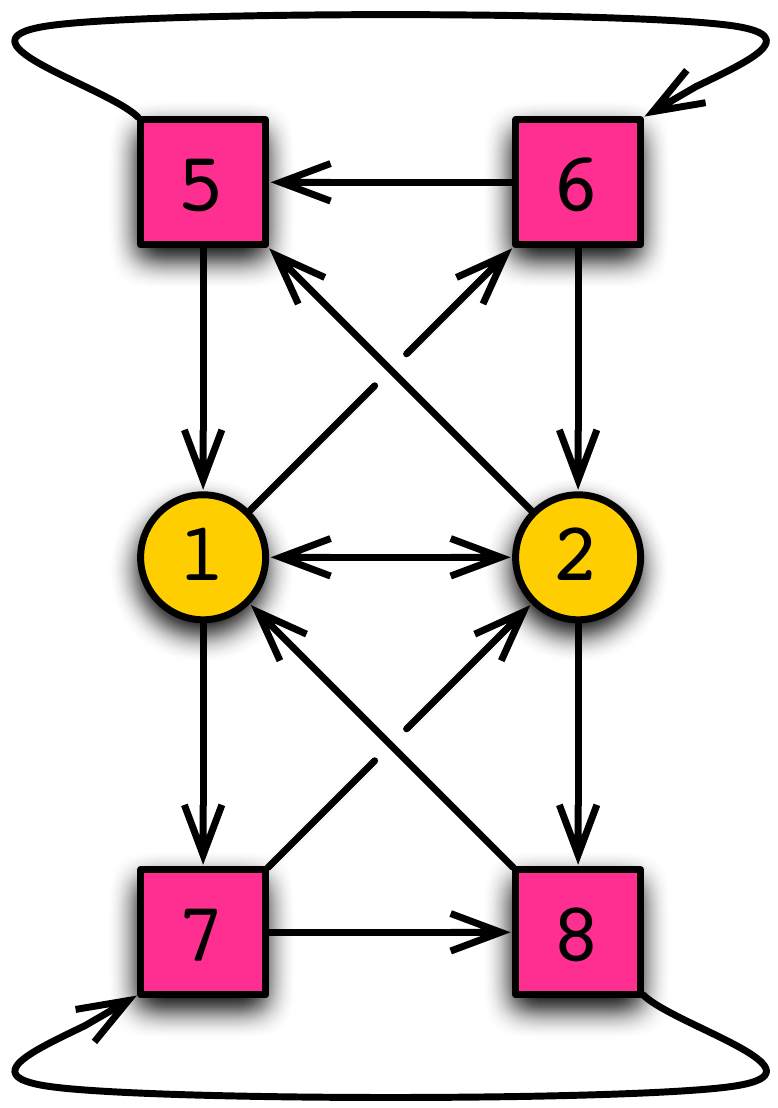}
\\ 
\hspace{0cm}\mbox{(a)} &  \hspace{2cm} & \hspace{0cm}\mbox{(b)}
 \end{tabular}
  \caption{\textit{The bipartite graph and quiver for $C'_2$.}
  \label{fm2b}}
 \end{center}
 \end{figure}

This model is obtained from $C_2$ by giving vevs to $X_5$ and $X_7$. The corresponding bipartite graph and quiver diagram are shown in \fref{fm2b}. 

Higgsing results in the removal of 9 perfect matchings. The toric diagram for the moduli space corresponds to

{\footnotesize
\beq
G_{C'_2}=\left(
\begin{array}{C{0.35cm}C{0.35cm}C{0.35cm}C{0.35cm}C{0.35cm}C{0.35cm}C{0.35cm}C{0.35cm}C{0.35cm}C{0.3cm}}
 0 & -1 & 1 & 0 & 0 & 0 & 1 & -1 & 0 & 0 \\
 1 & -1 & 0 & -1 & 0 & 1 & 1 & 0 & 0 & -1 \\
 0 & 1 & 0 & 0 & 0 & -1 & -1 & 0 & 0 & 1 \\
 0 & 1 & 0 & 1 & 1 & 1 & 0 & 1 & 0 & 1 \\
 0 & 1 & 0 & 1 & 0 & 0 & 0 & 1 & 1 & 0 \\ \hline
\bf{2}&\bf{1}&\bf{1}&\bf{2}&\bf{2}&\bf{1}
&\bf{1}&\bf{1}&\bf{1}&\bf{1} \\ 
\hline
\end{array}
\right) .
\label{G_matrix_multiplicities_C2'}
\eeq}
Comparing to \eref{G_matrix_multiplicities_C20}, one notices that the multiplicity of 4 points is reduced and 3 points disappear completely.

\bigskip

\paragraph{$C'_3$ Model.} 

\begin{figure}[htt!!]
\begin{center}
 \begin{tabular}[c]{ccc}
\includegraphics[trim=0cm 0cm 0cm 0cm,totalheight=4 cm]{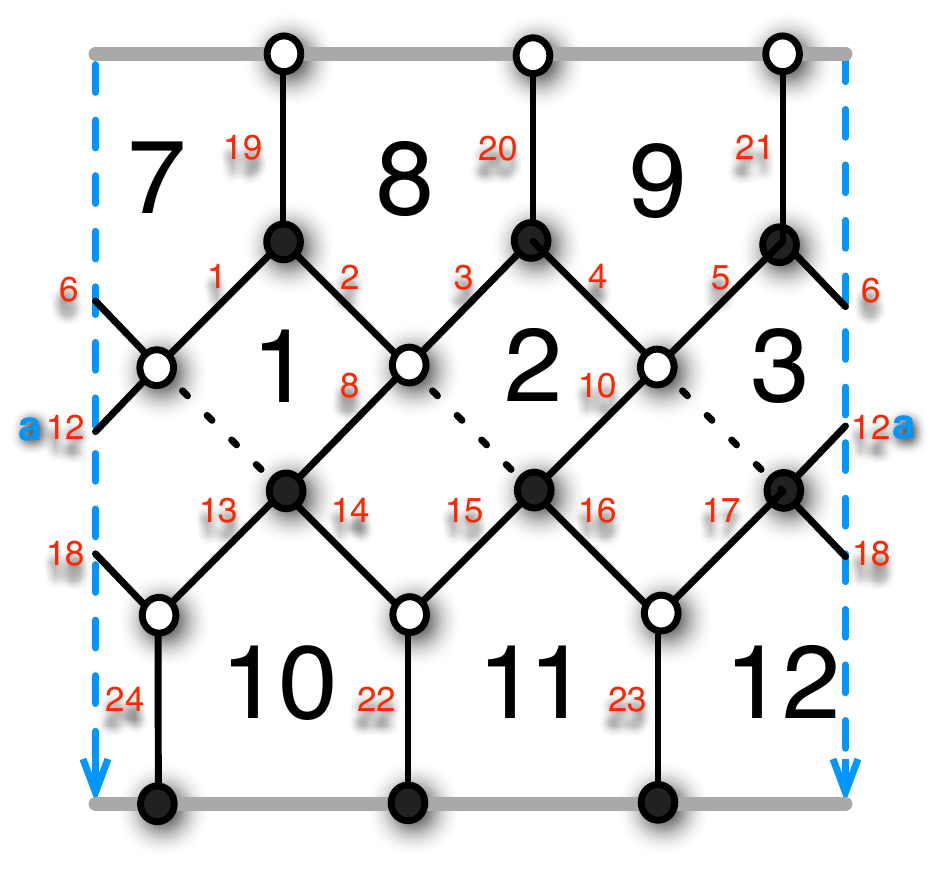}
& \hspace{2cm} &
\includegraphics[trim=0cm 0cm 0cm 0cm,totalheight=4 cm]{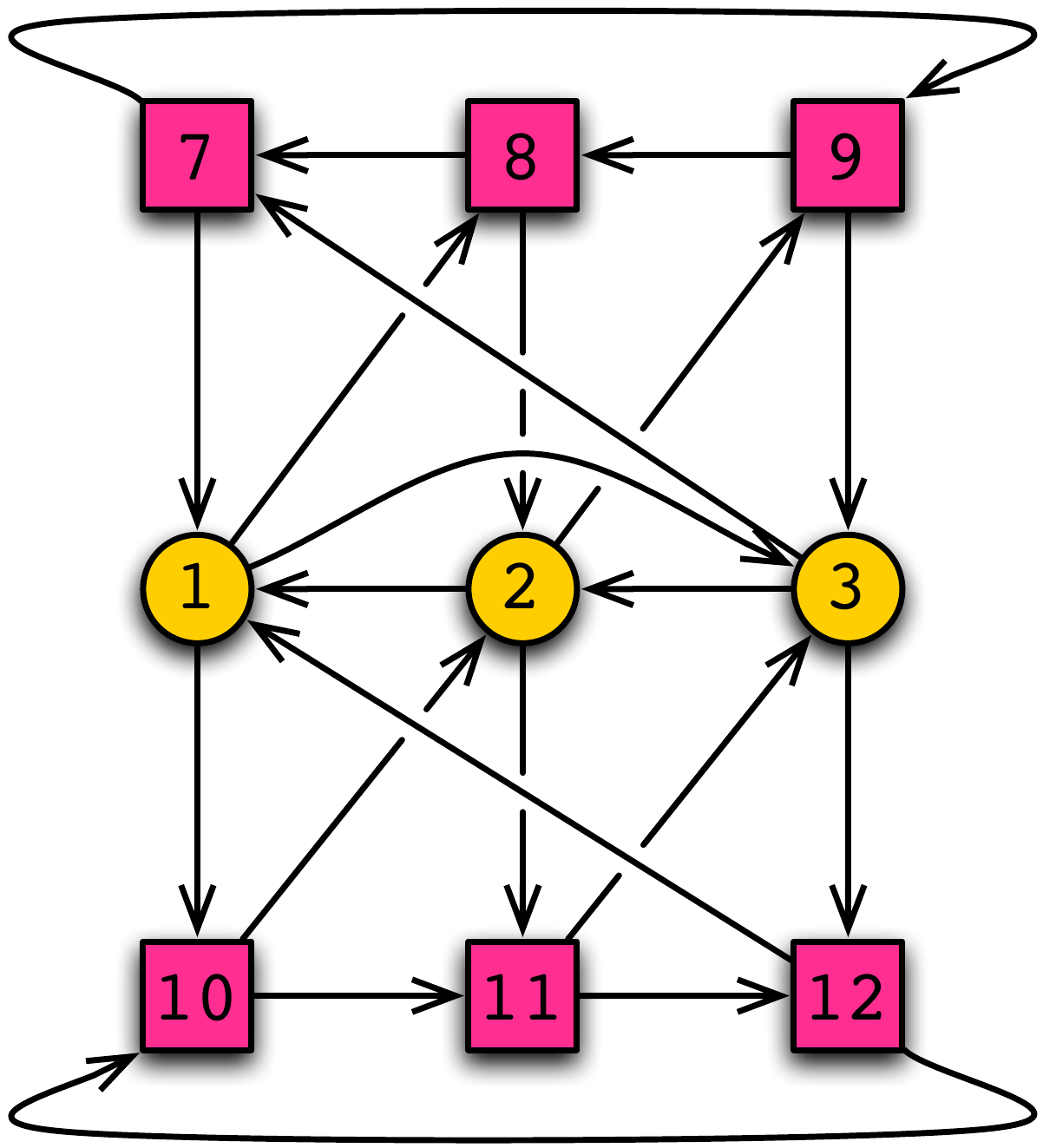}
\\ 
\hspace{0cm}\mbox{(a)} &  \hspace{2cm} & \hspace{0cm}\mbox{(b)}
 \end{tabular}
  \caption{\textit{The bipartite graph and quiver for $C'_3$.}
  \label{fm3b}}
 \end{center}
 \end{figure}

Starting from $C_3$, vevs are given to $X_7$, $X_9$ and $X_{11}$. The resulting theory is shown in \fref{fm3b}. 

The toric diagram of the moduli space corresponds to

\bigskip

{\scriptsize
\beq
G_{C_3} =
\left(
\begin{array}{C{0.35cm}C{0.35cm}C{0.35cm}C{0.35cm}C{0.35cm}C{0.35cm}C{0.35cm}C{0.35cm}C{0.35cm}C{0.35cm}C{0.35cm}C{0.35cm}C{0.35cm}C{0.35cm}}
 -2 & -1 & 0 & -1 & 0 & 0 & -1 & 0 & 0 & -1 & -1 & -1 & -1 & 0 \\
 1 & 0 & 0 & 1 & 1 & 0 & 1 & -1 & 0 & 0 & 1 & 0 & 0 & -1 \\
 1 & 1 & 0 & 0 & 1 & 0 & 1 & 0 & -1 & 0 & 2 & 0 & 1 & -1 \\
 0 & 0 & 0 & 0 & -1 & 0 & 0 & 1 & 1 & 1 & -1 & 1 & 0 & 1 \\
 0 & 0 & -1 & 0 & 0 & 1 & 0 & -1 & 1 & -1 & -1 & 1 & -1 & 0 \\
 1 & 1 & 1 & 1 & 0 & 0 & 0 & 1 & 0 & 1 & 1 & 0 & 1 & 1 \\
 0 & 0 & 1 & 0 & 0 & 0 & 0 & 1 & 0 & 1 & 0 & 0 & 1 & 1 \\
\hline
 \bf{3}  &  \bf{2}  &  \bf{3}  &  \bf{2}  &  \bf{2}  &  \bf{3}  &  \bf{1}  &  \bf{1}  &  \bf{1}  &  \bf{1}  &  \bf{1}  &  \bf{2}  &  \bf{2}  &  \bf{2}  \\
\hline
\end{array}
\right. \cdots 
\nonumber
\eeq

\beq
\cdots
\left. \begin{array}{C{0.35cm}C{0.35cm}C{0.35cm}C{0.35cm}C{0.35cm}C{0.35cm}C{0.35cm}C{0.35cm}C{0.35cm}C{0.35cm}C{0.35cm}C{0.35cm}C{0.35cm}C{0.2cm}}
 -1 & -1 & 0 & 1 & -1 & -1 & -1 & -1 & -1 & -1 & 0 & 0 & 0 & 0 \\
 2 & 1 & 0 & -1 & 1 & -1 & -1 & 2 & 0 & 0 & -1 & -1 & 0 & 0 \\
 1 & 0 & 0 & -1 & 1 & -1 & 0 & 2 & -1 & 0 & -1 & 0 & -1 & 0 \\
 -1 & 0 & 0 & 1 & -1 & 2 & 1 & -2 & 1 & 0 & 2 & 1 & 1 & 0 \\
 1 & 1 & 0 & 0 & 0 & 0 & 0 & 0 & 0 & 0 & 0 & 0 & 0 & 0 \\
 0 & 0 & 1 & 0 & 1 & 1 & 1 & 1 & 1 & 1 & 0 & 0 & 0 & 0 \\
 -1 & 0 & 0 & 1 & 0 & 1 & 1 & -1 & 1 & 1 & 1 & 1 & 1 & 1 \\
\hline
\bf{1}  &  \bf{2}  &  \bf{1}  &  \bf{2}  &  \bf{2}  &  \bf{1}  &  \bf{1}  &  \bf{1}  &  \bf{1}  &  \bf{1}  &  \bf{1}  &  \bf{1}  &  \bf{1}  &  \bf{1} \\ 
\hline
\end{array}
\right) .
\label{G_matrix_multiplicities_C3'}
\eeq}

\smallskip

\noindent Higgsing reduces the original 96 perfect matchings down to 44. A comparison with \eref{G_matrix_multiplicities_C30} reveals that multiplicities of some points in the toric diagram decrease and that 10 points disappear.

\bigskip

\subsection{Unhiggsing $\widetilde{C_3}$}

\label{section_unhiggsing_tildeC3}

Unhiggsing effectively splits a face in the bipartite graph into two separate faces. Such splitting is achieved by adding a diagonal edge and increasing the valence of a white and black node adjacent to the split face. The dimension of the moduli space remains constant in this process.

\begin{figure}[h]
\begin{center}
\includegraphics[height=7cm]{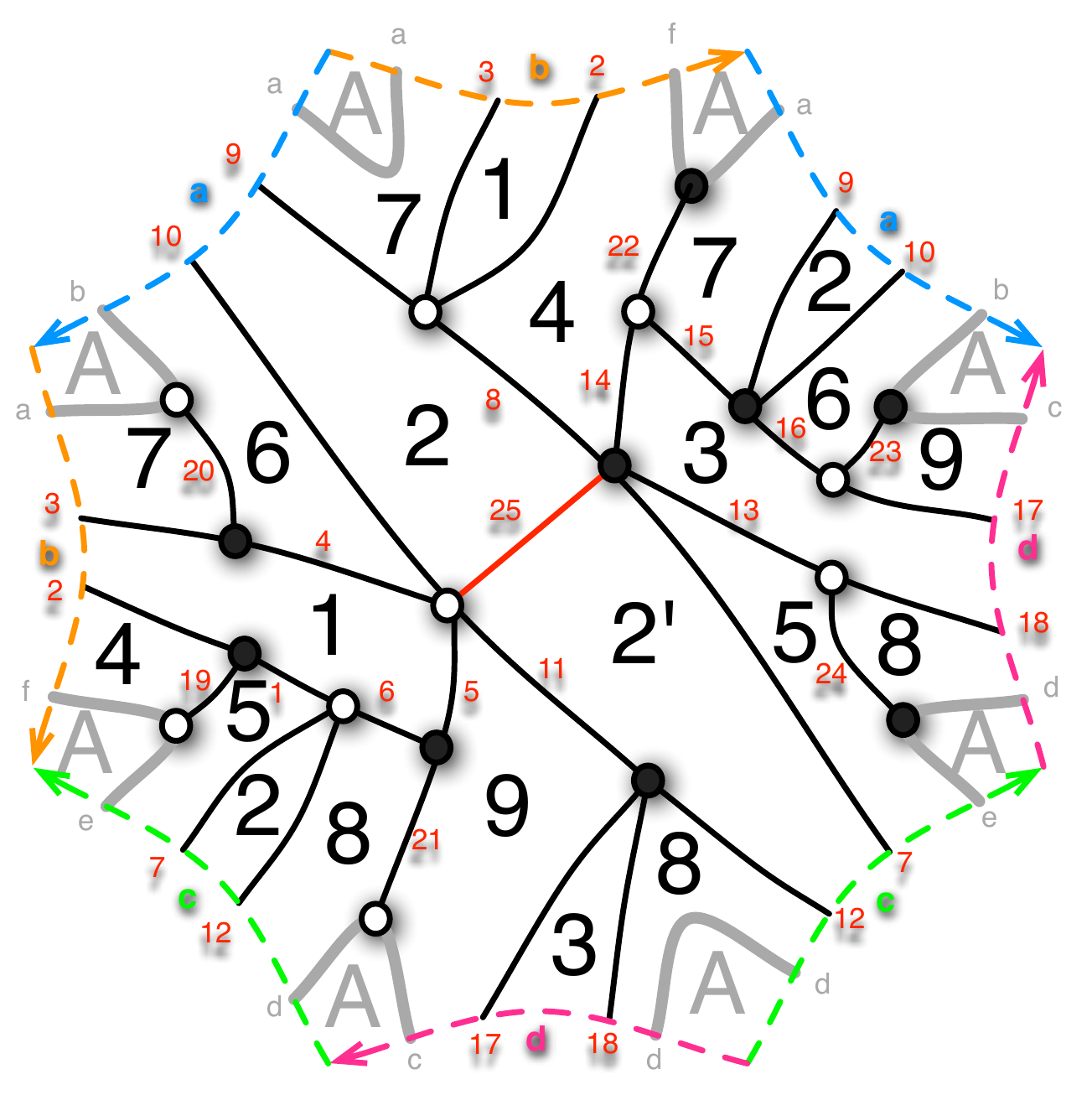}
\caption{\textit{Bipartite graph for a theory obtained by unhiggsing the $\widetilde{C_3}$ model.}}
\label{unhiggsing_tildeC3}
\end{center}
\end{figure}

As an example, let us consider the $\widetilde{C_3}$ model. \fref{unhiggsing_tildeC3} shows one possible unhiggsing of this theory, which has a couple of square faces. \sref{section_Seiberg} revisits this model in order to use it to illustrate Seiberg duality in a genus 2 BFT and to discuss its moduli space in detail. 

In this example, the moduli space of the new theory is different from that of the original one. The addition of {\bf BCF bridges} is a particular case of unhiggsing that becomes extremely useful for generating identities in the context of scattering amplitudes \cite{Britto:2004ap,Britto:2005fq,Nima}. Of course, reversing the arguments in \sref{section_reduction_by_higgsing}, it is also possible to find unhiggsed theories that share the moduli space of their parents. 

\bigskip

\section{Sewing \label{ssew}}

This section introduces an operation called {\bf sewing}, which corresponds to the identification of two boundary components of the graph containing the same number of external legs terminating on them.\footnote{Of course we can, more broadly, also consider the identification of individual pairs of external legs.}  The two glued boundaries might belong to the same Riemann surface or, more generally, correspond to individual Riemann surfaces that are merged into a single one. The sewing process is not unique, since there exist a discrete analogue of a Dehn twist, to which we refer as the {\bf sewing twist}, controlling how edges on the two boundaries are identified. As long as one keeps track of which external edges are merged, the process of sewing boundaries commutes with the untwisting move on a bipartite graph on $\Sigma$ which was discussed in \sref{section_untwisting}. The following section illustrates this feature with explicit examples.

\bigskip

\subsection{Orbifold Theories from Sewing Cylinder BFTs}

\begin{figure}[ht!]
\begin{center}
\includegraphics[trim=0cm 0cm 0cm 0cm,totalheight=4.5 cm]{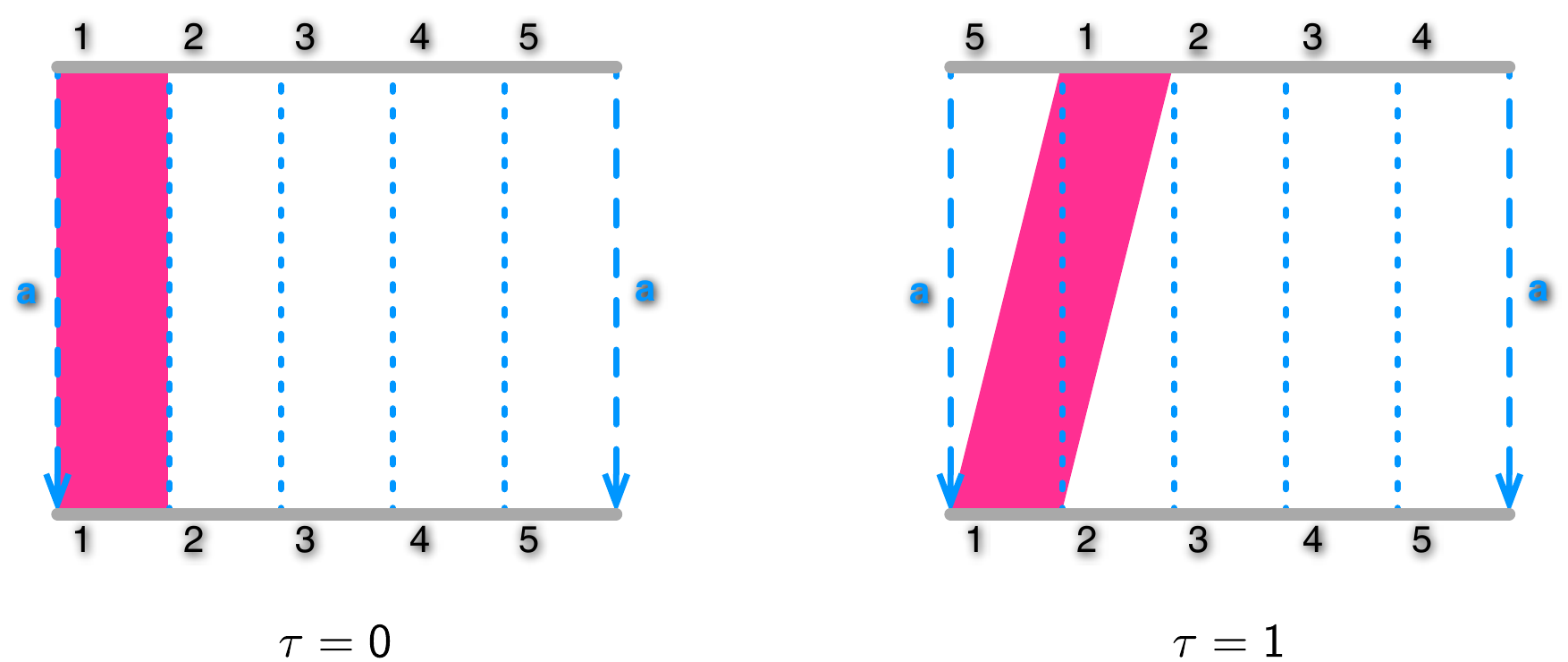}
  \caption{\textit{Schematic representation of the sewing operation along the two boundaries of a BFT on a cylinder.} The resulting theory corresponds to an orbifold of order 5, whose geometric action is controlled by the sewing twist, which can take the values $\tau=0,\dots,4$.
  \label{fsewing}}
 \end{center}
 \end{figure}

This section illustrates the effect of sewing in an interesting class of examples. Let us consider BFTs defined on a cylinder with an equal number of external nodes on each of the two boundaries. These are sewed together to obtain theories on $T^2$. The original theories are further restricted in order to be of a very specific type, in which the bipartite graph corresponds to the repetition of $n$ copies of a more elementary graph along the periodic direction. For this class of models, sewing results on BFTs associated to orbifolds of CY 3-folds. The sewing twist controls how unit cells are identified in the resulting theory on $T^2$ which, in turn, is in one-to-one correspondence with the choice of orbifold action \cite{Hanany:2010cx,Davey:2011dd,Hanany:2010ne,Davey:2010px,Hanany:2011iw}. \fref{fsewing} provides a schematic representation of the situation under consideration.

Let us focus on the $C_n$ class of models introduced in \sref{section_Cn} and their corresponding untwisted theories $\widetilde{C_n}$. The fundamental domain for $C_n$ is given in \fref{fm4a}. For $C_n$, the sewing parameter can take values $\tau=0,\dots,n-1$. The sewed theory is called $\sigma_\tau(C_n)$. The perfect matching matrices for the models are collected in appendix \ref{appendix_sewed_models}.

\bigskip

\paragraph{$\sigma_{0}(C_1)$ Model and its Untwisting.}

$\sigma_{0}(C_1)$ is obtained by sewing the external edges 7 and 8 in the bipartite graph for $C_1$ given in \fref{fm1a}. The resulting graph, shown in \fref{fm1c} (a), is the brane tiling for D3-branes over a complex cone over the suspended pinch point (SPP) \cite{Franco:2005rj}. Indeed, the toric diagram for the moduli space becomes
{\footnotesize
\beq
G_{\sigma_0(C_1)}=\left(
\begin{array}{C{0.35cm}C{0.35cm}C{0.35cm}C{0.35cm}C{0.2cm}}
 -1 & -2 & 0 & 0 & 1 \\
 1 & 2 & 0 & 1 & 0 \\
 1 & 1 & 1 & 0 & 0 \\
\hline
 \bf{2}  &  \bf{1}  &  \bf{1}  &  \bf{1}  &  \bf{1} \\ 
\hline
\end{array}
\right),
\label{G_matrix_multiplicities_sewed_C1}
\eeq}
which corresponds to the complex cone over SPP.

\begin{figure}[h]
\begin{center}
 \begin{tabular}[c]{ccc}
\includegraphics[trim=0cm 0cm 0cm 0cm,totalheight=4.5 cm]{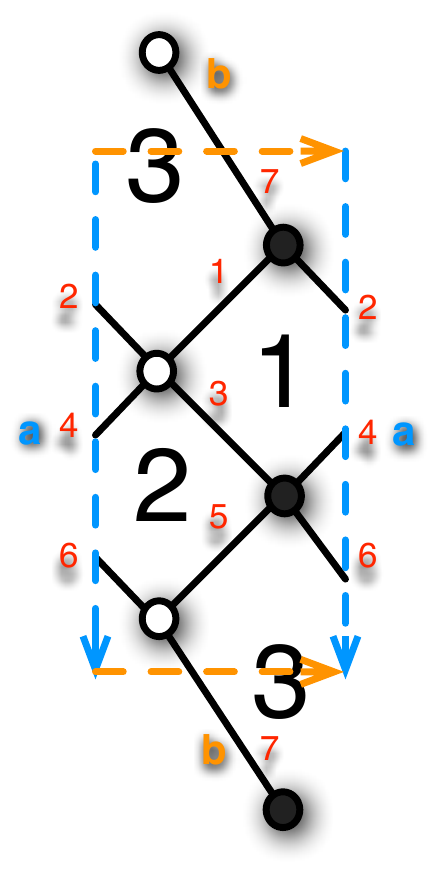}
& \hspace{2cm} &
\includegraphics[trim=0cm 0cm 0cm 0cm,totalheight=4.5 cm]{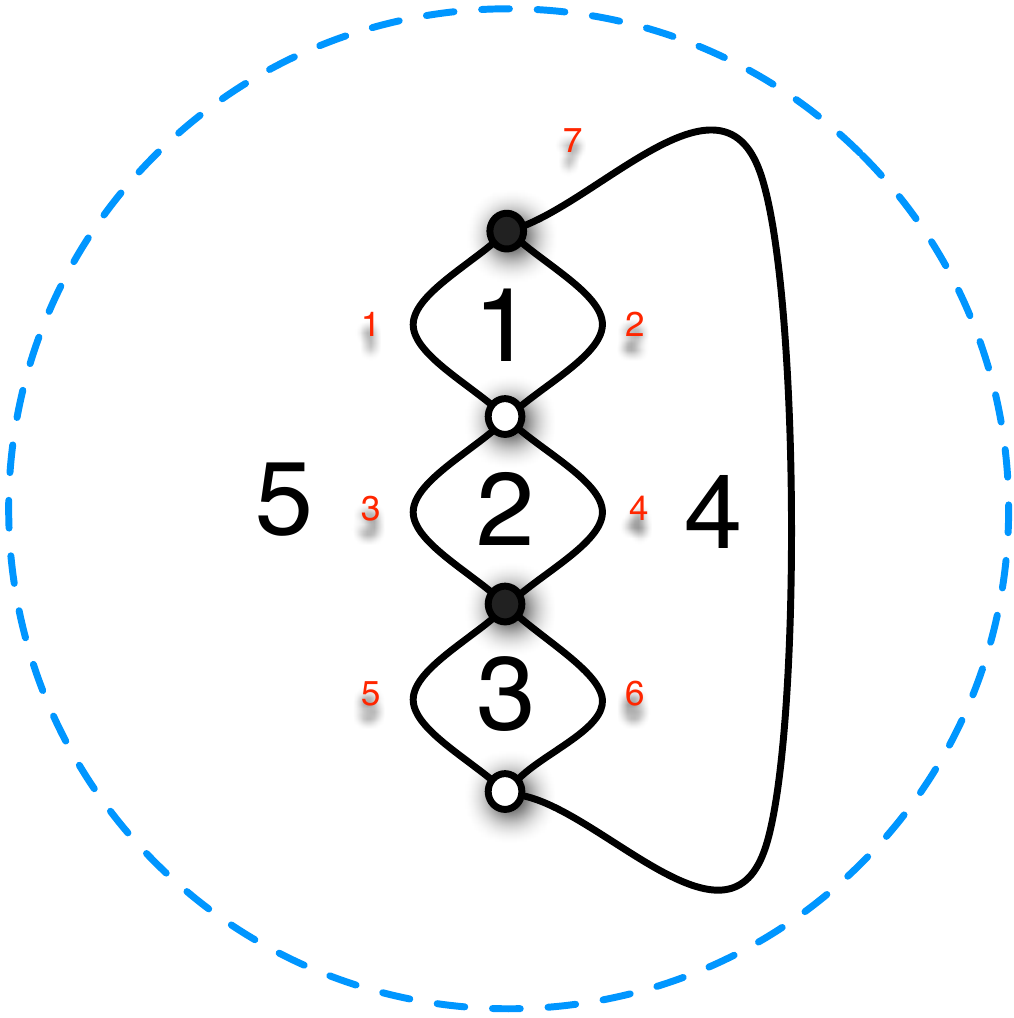}
\\ 
\hspace{0cm}\mbox{(a)} &  \hspace{2cm} & \hspace{0cm}\mbox{(b)}
 \end{tabular}
  \caption{\textit{The sewed model $\sigma_0(C_1)$ and its untwisting, $\widetilde{\sigma_0(C_1)}$.}
  \label{fm1c}}
 \end{center}
 \end{figure}

Untwisting $\sigma_0(C_1)$, one obtains the graph on a sphere with no boundaries shown in \fref{fm1c} (b), which is clearly highly reducible. The moduli space is 1-dimensional and its toric diagram is given by
{\footnotesize
\beq
G_{\widetilde{\sigma_0(C_1)}}=\left(
\begin{array}{c}
1 \\
\hline
\ \bf{6} \ \\ 
\hline
\end{array}
\right),
\label{G_matrix_multiplicities_untwisted_sewed_C1}
\eeq}
i.e. the six perfect matchings of this theory collapse onto a single point.

\bigskip

\paragraph{$\sigma_{0}(C_2)$ Model and its Untwisting.}

\begin{figure}[ht!]
\begin{center}
 \begin{tabular}[c]{ccc}
\includegraphics[trim=0cm 0cm 0cm 0cm,totalheight=4.5 cm]{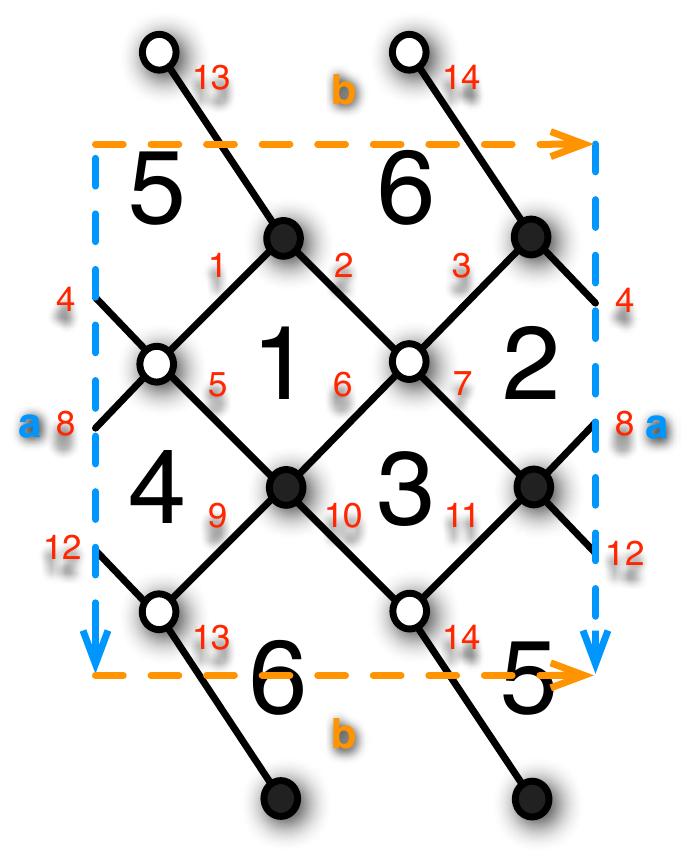}
& \hspace{2cm} &
\includegraphics[trim=0cm 0cm 0cm 0cm,totalheight=4.5 cm]{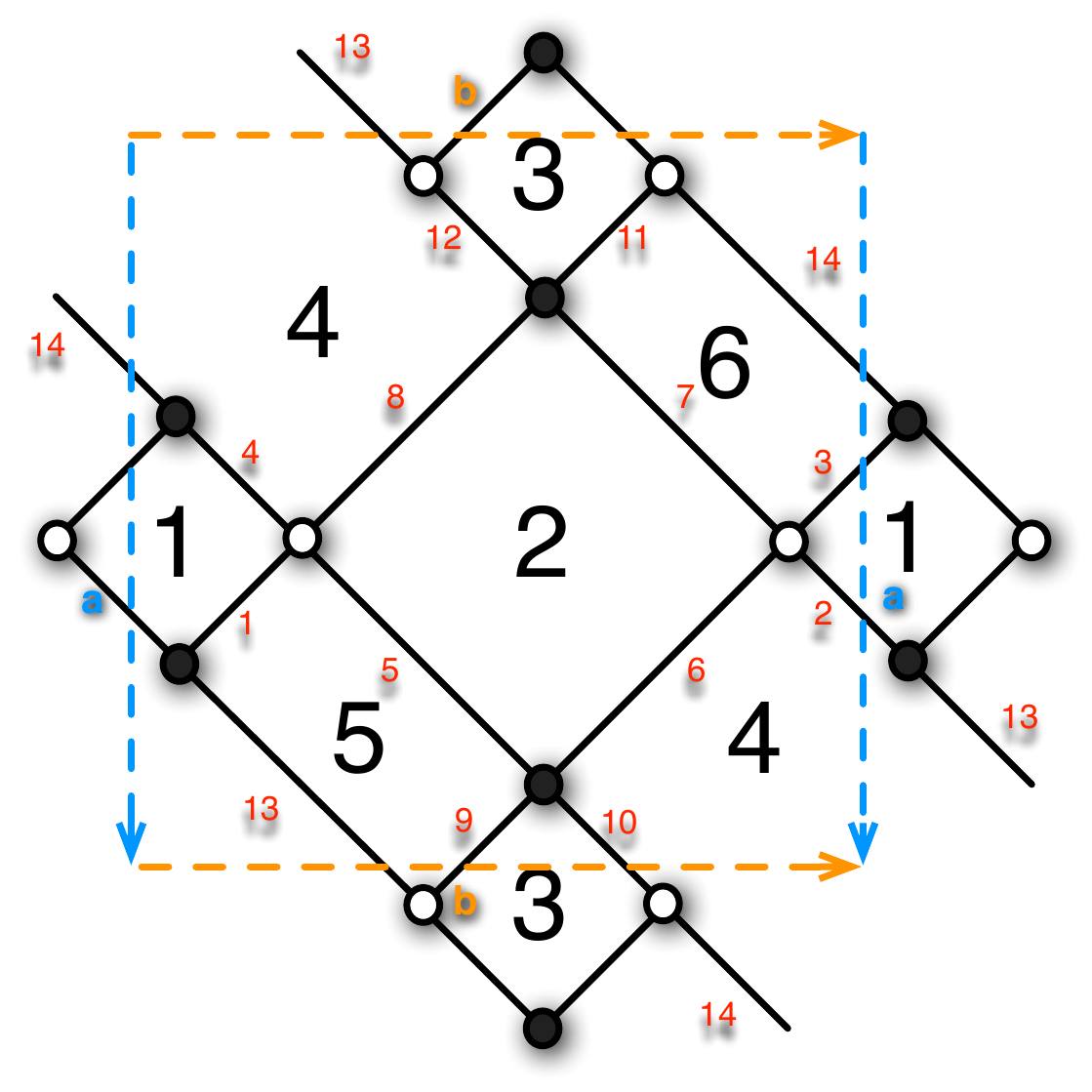}
\\ 
\hspace{0cm}\mbox{(a)} &  \hspace{2cm} & \hspace{0cm}\mbox{(b)}
 \end{tabular}
  \caption{\textit{The sewed model $\sigma_0(C_2)$ and its untwisting, $\widetilde{\sigma_0(C_2)}$.}
  \label{fm2c}}
 \end{center}
 \end{figure}

The bipartite graph for $\sigma_{0}(C_2)$ is shown in \fref{fm2c} (a). The moduli space of this theory is again a 3d toric CY with the toric diagram given by the matrix

{\footnotesize
\beq
G_{\sigma_0(C_2)}=\left(
\begin{array}{C{0.35cm}C{0.35cm}C{0.35cm}C{0.35cm}C{0.35cm}C{0.35cm}C{0.3cm}}
 1 & 0 & 0 & 1 & 1 & 2 & -1 \\
 0 & 0 & 1 & -1 & 1 & 0 & 1 \\
 0 & 1 & 0 & 1 & -1 & -1 & 1 \\
\hline
 \bf{6}  &  \bf{2}  &  \bf{2}  &  \bf{1}  &  \bf{1}  &  \bf{1}  &  \bf{1}  \\ 
\hline
\end{array}
\right).
\label{G_matrix_multiplicities_sewed_C2}
\eeq}
\medskip

\noindent This is in fact an Abelian orbifold of the form $\text{SPP}/\mathbb{Z}_2$, corresponding to the orbifold action $(0,1,1,1)$.\footnote{An Abelian orbifold action of the form $(a_1,a_2,...,a_m)$ specifies the action of the quotienting group $\mathbb{Z}_N$ on $\mathcal{M}$ for an orbifold of the form $\mathcal{M}/\mathbb{Z}_N$. The entries $a_i$ relate to the $m$ generators $z_i$ of the space $\mathcal{M}$ such that $z_i \sim \omega^a_i z_i$ and $\omega^N=1$\cite{Hanany:2010cx,Davey:2011dd,Hanany:2010ne,Davey:2010px,Hanany:2011iw}.}

The untwisted theory $\widetilde{\sigma_{0}(C_2)}$ is shown in \fref{fm2c} (b). The theory lives on a $T^2$, so its moduli space is also a CY 3-fold. Its toric diagram corresponds to

{\footnotesize
\beq
G_{\widetilde{\sigma_0(C_2)}}=\left(
\begin{array}{C{0.35cm}C{0.35cm}C{0.35cm}C{0.35cm}C{0.35cm}C{0.35cm}C{0.3cm}}
 0 & 1 & 0 & -1 & -1 & 0 & 1 \\
 1 & 1 & 2 & 2 & 1 & 0 & 0 \\
 0 & -1 & -1 & 0 & 1 & 1 & 0 \\
\hline
 \bf{8}  &  \bf{1}  &  \bf{1}  &  \bf{1}  &  \bf{1}  &  \bf{1}  &  \bf{1} \\ 
\hline
\end{array}
\right).
\label{G_matrix_multiplicities_untwisted_sewed_C2}
\eeq}

\bigskip

\paragraph{$\sigma_{0}(C_3)$ Model and its Untwisting.}

\begin{figure}[ht!]
\begin{center}
 \begin{tabular}[c]{ccc}
\raisebox{0.73cm}{\includegraphics[trim=0cm 0cm 0cm 0cm,totalheight=4.5 cm]{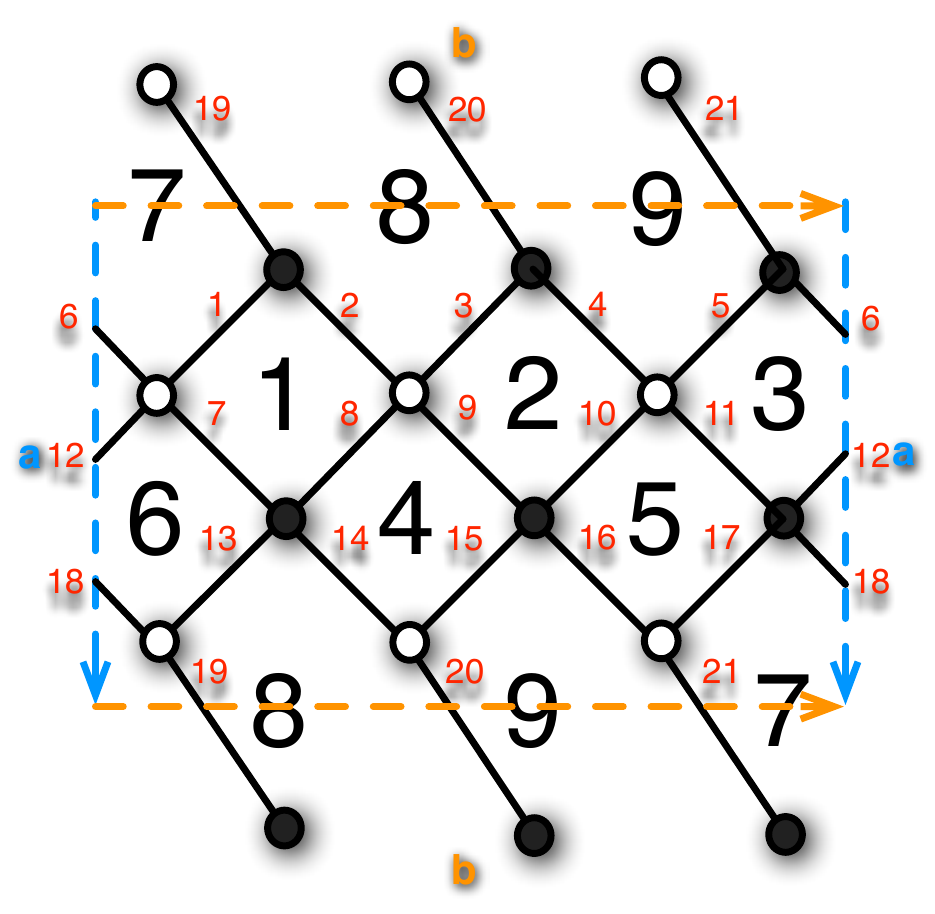}}
& \hspace{2cm} &
\includegraphics[trim=0cm 0cm 0cm 0cm,totalheight=6.3 cm]{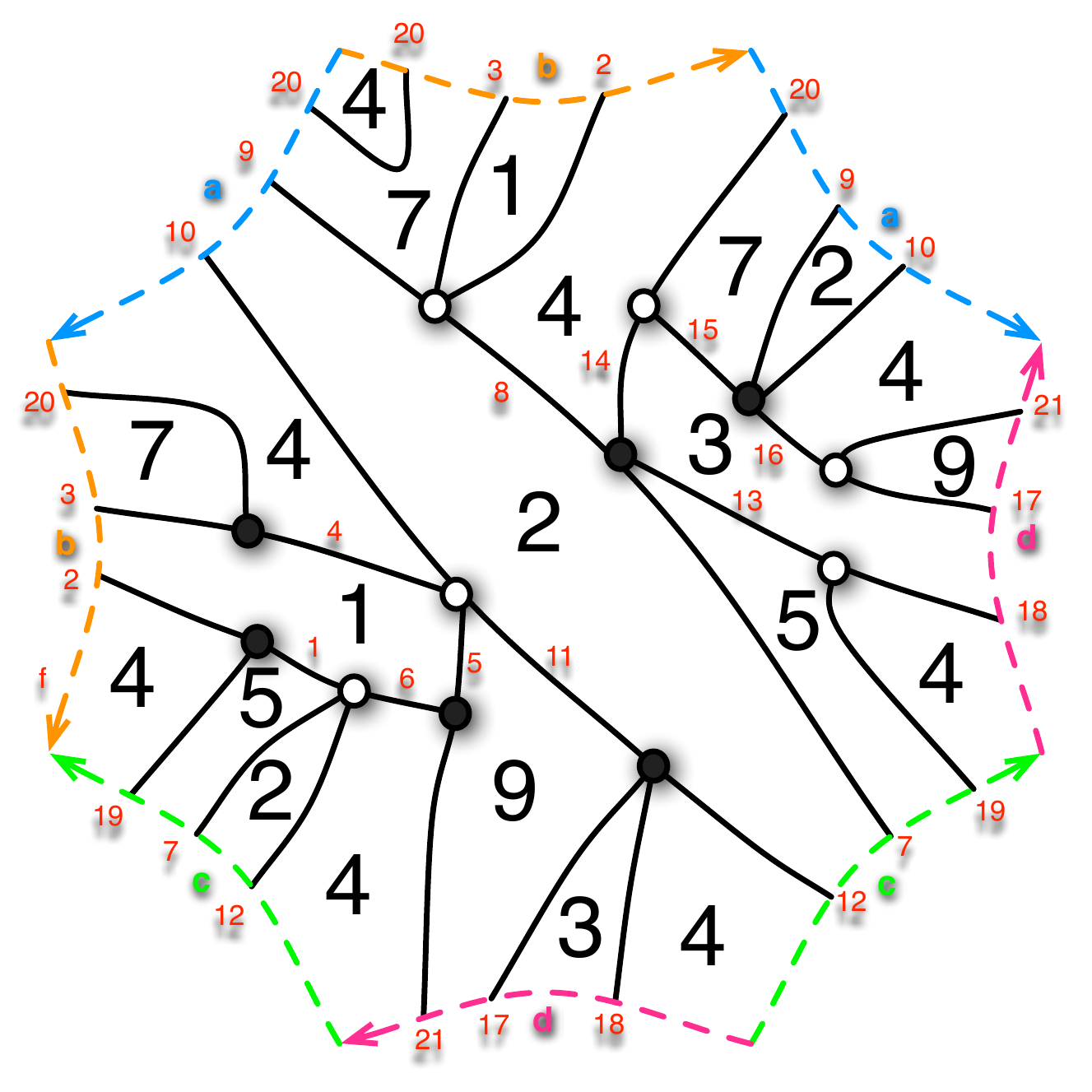}
\\ 
\hspace{0cm}\mbox{(a)} &  \hspace{2cm} & \hspace{0cm}\mbox{(b)}
 \end{tabular}
  \caption{\textit{The sewed model $\sigma_0(C_3)$ and its untwisting, $\widetilde{\sigma_0(C_3)}$.}
  \label{fm3c}}
 \end{center}
 \end{figure}

\fref{fm3c} (a) shows the $\sigma_0(C_3)$ theory. Its moduli space is a 3d CY with toric diagram given by

{\footnotesize
\beq
G_{\sigma_0(C_3)}=\left(
\begin{array}{C{0.35cm}C{0.35cm}C{0.35cm}C{0.35cm}C{0.35cm}C{0.35cm}C{0.35cm}C{0.35cm}C{0.3cm}}
 0 & -1 & 2 & 1 & 1 & 3 & -1 & -2 & 0 \\
 0 & 1 & -1 & 0 & -1 & -2 & 0 & 2 & 1 \\
 1 & 1 & 0 & 0 & 1 & 0 & 2 & 1 & 0 \\
\hline
 \bf{15}  &  \bf{9}  &  \bf{3}  &  \bf{3}  &  \bf{2}  &  \bf{1}  &  \bf{1}  &  \bf{1}  &  \bf{1} \\ 
\hline
\end{array}
\right).
\label{G_matrix_multiplicities_sewed_C3}
\eeq}
This is an Abelian orbifold $\text{SPP}/\mathbb{Z}_3$, with orbifold action $(0,1,2,1)$.

The untwisted theory, shown in \fref{fm3c} (b), lives on a genus 2 Riemann surface and its moduli space is a 5d CY, with toric diagram given by

{\scriptsize
\beq
G_{\widetilde{\sigma_0(C_3)}}=\left(
\begin{array}{C{0.35cm}C{0.35cm}C{0.35cm}C{0.35cm}C{0.35cm}C{0.35cm}C{0.35cm}C{0.35cm}C{0.35cm}C{0.35cm}C{0.35cm}C{0.35cm}C{0.35cm}C{0.35cm}C{0.35cm}C{0.35cm}C{0.35cm}C{0.35cm}C{0.35cm}C{0.35cm}C{0.35cm}C{0.35cm}C{0.35cm}C{0.35cm}C{0.2cm}}
 1 & 2 & 1 & 1 & 1 & 2 & 1 & 2 & 2 & 1 & 1 & 2 & 1 & 2 & 1 & 0 & 0 & 1 & 1 & 1 & 1 &
   0 & 0 & 0 & 0 \\
 0 & 0 & 1 & 1 & -1 & -1 & 0 & 0 & 1 & 1 & 2 & 0 & -1 & -1 & 0 & -1 & -1 & -1 & -1 &
   0 & 1 & 0 & 0 & 0 & 1 \\
 0 & 0 & -1 & 0 & 1 & 1 & 1 & 0 & -1 & 0 & -1 & -1 & 0 & 0 & -1 & 0 & 1 & 0 & 1 & 0 &
   -1 & 0 & 0 & 1 & 0 \\
 0 & 0 & 1 & 0 & 0 & -1 & -1 & -1 & 0 & -1 & 0 & 1 & 1 & 0 & 1 & 1 & 0 & 0 & -1 & -1
   & 0 & 0 & 1 & 0 & 0 \\
 0 & -1 & -1 & -1 & 0 & 0 & 0 & 0 & -1 & 0 & -1 & -1 & 0 & 0 & 0 & 1 & 1 & 1 & 1 & 1
   & 0 & 1 & 0 & 0 & 0 \\
\hline
 \bf{12}  &  \bf{1}  &  \bf{1}  &  \bf{1}  &  \bf{1}  &  \bf{1}  &  \bf{1}  &  \bf{1}  &  \bf{1}  &  \bf{1}  &  \bf{1}  &  \bf{1}  &  \bf{1}  &  \bf{1}  &  \bf{1}  &  \bf{1}  &  \bf{1}  &  \bf{1}  &  \bf{1}  &  \bf{1}  &  \bf{1}  &  \bf{1}  &  \bf{1}  &  \bf{1}  &  \bf{1}  \\ 
\hline
\end{array}
\right).
\label{G_matrix_multiplicities_untwisted_sewed_C3}
\eeq}

\section{Puncturing}

\begin{figure}[h!!]
\begin{center}
\resizebox{0.6\hsize}{!}{
\includegraphics[trim=0cm 0cm 0cm 0cm,totalheight=10 cm]{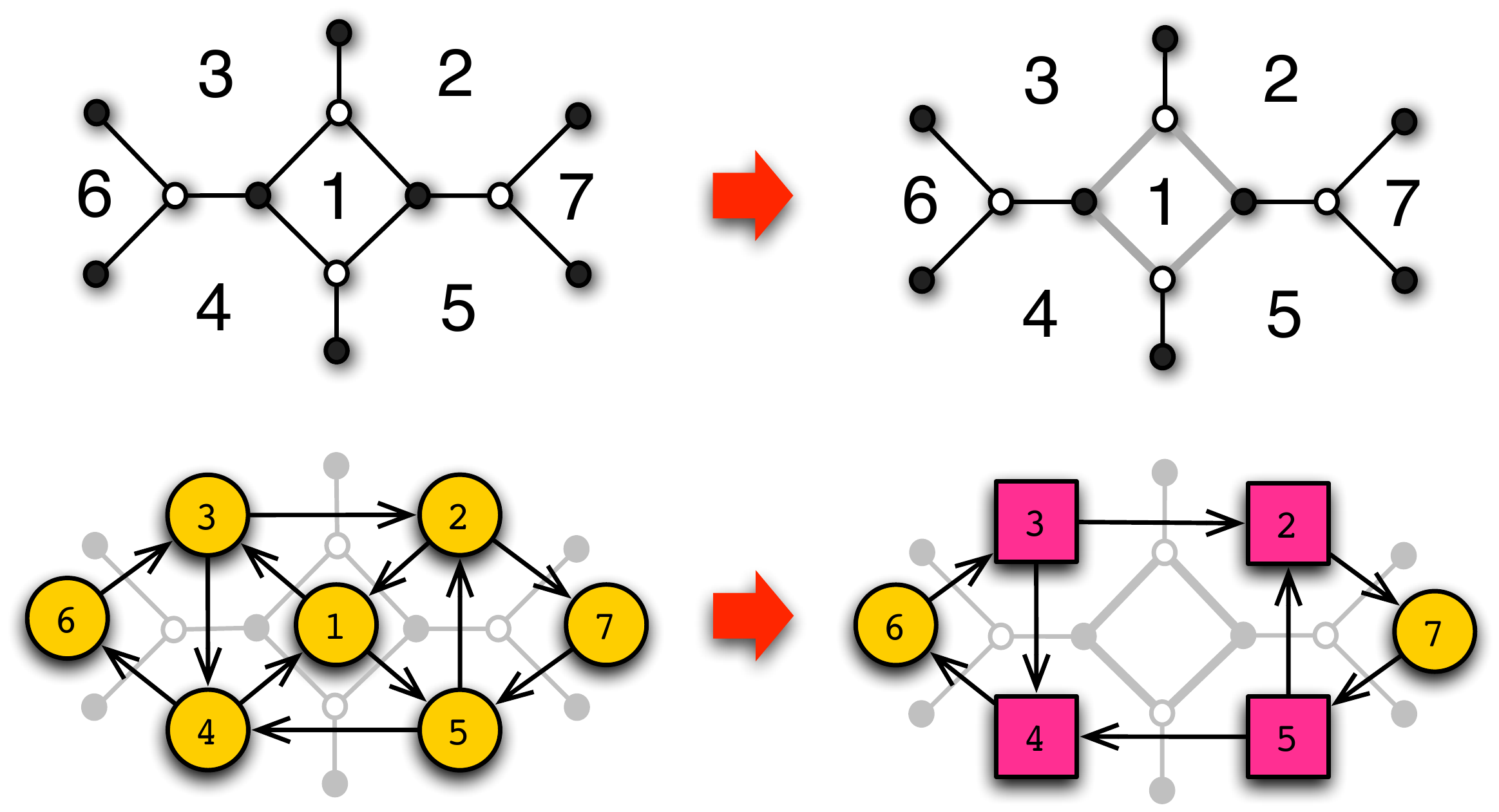}
}
  \caption{\textit{Puncturing a bipartite graph.} An internal face of the bipartite graph, in the example above face 1, is replaced by a boundary. The adjacent faces become external along with the edges between them.
  \label{fpunct}}
 \end{center}
 \end{figure}

Continuing with the general discussion of basic transformations of bipartite graphs and their BFT counterparts, this section discusses a procedure called {\bf puncturing}. It corresponds to replacing an internal face of a bipartite graph by a closed boundary.\footnote{Notice that we define puncturing as introducing a boundary rather than a puncture. The name has been chosen for simplicity and we expect it is not going to cause the reader any confusion.} Consequently, all the faces that are adjacent to the removed one become external and the edges terminating on it become external legs, as illustrated in \fref{fpunct}.

\subsection{Examples}

Let us consider the theory in \fref{fm1c}, which corresponds to the worldvolume theory on D3-branes probing a $\mathbb{Z}_2\times\mathbb{Z}_2$ orbifold of the conifold $\mathcal{C}$. In fact there are four BFTs on $T^2$ whose mesonic moduli space is $\mathcal{C}/(\mathbb{Z}_2\times\mathbb{Z}_2)$. The one for \fref{fm1c} is also known as phase (d) of $\text{PdP}_5$ \cite{Feng:2002fv}. They are all related by Seiberg duality transformations.

 \begin{figure}[ht!]
\begin{center}
\includegraphics[trim=0cm 0cm 0cm 0cm,totalheight=5 cm]{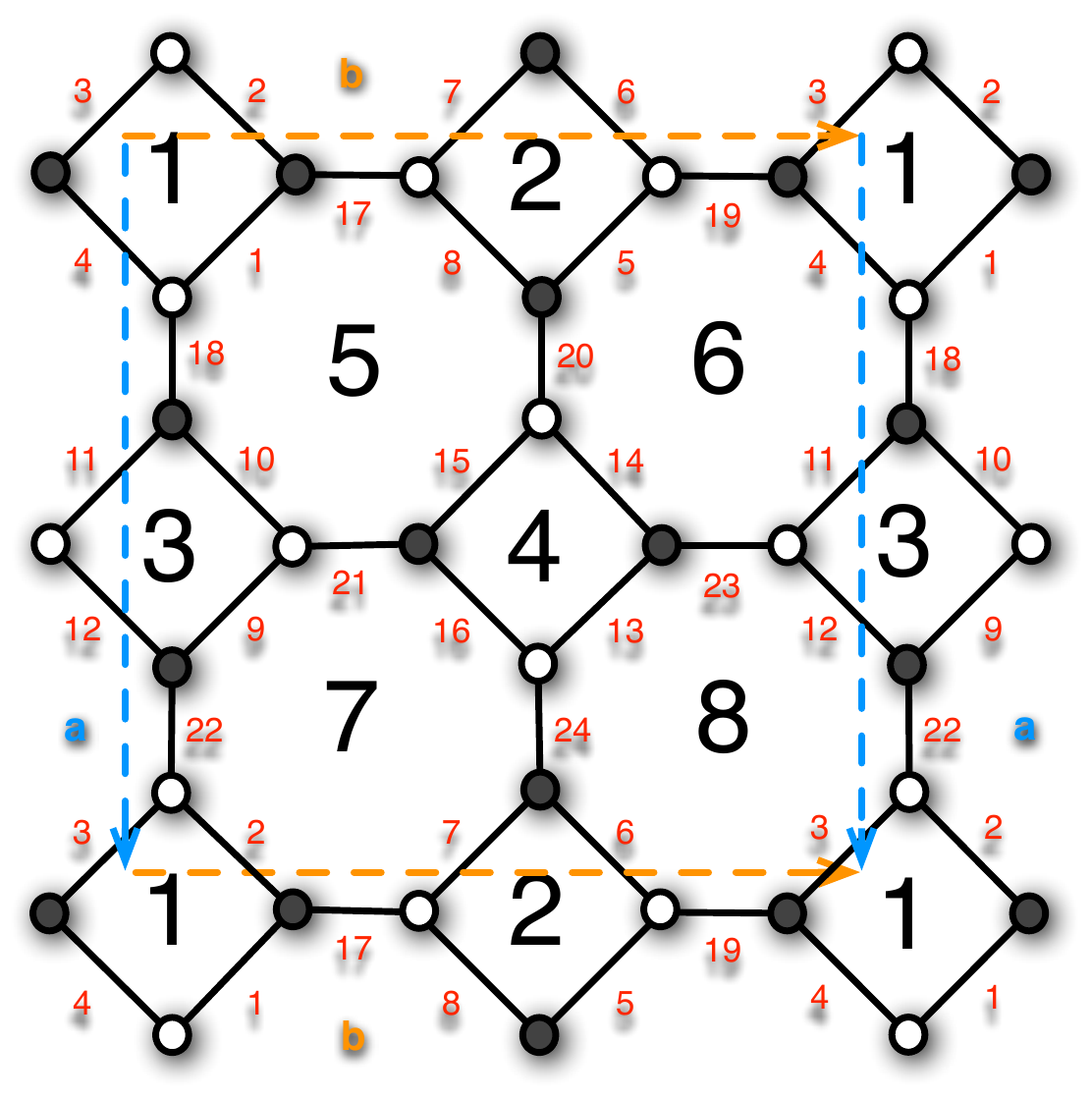}
  \caption{\textit{BFT on D3-branes over $\mathcal{C}/(\mathbb{Z}_2\times\mathbb{Z}_2)$}. It is also known as phase (d) of $\text{PdP}_5$.
  \label{fm1c}}
 \end{center}
 \end{figure}

The moduli space for this theory is indeed $\mathcal{C}/(\mathbb{Z}_2\times\mathbb{Z}_2)$ and its toric diagram is given by

{\footnotesize
\beq
G_{\mathcal{C}/(\mathbb{Z}_2\times\mathbb{Z}_2)}=\left(
\begin{array}{C{0.35cm}C{0.35cm}C{0.35cm}C{0.35cm}C{0.35cm}C{0.35cm}C{0.35cm}C{0.35cm}C{0.3cm}}
 1 & 0 & 0 & 2 & 2 & -1 & 1 & 1 & 3 \\
 0 & 0 & 1 & -1 & 0 & 1 & -1 & 1 & -1 \\
 0 & 1 & 0 & 0 & -1 & 1 & 1 & -1 & -1 \\
 \hline
 \bf{21} &  \bf{2}  &  \bf{2}  &  \bf{2}  &  \bf{2}  &  \bf{1}  &  \bf{1}  &  \bf{1}  &  \bf{1} \\
\hline
\end{array}
\right).
\label{G_matrix_multiplicities_PdP5}
\eeq}

\noindent The toric diagrams for the moduli spaces of the dual theories differ in the multiplicities of perfect matchings for each toric point.

\begin{figure}[H]
\begin{center}
 \begin{tabular}[c]{ccc}
\includegraphics[trim=0cm 0cm 0cm 0cm,totalheight=5 cm]{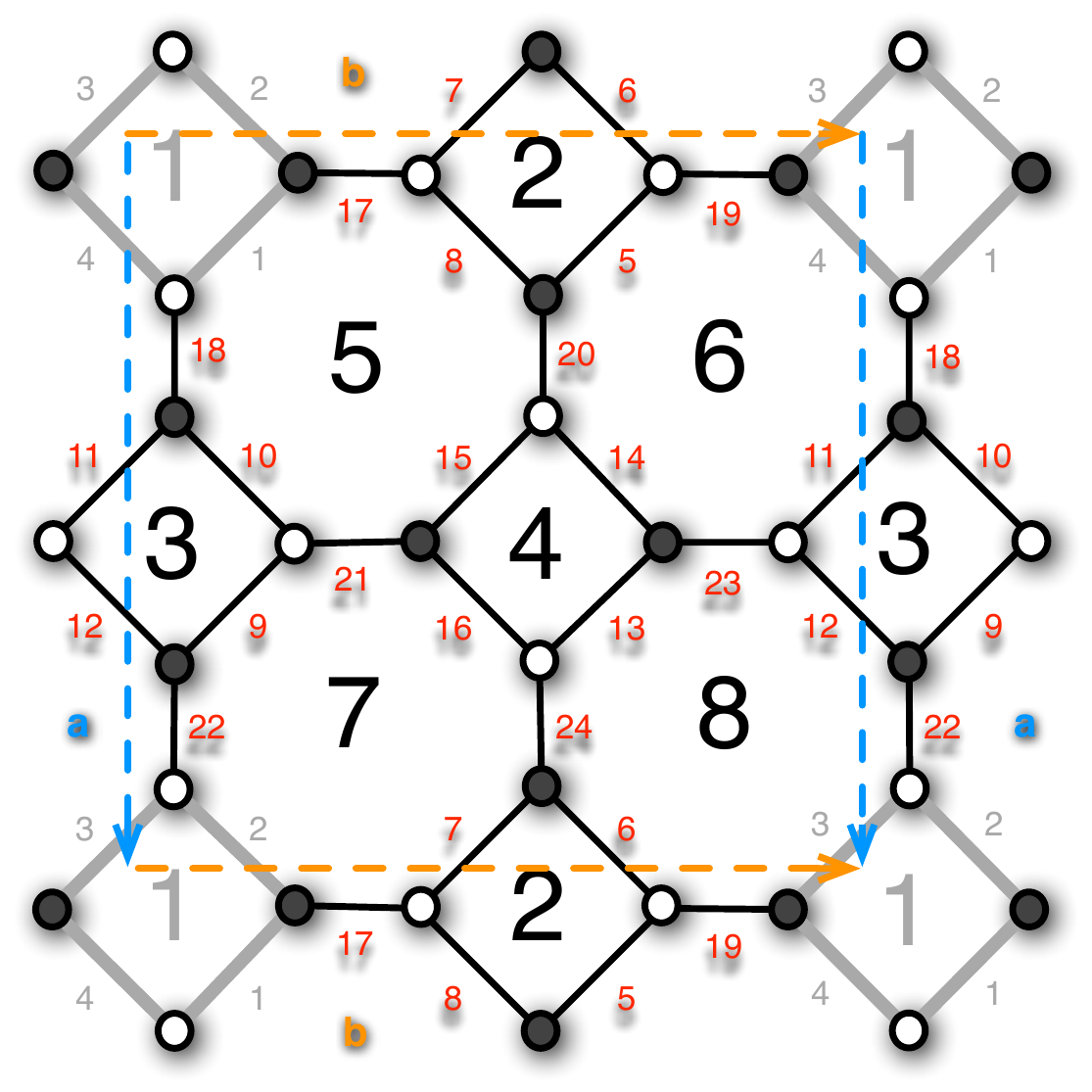}
& \hspace{1cm} &
\includegraphics[trim=0cm 0cm 0cm 0cm,totalheight=5 cm]{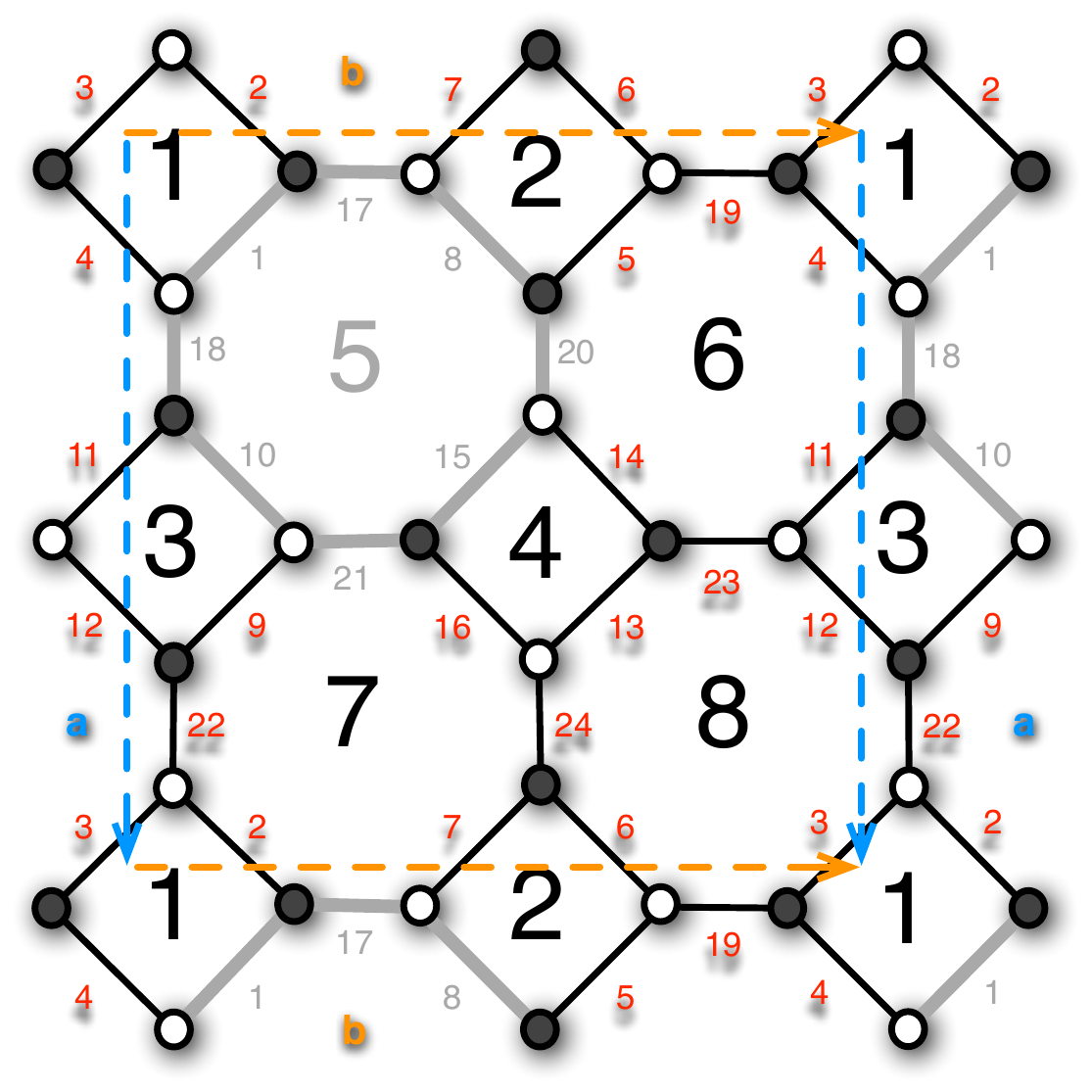}
\\ 
\hspace{0cm}\mbox{(a)} &  \hspace{1cm} & \hspace{0cm}\mbox{(b)}
 \end{tabular}
  \caption{\textit{Two possible ways of puncturing the theory in \fref{fm1c}.}\label{fm1ccc}}
 \end{center}
 \end{figure}

The effect on the moduli space of puncturing face 1 (theory (a)) or face 5 (theory (b)) is now investigated. The corresponding graphs are shown in \fref{fm1ccc}. 

The resulting moduli spaces are given by the toric diagrams:

{\scriptsize
\bea
&&
G_{(a)}=\left(
\begin{array}{C{0.35cm}C{0.35cm}C{0.35cm}C{0.35cm}C{0.35cm}C{0.35cm}C{0.35cm}C{0.35cm}C{0.35cm}C{0.35cm}C{0.35cm}C{0.35cm}C{0.35cm}C{0.35cm}C{0.35cm}C{0.35cm}C{0.35cm}C{0.35cm}}
 -1 & 0 & 0 & 1 & -3 & -2 & -2 & -2 & -2 & -2 & -2 & -2 & -2 & -2 & -2 & -1 & -1 & -1
   \\
 1 & 0 & 1 & 0 & 1 & 0 & 1 & 1 & 1 & 1 & 1 & 1 & 1 & 1 & 1 & 0 & 0 & 0 \\
 1 & 1 & 0 & 0 & 1 & 1 & 0 & 1 & 1 & 1 & 1 & 1 & 1 & 1 & 1 & 0 & 1 & 1 \\
 0 & 0 & 0 & 0 & 0 & 0 & 0 & -1 & -1 & 0 & 0 & 0 & 0 & 1 & 1 & 0 & -1 & -1 \\
 0 & 0 & 0 & 0 & 1 & 1 & 1 & 1 & 1 & 0 & 0 & 1 & 1 & 0 & 0 & 1 & 1 & 1 \\
 0 & 0 & 0 & 0 & 0 & 0 & 0 & 0 & 1 & 0 & 1 & -1 & 0 & -1 & 0 & 0 & 0 & 1 \\
 0 & 0 & 0 & 0 & 1 & 1 & 1 & 1 & 0 & 1 & 0 & 1 & 0 & 1 & 0 & 1 & 1 & 0 \\
 \hline
 \bf 4 & \bf 4 & \bf 4 & \bf 4 & \bf 1 & \bf 1 & \bf 1 & \bf 1 & \bf 1 & \bf 1 & \bf 1 & \bf 1 & \bf 1 & \bf 1 & \bf 1 & \bf 1 & \bf 1 & \bf 1\\
\hline
\end{array}
\right.
\dots
\nn
\\ \nn \\
&&
\hspace{2cm}
\dots
\left.
\begin{array}{C{0.35cm}C{0.35cm}C{0.35cm}C{0.35cm}C{0.35cm}C{0.35cm}C{0.35cm}C{0.35cm}C{0.35cm}C{0.35cm}C{0.35cm}C{0.35cm}C{0.35cm}C{0.35cm}C{0.35cm}C{0.35cm}C{0.35cm}C{0.35cm}C{0.35cm}C{0.35cm}C{0.35cm}C{0.2cm}}
 -1 & -1 & -1 & -1 & -1 & -1 & -1 & -1 & -1 & -1 & -1 & -1 & -1 & -1 & 0 & 0 & 0 & 0 &
   0 & 0 & 0 & 0 \\
 0 & 0 & 0 & 0 & 0 & 0 & 1 & 1 & 1 & 1 & 1 & 1 & 1 & 1 & 0 & 0 & 0 & 0 & 0 & 0 & 0 & 0
   \\
 1 & 1 & 1 & 1 & 1 & 1 & 0 & 0 & 0 & 0 & 0 & 0 & 0 & 0 & 0 & 0 & 0 & 0 & 0 & 0 & 0 & 0
   \\
 0 & 0 & 0 & 0 & 1 & 1 & -1 & -1 & 0 & 0 & 0 & 0 & 1 & 1 & -1 & -1 & 0 & 0 & 0 & 0 & 1
   & 1 \\
 0 & 0 & 1 & 1 & 0 & 0 & 1 & 1 & 0 & 0 & 1 & 1 & 0 & 0 & 1 & 1 & 0 & 0 & 1 & 1 & 0 & 0
   \\
 0 & 1 & -1 & 0 & -1 & 0 & 0 & 1 & 0 & 1 & -1 & 0 & -1 & 0 & 0 & 1 & 0 & 1 & -1 & 0 &
   -1 & 0 \\
 1 & 0 & 1 & 0 & 1 & 0 & 1 & 0 & 1 & 0 & 1 & 0 & 1 & 0 & 1 & 0 & 1 & 0 & 1 & 0 & 1 & 0
   \\
   \hline
 \bf 1 & \bf 1 & \bf 1 & \bf 1 & \bf 1 & \bf 1 & \bf 1 & \bf 1 & \bf 1 & \bf 1 & \bf 1 & \bf 1 & \bf 1 & \bf 1 & \bf 1 & \bf 1 & \bf 1 & \bf 1 & \bf 1 & \bf 1 & \bf 1 & \bf 1\\
 \hline
\end{array}
\right) ,
\nn\\
\eea
}

{\scriptsize
\bea
&&
G_{(b)}=\left(
\begin{array}{C{0.35cm}C{0.35cm}C{0.35cm}C{0.35cm}C{0.35cm}C{0.35cm}C{0.35cm}C{0.35cm}C{0.35cm}C{0.35cm}C{0.35cm}C{0.35cm}C{0.35cm}C{0.35cm}C{0.35cm}C{0.35cm}C{0.35cm}C{0.35cm}C{0.35cm}C{0.35cm}C{0.35cm}C{0.35cm}}
 -3 & -2 & -2 & -2 & -2 & -2 & -2 & -2 & -2 & -2 & -2 & -2 & -2 & -2 & -1 & -1 & -1
   & -1 & -1 & -1 & -1 & -1 \\
 1 & 0 & 0 & 0 & 0 & 1 & 1 & 1 & 1 & 1 & 1 & 1 & 1 & 1 & 0 & 0 & 0 & 0 & 0 & 0 & 0 &
   0 \\
 0 & 0 & 1 & 1 & 1 & -1 & -1 & -1 & 0 & 0 & 0 & 0 & 0 & 0 & 0 & 0 & 0 & 0 & 0 & 0 &
   0 & 0 \\
 0 & 0 & 0 & 0 & 1 & -1 & 0 & 0 & 0 & 0 & 0 & 0 & 1 & 1 & -1 & -1 & 0 & 0 & 0 & 0 &
   0 & 1 \\
 1 & 1 & 0 & 1 & 0 & 2 & 1 & 1 & 0 & 0 & 1 & 1 & 0 & 0 & 1 & 2 & 0 & 0 & 1 & 1 & 1 &
   0 \\
 0 & 0 & 1 & 0 & 1 & -1 & -1 & 0 & 0 & 1 & 0 & 0 & 0 & 1 & 0 & -1 & 0 & 1 & -1 & 0 &
   0 & 0 \\
 1 & 1 & 0 & 0 & 0 & 2 & 2 & 1 & 1 & 0 & 0 & 1 & 1 & 0 & 1 & 1 & 1 & 0 & 1 & 0 & 1 &
   1 \\
 1 & 1 & 1 & 1 & 0 & 1 & 1 & 1 & 1 & 1 & 1 & 0 & 0 & 0 & 1 & 1 & 1 & 1 & 1 & 1 & 0 &
   0 \\
 1 & 2 & 1 & 1 & 1 & 1 & 1 & 1 & 1 & 1 & 1 & 1 & 1 & 1 & 1 & 1 & 1 & 1 & 1 & 1 & 1 &
   1 \\
   \hline
 \bf 1 & \bf 1 & \bf 1 & \bf 1 & \bf 1 & \bf 1 & \bf 1 & \bf 1 & \bf 1 & \bf 1 & \bf 1 & \bf 1 & \bf 1 & \bf 1 & \bf 1 & \bf 1 & \bf 1 & \bf 1 & \bf 1 & \bf 1 & \bf 1 &
   \bf 1\\ 
   \hline
\end{array}
\right.
\dots
\nn
\\ \nn 
\\
&&
\hspace{4cm}
\dots
\left.
\begin{array}{C{0.35cm}C{0.35cm}C{0.35cm}C{0.35cm}C{0.35cm}C{0.35cm}C{0.35cm}C{0.35cm}C{0.35cm}C{0.35cm}C{0.35cm}C{0.35cm}C{0.35cm}C{0.35cm}C{0.35cm}C{0.35cm}C{0.35cm}C{0.2cm}}
 -1 & -1 & -1 & -1 & -1 & -1 & -1 & -1 & -1 & -1 & -1 & -1 & -1 & -1 & 0 & 0 & 0 & 0
   \\
 0 & 0 & 0 & 0 & 0 & 1 & 1 & 1 & 1 & 1 & 1 & 1 & 1 & 1 & 0 & 0 & 0 & 0 \\
 0 & 1 & 1 & 1 & 1 & -1 & -1 & -1 & -1 & 0 & 0 & 0 & 0 & 0 & 0 & 0 & 0 & 0 \\
 1 & 0 & 0 & 0 & 1 & -1 & -1 & -1 & 0 & 0 & 0 & 0 & 0 & 1 & -1 & 0 & 0 & 0 \\
 0 & 0 & 0 & 1 & 0 & 1 & 1 & 2 & 1 & 0 & 0 & 0 & 1 & 0 & 1 & 0 & 0 & 0 \\
 1 & 0 & 1 & 0 & 0 & -1 & 0 & -1 & -1 & 0 & 0 & 1 & 0 & 0 & -1 & 0 & 0 & 1 \\
 0 & 0 & 0 & 0 & 0 & 2 & 1 & 1 & 1 & 0 & 1 & 0 & 0 & 0 & 1 & 0 & 1 & 0 \\
 0 & 1 & 0 & 0 & 0 & 1 & 1 & 1 & 1 & 1 & 0 & 0 & 0 & 0 & 1 & 1 & 0 & 0 \\
 1 & 1 & 1 & 1 & 1 & 1 & 1 & 1 & 1 & 1 & 1 & 1 & 1 & 1 & 1 & 1 & 1 & 1 \\
 \hline
\bf 1 & \bf 1 & \bf 1 & \bf 1 & \bf 1 & \bf 1 & \bf 1 & \bf 1 & \bf 1 & \bf 1 & \bf 1 & \bf 1 & \bf 1 & \bf 1 & \bf 1 & \bf 1 & \bf 1 & \bf 1\\
\hline
\end{array}
\right) .
\nn\\
\eea
}
Interestingly, for theory (b) the multiplicities of all points in the toric diagram are equal to 1. This is a consequence of the fact that there are no pairs of perfect matchings whose difference is given by a loop enclosing the only surviving internal face, namely face 8.

\bigskip

\section{Edge Splitting \label{section_edge_splitting}
}

Having considered the effect of sewing edges, we now move in the opposite direction and consider the {\bf splitting} of an internal edge. Edge splitting can either increase the planarity of a graph (this application has indeed been considered in \cite{Nima}) or decrease it, by introducing new boundaries.

It is possible to discuss in rather general terms how the master and moduli spaces of the corresponding BFT are affected by this operation. The split edge can be of three types, internal/internal, internal/external or external/external, depending on the two types of faces it separates. The external/external case occurs when the split edge is an external leg or an edge connected to an external leg by a number of massive fields. In this case, the result of the splitting is a graph that is equivalent to the original one, plus a decoupled edge that connects two external nodes. The geometry associated to this new graph is trivially related to the original one, so we are not going to consider the external/external case any further.

\begin{figure}[h]
\begin{center}
\includegraphics[width=10.5cm]{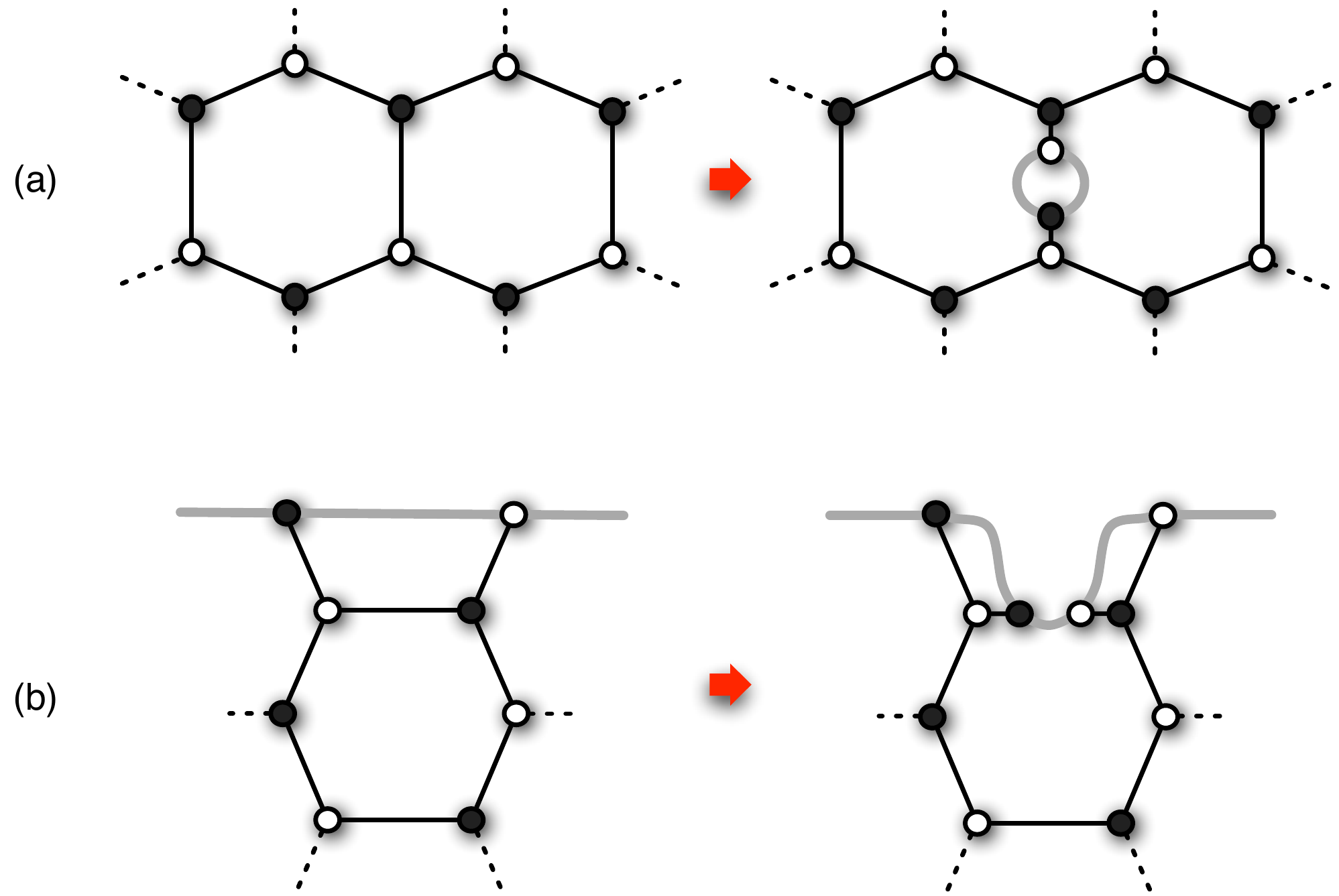}
\caption{\textit{Edge Splitting.} The effect of splitting on the two possible classes of edges: (a) internal/internal and (b) internal/external.}
\label{edge_splitting}
\end{center}
\end{figure}

\fref{edge_splitting} shows how a generic bipartite graph is modified by splitting, resulting in the following changes in the number of boundaries and faces.

\beq
\begin{array}{ccrcl}
\mbox{{\bf Split edge}} & & & & \\ \hline
\mbox{internal/internal} & \ \ \ & B & \to & B+1 \\
& & F_e & \to & F_e + 2 \\
& & F_i & \to & F_i -2  \\ \hline
\mbox{internal/external} & \ \ \ & B & \to & B \\
& & F_e & \to & F_e + 2 \\
& & F_i & \to & F_i -1  \\ \hline
 \end{array}
\eeq
\smallskip

\noindent These changes result in an increased dimension for the master and moduli spaces, except for the case of the master space for originally $B=0$ theories, whose dimension remains constant. The table below summarizes the changes in dimensions depending on the type of split edge and the initial number of boundaries. It is assumed that the genus of the Riemann surface remains constant, although generally this may not be the case.

\beq
\begin{array}{cccccc}
\mbox{{\bf Split edge}} & & & & \ \Delta d_{master} \ & \ \Delta d_{moduli} \ \\ \hline
\mbox{internal/internal} & \ \ \ & B=0 & \ \ \ & 0 & 1 \\
& & B\neq 0 & & 1 & 3 \\ \hline
\mbox{internal/external} & \ \ \ & B \neq 0 & & 1 & 2 \\ \hline
 \end{array}
\eeq
\smallskip

\noindent Generally, new perfect matchings arise when splitting edges. Every broken edge produces two new edges.  For any perfect matching of the initial graph, the original edge can be either occupied (O) or empty (E). After splitting, the two new edges can be in one of four combinations O-O, O-E, E-O or E-E. Original perfect matchings in which the edge under consideration is O or E, are in one-to-one correspondence with perfect matchings of the new graph in which the new edges are O-O and E-E, respectively. New perfect matchings are associated to the O-E and E-O combinations and whether they actually appear depends on the detailed structure of the graph. 

\bigskip

\subsection{Examples}

\paragraph{$F_0$ Theory.}

\begin{figure}[H]
\begin{center}
\includegraphics[width=7.5cm]{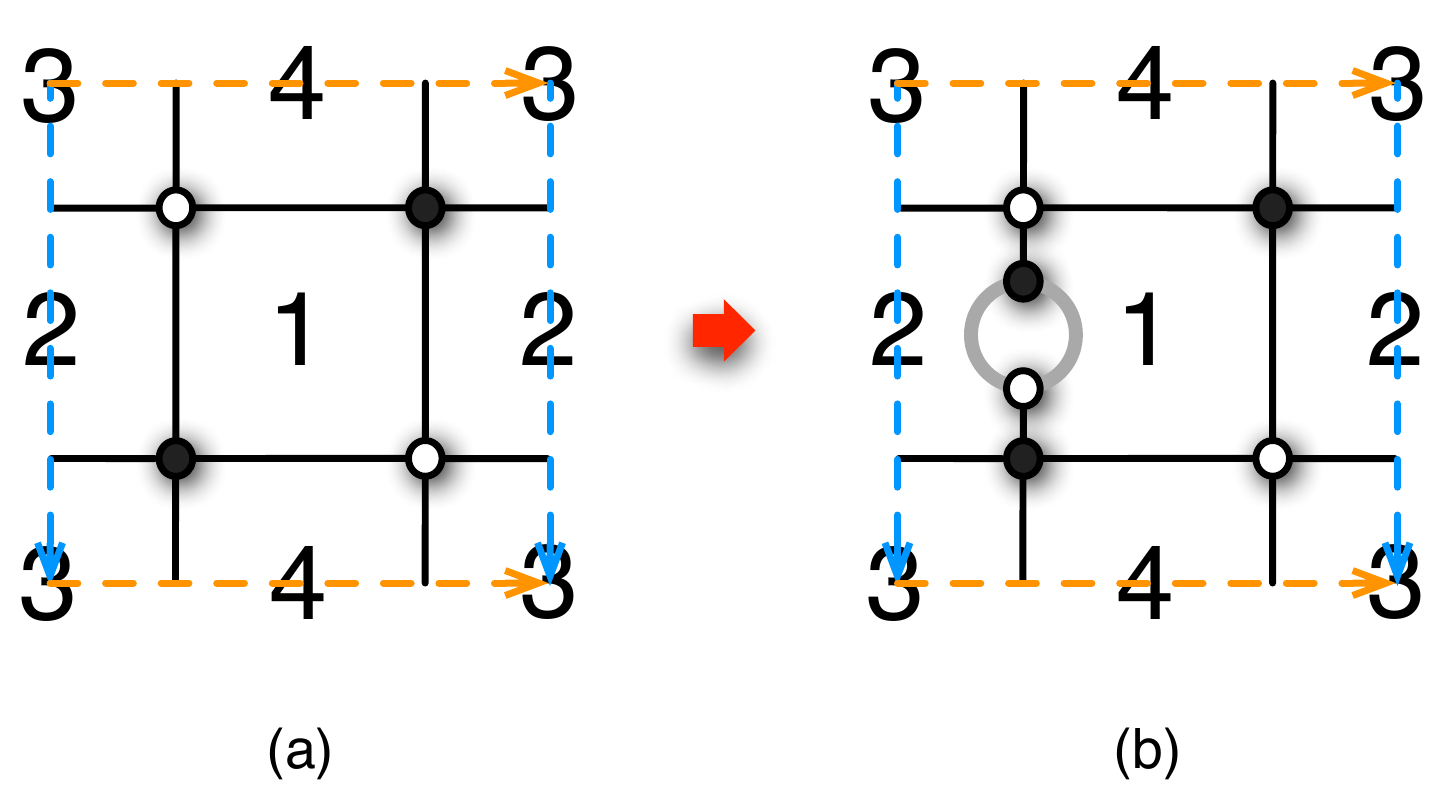}
\caption{\textit{Edge Splitting in a $B=0$ Theory.} (a) A BFT for $F_0$ and (b) theory resulting from splitting an internal/internal edge.}
\label{splitf1}
\end{center}
\end{figure}

Let us consider the theory in \fref{splitf1} (a) which corresponds to D3-branes probing the complex cone over $F_0$ \cite{Franco:2005rj}. This is a $B=0$ theory and we split an internal/internal edge as shown in \fref{splitf1} (b). 

In agreement with the general analysis, the moduli spaces are respectively 3d and 4d CYs with toric diagrams given by

{\footnotesize
\beq
G_{F_0}=\left(
\begin{array}{C{0.35cm}C{0.35cm}C{0.35cm}C{0.35cm}C{0.35cm}}
 1 & 2 & 0 & 2 & 0 \\
 0 & 0 & 0 & -1 & 1 \\
 0 & -1 & 1 & 0 & 0 \\  \hline

 \bf{4} & \bf{1}  &  \bf{1}  &  \bf{1}  &  \bf{1} \\
\hline
\end{array}
\right) , \ \ \ \ \ \ \ \ 
G_{F_0'}=\left(
\begin{array}{C{0.35cm}C{0.35cm}C{0.35cm}C{0.35cm}C{0.35cm}C{0.35cm}}
 0 & -1 & 0 & 0 & 0 & 1 \\
 1 & 1 & 0 & 2 & 0 & 0 \\
 0 & 0 & 0 & -1 & 1 & 0 \\
 0 & 1 & 1 & 0 & 0 & 0 \\ \hline

 \bf{3} & \bf{1}  & \bf{1} & \bf{1}  & \bf{1}  & \bf{1} \\
\hline
\end{array}
\right).
\label{G_matrix_multiplicities_internal_internal}
\eeq}
In this example, the total number of perfect matchings is preserved.

\bigskip

\paragraph{Top Dimensional Cell of $G(2,5)$.}

\begin{figure}[h]
\begin{center}
\includegraphics[width=13.5cm]{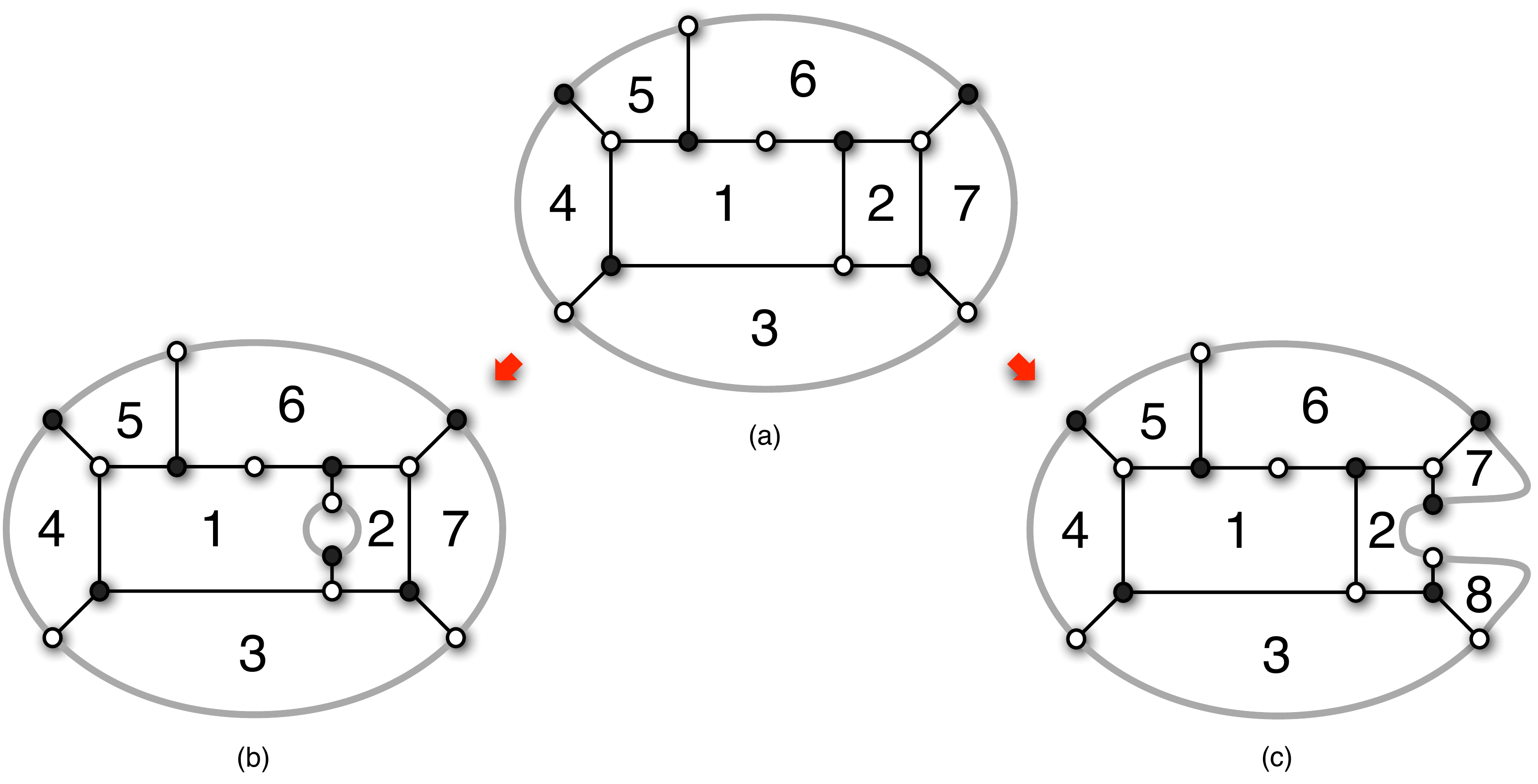}
\caption{\textit{Edge Splitting in a $B=1$ Theory.} (a) BFT for the top dimensional cell of $G(2,5)$ (b) splitting of an internal/internal edge and (c) splitting of an internal/external edge.}
\label{splitf2}
\end{center}
\end{figure}

Next, let us study the theory in \fref{splitf2} (a), which corresponds to the top dimensional cell of the Grassmannian $G(2,5)$. This theory has $B=1$ and its moduli space is a 5d CY with toric diagram

{\scriptsize
\beq
G_{a}=\left(
\begin{array}{C{0.35cm}C{0.35cm}C{0.35cm}C{0.35cm}C{0.35cm}C{0.35cm}C{0.35cm}C{0.35cm}C{0.35cm}C{0.35cm}}
 1 & 0 & 0 & 0 & -1 & 0 & -1 & 0 & -1 & 0 \\
 0 & 1 & 0 & 1 & 1 & 0 & 0 & 0 & 1 & 0 \\
 0 & 0 & 1 & 1 & 1 & 0 & 1 & 0 & 1 & 1 \\
 0 & 0 & 1 & 0 & 1 & 0 & 1 & 1 & 0 & 0 \\
 0 & 0 & -1 & -1 & -1 & 1 & 0 & 0 & 0 & 0  \\ \hline

 \bf{3} & \bf{2}  & \bf{2} & \bf{1}  & \bf{1}  & \bf{1} & \bf{1}  & \bf{1} & \bf{1} & \bf{1} \\
\hline
\end{array}
\right).
\label{G_matrix_multiplicities_G25}
\eeq}

Splitting an internal-internal edge as in \fref{splitf2} (b) results in a 8d moduli space with toric diagram given by

{\scriptsize
\beq
G_{b}=\left(
\begin{array}{C{0.35cm}C{0.35cm}C{0.35cm}C{0.35cm}C{0.35cm}C{0.35cm}C{0.35cm}
C{0.35cm}C{0.35cm}C{0.35cm}C{0.35cm}C{0.35cm}C{0.35cm}C{0.35cm}C{0.35cm}C{0.35cm}C{0.35cm}C{0.35cm}C{0.35cm}C{0.35cm}
C{0.35cm}C{0.35cm}C{0.35cm}C{0.35cm}C{0.35cm}C{0.35cm}}
 1 & 0 & 0 & -1 & 0 & 0 & -1 & -1 & 0 & 0 & -2 & -1 & -1 & 0 & -1 & 0 & -1 & 0 & -2 &
   -1 & -1 & 0 & -1 & 0 & 0 & 1 \\
 -1 & 0 & 0 & 1 & 0 & -1 & 0 & 0 & 0 & -1 & 1 & 0 & 1 & 0 & 0 & 0 & 0 & 0 & 1 & 0 & 1
   & 0 & 1 & 0 & 1 & 0 \\
 -1 & 0 & -1 & 0 & 0 & -1 & 0 & -1 & -1 & 0 & 0 & 0 & 0 & 0 & 1 & 0 & 0 & 0 & 1 & 1 &
   0 & 0 & 1 & 1 & 0 & 0 \\
 -1 & -1 & 0 & 0 & 0 & 0 & 0 & 1 & 0 & 0 & 1 & 1 & 0 & 0 & 0 & 0 & 1 & 0 & 1 & 1 & 1
   & 1 & 0 & 0 & 0 & 0 \\
 1 & 0 & 1 & 0 & 0 & 1 & 0 & 1 & 1 & 1 & 0 & 0 & 0 & 0 & 0 & 0 & 1 & 1 & 0 & 0 & 0 &
   0 & 0 & 0 & 0 & 0 \\
 1 & 1 & 0 & 0 & 0 & 1 & 1 & 0 & 0 & 1 & 0 & 0 & 0 & 0 & 1 & 1 & 0 & 0 & 0 & 0 & 0 &
   0 & 0 & 0 & 0 & 0 \\
 0 & 0 & 0 & 0 & 0 & 1 & 1 & 1 & 1 & 0 & 1 & 1 & 1 & 1 & 0 & 0 & 0 & 0 & 0 & 0 & 0 &
   0 & 0 & 0 & 0 & 0 \\
 1 & 1 & 1 & 1 & 1 & 0 & 0 & 0 & 0 & 0 & 0 & 0 & 0 & 0 & 0 & 0 & 0 & 0 & 0 & 0 & 0 &
   0 & 0 & 0 & 0 & 0 \\\hline

 \bf{1} & \bf{1}  & \bf{1} & \bf{1}  & \bf{1}  & \bf{1} & \bf{1}  & \bf{1} & \bf{1} & \bf{1} & \bf{1} & \bf{1}  & \bf{1} & \bf{1}  & \bf{1}  & \bf{1} & \bf{1}  & \bf{1} & \bf{1} & \bf{1} & \bf{1}  & \bf{1} & \bf{1}  & \bf{1} & \bf{1} & \bf{1} \\
\hline
\end{array}
\right).
\label{G_matrix_multiplicities_G25_int_int}
\eeq}
The multiplicity of all points is equal to 1 because the theory we obtain does not have any gauge group. If an internal/external edge is split as in \fref{splitf2} (c), the moduli space becomes a 7d CY with toric diagram

{\scriptsize
\beq
G_{c}=\left(
\begin{array}{C{0.35cm}C{0.35cm}C{0.35cm}C{0.35cm}C{0.35cm}C{0.35cm}C{0.35cm}
C{0.35cm}C{0.35cm}C{0.35cm}C{0.35cm}C{0.35cm}C{0.35cm}C{0.35cm}C{0.35cm}C{0.35cm}C{0.35cm}C{0.35cm}C{0.35cm}C{0.35cm}
C{0.35cm}C{0.35cm}}
 -2 & -2 & -1 & -1 & -1 & -1 & 0 & 0 & -1 & -1 & -1 & 0 & -1 & 0 & -1 & -1 & -1 & 0 & 0 & 0 & 0 & 1 \\
 0 & 1 & 0 & 1 & 0 & 1 & 0 & 1 & 0 & 0 & 1 & 0 & 0 & 0 & 0 & 0 & 1 & 0 & 0 & 0 & 1 & 0 \\
 1 & 1 & 1 & 1 & 1 & 1 & 1 & 1 & 0 & 0 & 0 & 0 & 0 & 0 & 0 & 1 & 0 & 0 & 0 & 1 & 0 & 0 \\
 1 & 0 & 1 & 0 & 1 & 0 & 1 & 0 & 0 & 1 & 0 & 0 & 0 & 0 & 1 & 0 & 0 & 0 & 1 & 0 & 0 & 0 \\
 1 & 1 & 0 & 0 & 1 & 1 & 0 & 0 & 0 & 0 & 0 & 0 & 1 & 0 & 1 & 1 & 1 & 1 & 0 & 0 & 0 & 0 \\
 0 & 0 & 0 & 0 & -1 & -1 & -1 & -1 & 1 & 0 & 0 & 0 & 1 & 1 & 0 & 0 & 0 & 0 & 0 & 0 & 0 & 0 \\
 0 & 0 & 0 & 0 & 0 & 0 & 0 & 0 & 1 & 1 & 1 & 1 & 0 & 0 & 0 & 0 & 0 & 0 & 0 & 0 & 0 & 0 \\ \hline
\bf{2} & \bf{2}  & \bf{2} & \bf{2}  & \bf{1}  & \bf{1} & \bf{1}  & \bf{1} & \bf{1} & \bf{1} & \bf{1} & \bf{1}  & \bf{1} & \bf{1}  & \bf{1}  & \bf{1} & \bf{1}  & \bf{1} & \bf{1} & \bf{1} & \bf{1}  & \bf{1} \\
\hline
\end{array}
\right).
\label{G_matrix_multiplicities_G25_int_ext}
\eeq}
In both of the cases above, the number of perfect matchings increases after splitting and edge.

\bigskip

\section{Seiberg/Toric Duality for Higher Genus BFTs}

\label{section_Seiberg}

Seiberg duality of quiver gauge theories has been extensively investigated in the past \cite{Feng:2000mi,Feng:2001xr,Feng:2002zw,Seiberg:1994pq,Feng:2001bn,2001JHEP...12..001B,Franco:2003ea}. In the context of gauge theories on D3-branes over toric CY 3-folds, which are BFTs on $T^2$'s, the duality is also known as {\bf toric duality}.\footnote{More generally the coincidence of the Abelian moduli space for theories in which it is toric, regardless of its dimension and the dimension in which the gauge theory lives, is also often referred to as Seiberg duality.}
As explained in \sref{section_moves}, in the context of BFTs, Seiberg duality is implemented by the square move shown in \fref{reduction_moves}. We have already emphasized that the Abelian moduli space is invariant under square moves and hence serves as an ideal diagnostic for identifying potentially dual theories. Furthermore, it is a sufficient condition for two BFTs to be equivalent from a leading singularity perspective. An explicit example illustrating this invariance for a BFT on a genus 2 Riemann surface is presented in the following discussion.

\begin{figure}[H]
\begin{center}
 \begin{tabular}[c]{ccc}
\includegraphics[trim=0cm 0cm 0cm 0cm,totalheight=7 cm]{m3aay2.pdf}
& \hspace{0.2cm} &
\includegraphics[trim=0cm 0cm 0cm 0cm,totalheight=7 cm]{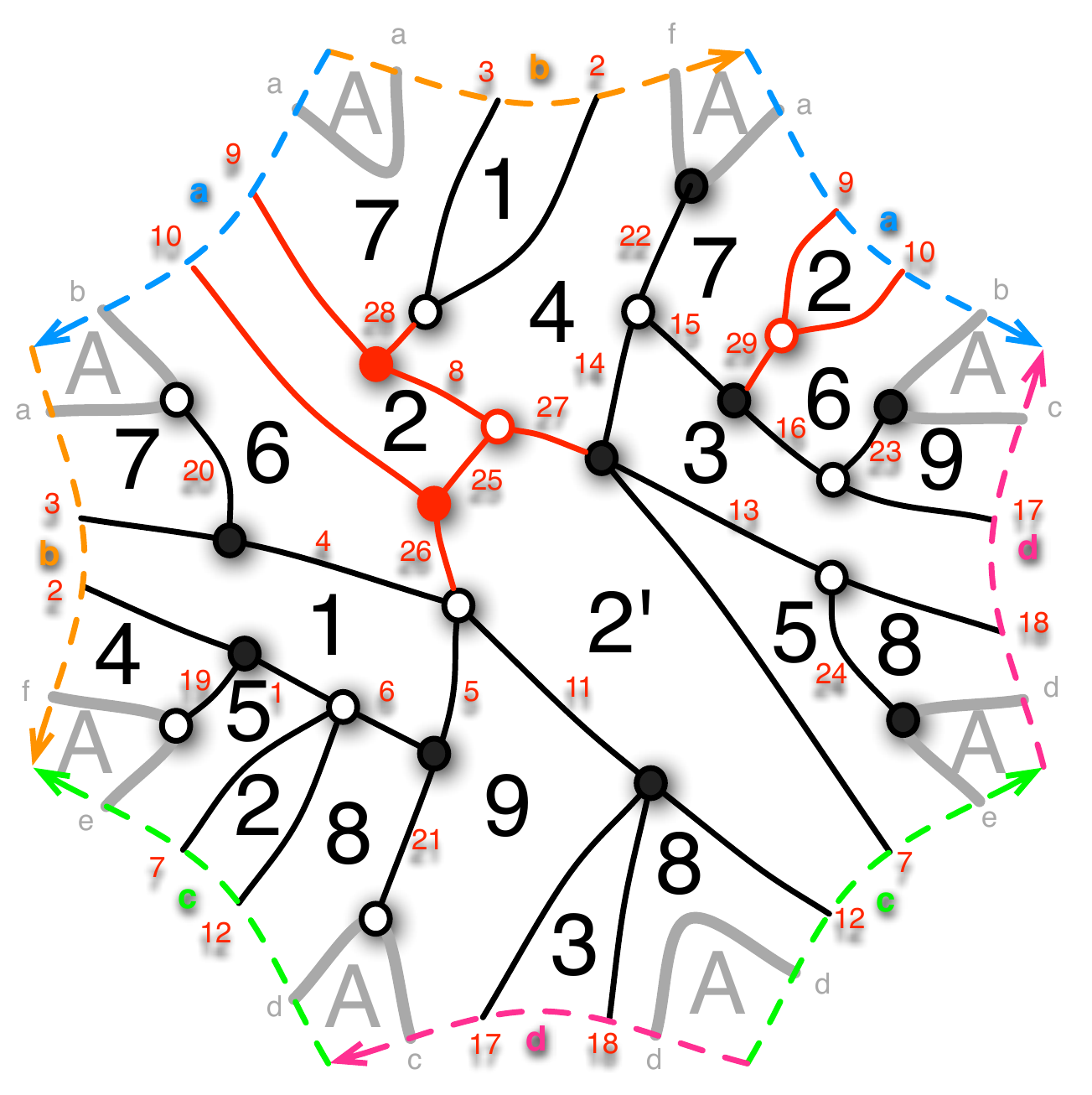}
\\ 
\hspace{0cm}\mbox{(a)} &  \hspace{0.2cm} & \hspace{0cm}\mbox{(b)}
 \end{tabular}
  \caption{\textit{Seiberg duality for BFTs living on a genus 2 Riemann surface.} (a) shows the theory obtained in \sref{section_unhiggsing_tildeC3} by unhiggsing the $\widetilde{C_3}$ model and (b) shows the dual theory obtained by a square move on face 2.
  \label{fmdual}}
 \end{center}
 \end{figure} 

Let us consider the theory introduced in \sref{section_unhiggsing_tildeC3}, which was obtained by unhiggsing the $\widetilde{C_3}$ model. For convenience, it is presented again in \fref{fmdual} (a). The master Kasteleyn matrix for this theory is

{\footnotesize
\beq
K_0^{(a)} = 
\left(
\begin{array}{cccccc|ccc}
 X_{1} & 0 & X_{6} & X_{7} & 0 & X_{12} & 0 & 0 & 0 \\
 X_{2} & X_{3} & 0 & X_{8} & X_{9} & 0 & 0 & 0 & 0 \\
 0 & X_{4} & X_{5} & X_{25} & X_{10} & X_{11} & 0 & 0 & 0 \\
 0 & 0 & 0 & X_{13} & 0 & X_{18} & X_{24} & 0 & 0 \\
 0 & 0 & 0 & X_{14} & X_{15} & 0 & 0 & X_{22} & 0 \\
 0 & 0 & 0 & 0 & X_{16} & X_{17} & 0 & 0 & X_{23} \\
 \hline
 X_{19} & 0 & 0 & 0 & 0 & 0 & 0 & 0 & 0 \\
 0 & X_{20} & 0 & 0 & 0 & 0 & 0 & 0 & 0 \\
 0 & 0 & X_{21} & 0 & 0 & 0 & 0 & 0 & 0
\end{array}
\right),
\eeq}
from which one can determine that this theory has a total of 114 perfect matchings. The moduli space is obtained by demanding invariance under the gauge symmetries associated to all the internal faces, i.e. under $Q_1$, $Q_2$, $Q_2'$ and $Q_3$. As for $\widetilde{C_3}$, the moduli space is a 10d Calabi-Yau, whose toric diagram consists of 101 distinct lattice points. For practical purposes, we omit presenting the explicit list of these points, although we emphasize that it is straightforward to determine the toric diagram using the tools discussed in the paper.

Let us now investigate the moduli space of the dual theory presented in \fref{fmdual} (b), which is obtained from the previous one by performing a square move on face 2. The master Kasteleyn matrix for this theory is

{\footnotesize
\beq
K_0^{(b)} = 
\left(
\begin{array}{cccccc|cc|ccc}
 X_{1} & 0 & X_{6} & X_{7} & 0 & X_{12} & 0 & 0 & 0 & 0 & 0 \\
 X_{2} & X_{3} & 0 & 0 & 0 & 0 & X_{28} & 0 & 0 & 0 & 0 \\
 0 & X_{4} & X_{5} & 0 & 0 & X_{11} & 0 & X_{26} & 0 & 0 & 0 \\
 0 & 0 & 0 & X_{13} & 0 & X_{18} & 0 & 0 & X_{24} & 0 & 0 \\
 0 & 0 & 0 & X_{14} & X_{15} & 0 & 0 & 0 & 0 & X_{22} & 0 \\
 0 & 0 & 0 & 0 & X_{16} & X_{17} & 0 & 0 & 0 & 0 & X_{23} \\
 \hline
 0 & 0 & 0 & 0 & X_{29} & 0 & X_{9} & X_{10} & 0 & 0 & 0 \\
 0 & 0 & 0 & X_{27} & 0 & 0 & X_{8} & X_{25} & 0 & 0 & 0 \\
 \hline
 X_{19} & 0 & 0 & 0 & 0 & 0 & 0 & 0 & 0 & 0 & 0 \\
 0 & X_{20} & 0 & 0 & 0 & 0 & 0 & 0 & 0 & 0 & 0 \\
 0 & 0 & X_{21} & 0 & 0 & 0 & 0 & 0 & 0 & 0 & 0
\end{array}
\right),
\eeq}

\smallskip

\noindent where we have indicated the rows and columns associated to the new superpotential terms in the dual theory. The total number of perfect matchings is now 144. Once again, the moduli space is obtained by quotienting the master space by $Q_1$, $Q_2$, $Q_2'$ and $Q_3$. Its toric diagram is indeed identical to the one for theory (a), being 10d and consisting of 101 different points, as expected from our general arguments. The different number of perfect matchings in the two theories is reflected by the different multiplicities for the points in the respective toric diagrams.

\bigskip

\section{Remarks on Gauging 2}

\label{section_gauging_2}

As discussed in \sref{section_two_gaugings}, two natural gaugings can be associated to a bipartite graph. The two resulting classes of theories can be analyzed using identical tools. For example, \sref{section_moduli_spaces} explains the identification of moduli spaces in completely general terms, independent of the choice of gauging. Other sections focused on illustrating our methods for the case of gauging 1. 

The most prominent feature of the theories resulting from gauging 2 is that, in the Abelian case, they are independent of any embedding into a Riemann surface.  This property is related to the relevance of bipartite graphs for scattering amplitudes. This section collects various examples of BFT$_2$'s, emphasizing some of the main differences with respect to BFT$_1$'s.

\bigskip

\subsection{Implications for Surfaces with No Boundaries}

Bipartite graphs on surfaces $\Sigma$ with no boundaries exhibit a special feature if the corresponding BFT admits extra gaugings. The corresponding BFT$_2$ has precisely $n$ extra gaugings, where $n=2g$ is the number of fundamental cycles of the genus $g$ Riemann surface. This in return has the effect that the BFT$_2$ has always, independent of the number of fields, gauge symmetries and superpotential, a mesonic moduli space of dimension 1. The toric diagram of the mesonic moduli space is a single point, i.e. it corresponds to the complex plane, with multiplicity equal to the number of perfect matchings of the bipartite graph. 

This phenomenon is a straightforward consequence of our discussion in \sref{section_geometry_from_BFT}. Mesonic operators are given by closed loops on the graph, which in turn can be expressed in terms of a basis of cycles. The coordinates in the toric diagram are obtained by considering the coordinates in this basis and eliminating those associated to gauge groups. In BFT$_2$'s all closed loops are gauged, which implies that the moduli space has a trivial toric diagram. The moduli space of BFT$_2$ theories becomes non-trivial when introducing boundaries.

\bigskip

\subsection{Untwisting}

\label{section_untwisting_gauging_2}

One of the main differences between BFT$_1$'s and BFT$_2$'s is their behavior under untwisting. Untwisting does not alter the bipartite graph itself, but generically changes the Riemann surface on which the graph is embedded. Since BFT$_2$'s are independent of any graph embedding into a Riemann surface, we conclude that they are insensitive to untwisting.  A consequence of this statement is that while untwisting plays a central role in various contexts in which bipartite graphs appear, it does not play any non-trivial role in the study of scattering amplitudes. Below, explicit examples of this invariance are presented by revisiting the $C_n$ and $\widetilde{C_n}$ theories under gauging 2.\footnote{Strictly speaking, changing the gauging gives rise to new gauge theories. We continue using the $C_n$ and $\widetilde{C_n}$ names to indicate that the new theories are generated by the same graphs.}

\bigskip

\paragraph{$C_1$ and $\widetilde{C_1}$ Models.} 

The bipartite graphs for these two theories are given in Figures \ref{fm1a} and \ref{fm1aa}. Both of them have the same perfect matching matrix, which is listed in appendix \ref{appendix_Cn}. For the $C_1$ model, gauging 2 implies the additional gauging of the path $(X_1,X_2)$. The new charge matrix becomes

{\footnotesize
\beal{esmm1}
d_{C_1} &=&
\left(
\begin{array}{C{0.4cm}C{0.4cm}C{0.4cm}C{0.4cm}C{0.4cm}C{0.4cm}C{0.4cm}C{0.4cm}}
$X_1$ & $X_2$ & $X_3$ & $X_4$ & $X_5$ & $X_6$ & $X_7$ & $X_8$\\
\hline
 1 & -1 & -1 & 1 & 0 & 0 & 0 & 0 \\
 0 & 0 & 1 & -1 & -1 & 1 & 0 & 0 \\
 \hline
 1 & -1 & 0 & 0 & 0 & 0 & 0 & 0
\end{array}\right),
\eea}
where the last row corresponds to the new gauge symmetry. On the other hand, the $\widetilde{C^1}$ model has no further gauging beyond the one associated to internal faces. The charge matrix is

{\footnotesize
\beal{esmm2}
d_{\widetilde{C_1}} &=&
\left(
\begin{array}{C{0.4cm}C{0.4cm}C{0.4cm}C{0.4cm}C{0.4cm}C{0.4cm}C{0.4cm}C{0.4cm}}
$X_1$ & $X_2$ & $X_3$ & $X_4$ & $X_5$ & $X_6$ & $X_7$ & $X_8$\\
\hline
 1 & -1 & 0 & 0 & 0 & 0 & 0 & 0 \\
 0 & 0 & -1 & 1 & 0 & 0 & 0 & 0 \\
 0 & 0 & 0 & 0 & 1 & -1 & 0 & 0
\end{array}\right).
\eea}

Imposing the D-term charges arising from the matrices above, one can see that, for gauging 2, the moduli spaces for the $C_1$ and $\widetilde{C_1}$ models coincide  and have a toric diagram given by

{\footnotesize
\beq
G_{C_1}= G_{\widetilde{C_1}}= \left(
\begin{array}{C{0.35cm}C{0.2cm}}
 0 & 1 \\
 1 & 0 \\
\hline
 \bf{4}  &  \bf{2} \\
\hline
\end{array}\right).
\label{G_C1_gauging1}
\eeq}

\bigskip

\paragraph{$C_2$ and $\widetilde{C_2}$ Models.} 

Figures \ref{fm2a} and \ref{fm2aa} show the graphs for these models and the perfect matching matrix, which is the same for both theories, appears in appendix \ref{appendix_Cn}. For $C_2$, the extra gauging corresponds to the path $(X_1,X_2,X_3,X_4)$. The new gauge charge matrix is

{\footnotesize
\beal{esmm11}
d_{C_2} &=&
\left(
\begin{array}{C{0.5cm}C{0.5cm}C{0.5cm}C{0.5cm}C{0.5cm}C{0.5cm}C{0.5cm}C{0.5cm}C{0.5cm}C{0.5cm}C{0.5cm}C{0.5cm}C{0.5cm}C{0.5cm}C{0.5cm}C{0.5cm}}
$X_1$ & $X_2$ & $X_3$ & $X_4$ & $X_5$ & $X_6$ & $X_7$ & $X_8$ & $X_9$ & $X_{10}$
& $X_{11}$ & $X_{12}$ & $X_{13}$ & $X_{14}$ & $X_{15}$ & $X_{16}$
\\
\hline
 1 & -1 & 0 & 0 & -1 & 1 & 0 & 0 & 0 & 0 & 0 & 0 & 0 & 0 & 0 & 0 \\
 0 & 0 & 1 & -1 & 0 & 0 & -1 & 1 & 0 & 0 & 0 & 0 & 0 & 0 & 0 & 0 \\
 0 & 0 & 0 & 0 & 0 & -1 & 1 & 0 & 0 & 1 & -1 & 0 & 0 & 0 & 0 & 0 \\
 0 & 0 & 0 & 0 & 1 & 0 & 0 & -1 & -1 & 0 & 0 & 1 & 0 & 0 & 0 & 0 \\
 \hline
 1 & -1 & 1 & -1 & 0 & 0 & 0 & 0 & 0 & 0 & 0 & 0 & 0 & 0 & 0 & 0
\end{array}\right).
\eea}

$\widetilde{C^2}$ can be embedded into a torus with a single boundary. The two fundamental cycles of the torus give rise to the new gauge symmetries, which can be identified with the loops $(X_1,X_2,X_5,X_6)$ and $(X_5,X_8,X_9,X_{12})$. The resulting gauge charge matrix is

{\footnotesize
\beal{esmm12}
d_{\widetilde{C_2}} &=&
\left(
\begin{array}{C{0.5cm}C{0.5cm}C{0.5cm}C{0.5cm}C{0.5cm}C{0.5cm}C{0.5cm}C{0.5cm}C{0.5cm}C{0.5cm}C{0.5cm}C{0.5cm}C{0.5cm}C{0.5cm}C{0.5cm}C{0.5cm}}
$X_1$ & $X_2$ & $X_3$ & $X_4$ & $X_5$ & $X_6$ & $X_7$ & $X_8$ & $X_9$ & $X_{10}$
& $X_{11}$ & $X_{12}$ & $X_{13}$ & $X_{14}$ & $X_{15}$ & $X_{16}$
\\
\hline
 1 & -1 & 1 & -1 & 0 & 0 & 0 & 0 & 0 & 0 & 0 & 0 & 0 & 0 & 0 & 0 \\
 0 & 0 & 0 & 0 & -1 & 1 & -1 & 1 & 0 & 0 & 0 & 0 & 0 & 0 & 0 & 0 \\
 0 & 0 & 0 & 0 & 0 & 0 & 0 & 0 & 1 & -1 & 1 & -1 & 0 & 0 & 0 & 0 \\
 \hline
 1 & -1 & 0 & 0 & -1 & 1 & 0 & 0 & 0 & 0 & 0 & 0 & 0 & 0 & 0 & 0 \\
 0 & 0 & 0 & 0 & 1 & 0 & 0 & -1 & -1 & 0 & 0 & 1 & 0 & 0 & 0 & 0
\end{array}\right).
\eea}
Once again, the resulting moduli spaces are identical. The toric diagram is given by
{\footnotesize
\beq
G_{C_2}= G_{\widetilde{C_2}}= \left(
\begin{array}{C{0.35cm}C{0.35cm}C{0.35cm}C{0.35cm}C{0.35cm}C{0.35cm}}
 1 & 0 & 0 & 0 & -1 & 0\\
 0 & 1 & 0 & -1 & 0 & 0\\
 0 & 0 & 1 & 1 & 1 & 0\\
 0 & 0 & 0 & 1 & 1 & 1\\
 \hline
 \bf{4}  &  \bf{4}  &  \bf{4}  &  \bf{4}  &  \bf{4}  &  \bf{2} \\
\hline
\end{array}\right).
\label{G_C1_gauging2}
\eeq}

\bigskip

\paragraph{$C_3$ and $\widetilde{C_3}$ Models.} 

The graphs for these two models are in Figures \ref{fm3a} and \ref{fm3aa}.
For $C_3$, the extra gauging corresponds to the path $(X_1,X_2,X_3,X_4,X_5,X_6)$. The new gauge charges carried by the fields are
{\footnotesize
\beal{esmm21}
\left(
\begin{array}{C{0.4cm}C{0.4cm}C{0.4cm}C{0.4cm}C{0.4cm}C{0.4cm}}
$X_1$ & $X_2$ & $X_3$ & $X_4$ & $X_5$ & $X_6$ \\
\hline
1 & -1 & 1 & -1 & 1 & -1
\end{array}
\right).
\eea}

The bipartite graph of model $\widetilde{C_3}$ can be embedded into a $g=2$ Riemann surface with four fundamental cycles, which are
{\footnotesize
\beal{esmm22}
\left(
\begin{array}{C{0.5cm}C{0.5cm}C{0.5cm}C{0.5cm}}
$X_{5}$ & $X_{11}$ & $X_{12}$ & $X_{6}$ \\
\hline
1 & -1 & 1 & -1
\end{array}
\right)
~,~
\left(
\begin{array}{C{0.5cm}C{0.5cm}C{0.5cm}C{0.5cm}}
$X_{10}$ & $X_{16}$ & $X_{17}$ & $X_{11}$ \\
\hline
1 & -1 & 1 & -1
\end{array}
\right)~,~
\left(
\begin{array}{C{0.5cm}C{0.5cm}C{0.5cm}C{0.5cm}}
$X_{9}$ & $X_{15}$ & $X_{14}$ & $X_{8}$ \\
\hline
1 & -1 & 1 & -1
\end{array}
\right)
~,~
\left(
\begin{array}{C{0.5cm}C{0.5cm}C{0.5cm}C{0.5cm}}
$X_{2}$ & $X_{1}$ & $X_{7}$ & $X_{8}$ \\
\hline
1 & -1 & 1 & -1
\end{array}
\right),
\nn
\eea}
where we have given the edge charges for the new gauge symmetries, under which all other edges are neutral.

With gauging 2, the moduli spaces of models $C_3$ and $\widetilde{C_3}$ are identical, with the toric diagram given by
{\footnotesize
\beq
G_{C_3}= G_{\widetilde{C_3}}= \left(
\begin{array}{C{0.4cm}C{0.4cm}C{0.4cm}C{0.4cm}C{0.4cm}C{0.4cm}C{0.4cm}C{0.4cm}C{0.4cm}C{0.4cm}C{0.4cm}C{0.4cm}C{0.4cm}C{0.4cm}C{0.4cm}C{0.4cm}C{0.4cm}C{0.4cm}C{0.4cm}C{0.4cm}}
 -2 & -1 & -1 & -1 & -1 & 0 & 0 & 0 & 0 & -1 & -1 & -1 & -1 & -1 & 0 & 0 & 0 & 0 & 1
   & 0 \\
 1 & 0 & 1 & 1 & 1 & 0 & 0 & 1 & 1 & 0 & 0 & 1 & 1 & 1 & 0 & 0 & 0 & 1 & 0 & 0 \\
 1 & 1 & 0 & 0 & 2 & 0 & 0 & 1 & 1 & 0 & 0 & 1 & 1 & 1 & -1 & 1 & 1 & 0 & 0 & 0 \\
 0 & 0 & 0 & 0 & -1 & 0 & 0 & -1 & -1 & 1 & 1 & -1 & 0 & 0 & 1 & 0 & 0 & 0 & 0 & 1
   \\
 0 & 0 & 0 & 1 & -1 & 0 & 1 & -1 & 0 & 0 & 1 & 0 & -1 & 0 & 1 & -1 & 0 & 0 & 0 & 0
   \\
 1 & 1 & 1 & 0 & 1 & 1 & 0 & 1 & 0 & 1 & 0 & 1 & 1 & 0 & 0 & 1 & 0 & 0 & 0 & 0 \\
 \hline
 \bf{6}  &  \bf{6}  &  \bf{6}  &  \bf{6}  &  \bf{6}  &   \bf{6}   &   \bf{6}  &  \bf{6}  &  \bf{6}  &  \bf{4}  &  \bf{4}  &  \bf{4}  &  \bf{4}  &  \bf{4}  &  \bf{4}  &  \bf{4}  &  \bf{4} &  \bf{4} &  \bf{4} &  \bf{2} \\
\hline
\end{array}\right). \\
\label{G_C1_gauging3}
\eeq}

\bigskip

\subsection{Reduction by Higgsing: Further Examples}

\label{section_reduction_by_higgsing_examples}

Following the initial discussion in \sref{section_reduction_by_higgsing} on graph reduction, this section presents examples illustrating higgsings that do not modify the moduli space. Since one of the main applications of graph reduction is related to scattering amplitudes, it is rather natural to study it in the context of gauging 2.\footnote{It is also possible to study graph reduction for gauging 1. It would be interesting to determine whether, for this gauging, it is possible to reduce graphs beyond the planar case.}

We are first going to present a model exhibiting a non-planar to planar reduction and then reconsider $C_n$ and $C_n'$ theories under gauging 2. One of the effects of the additional gauging is to increase the multiplicity of points in the toric diagram of the moduli space which, heuristically, can be linked to an increased reducibility. Indeed, one is going to see that $C_n$ theories are reducible when gauging 2 is considered, and that in this case $C_n'$ theories are related to them by reductions. Finally, an example is included which illustrates how it is possible to systematically investigate all possible combinations of multiple vevs that lead to reduced graphs.

\bigskip

\subsubsection{Non-Planar to Planar Reduction}

\label{section_nonplanar_to_planar_reduction}

\begin{figure}[h]
\begin{center}
\includegraphics[width=5.5cm]{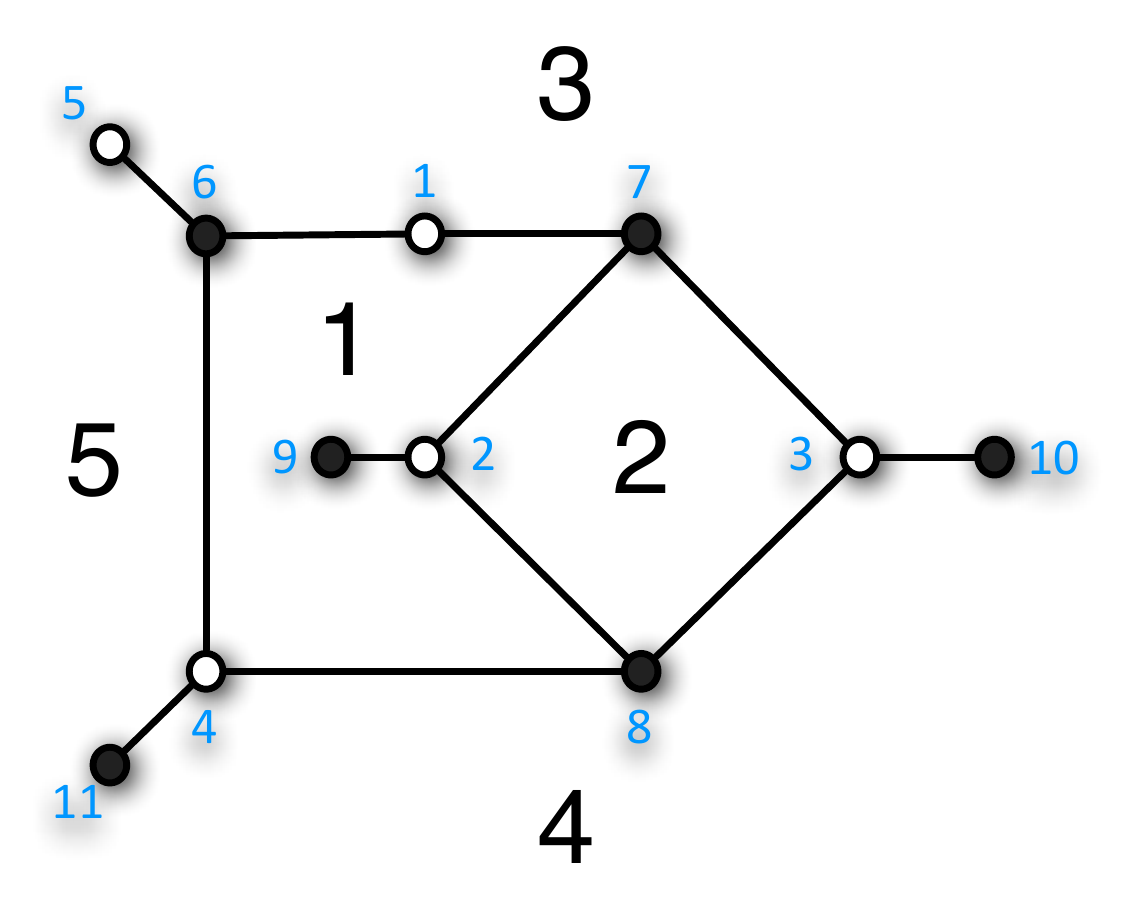}
\caption{\textit{A 4-leg non-planar graph.} We study it under gauging 2, which makes an embedding into a Riemann surface unnecessary.}
\label{non-planar_2loops_4legs}
\end{center}
\end{figure}

Let us consider the non-planar model shown in \fref{non-planar_2loops_4legs}. The corresponding master Kasteleyn matrix is

{\footnotesize
\beq
K_0 =
\left(
\begin{array}{c|ccc|ccc}
 & \ \ 6 \ \ & \ \ 7 \ \ & \ \ 8 \ \ & \ \ 9 \ \ & \ \ 10 \ \ & \ \ 11 \ \ \\ \hline
1 & \ X_{13} \ & \ X_{31} \ & 0 & 0 & 0 & 0 \\
2 & 0 & X_{12} & X_{21} & \ X_{11} \ & 0 & 0 \\
3 & 0 & X_{23} & X_{42} & 0 & \ X_{34} \ & 0 \\
4 & X_{51} & 0 & X_{14} & 0 & 0 & \ X_{45} \ \\ \hline
5 & X_{35} & 0 & 0 & 0 & 0 & 0
\end{array}
\right).
\eeq}

\noindent The perfect matching matrix is

{\scriptsize
\beq
P=\left(
\begin{array}{c|cccccccccccccccccccccc}
& \ p_1 \ & \ p_2 \ & \ p_3 \ & \ p_4 \ & \ p_5 \ & \ p_6 \ & \ p_7 \ & \ p_8 \ & \ p_9 \ \\ \hline
X_{11} & 1 & 1 & 1 & 1 & 0 & 0 & 0 & 0 & 0 \\
X_{13} & 1 & 0 & 0 & 0 & 1 & 1 & 1 & 0 & 0 \\
X_{14} & 1 & 1 & 0 & 0 & 1 & 0 & 0 & 0 & 0 \\
X_{23} & 1 & 0 & 0 & 0 & 0 & 1 & 0 & 0 & 0 \\
X_{12} & 0 & 0 & 0 & 0 & 1 & 0 & 1 & 0 & 0 \\
X_{34} & 0 & 1 & 0 & 0 & 1 & 0 & 0 & 1 & 1 \\
X_{31} & 0 & 1 & 1 & 1 & 0 & 0 & 0 & 1 & 1 \\
X_{35} & 0 & 1 & 1 & 0 & 0 & 0 & 0 & 1 & 0 \\
X_{21} & 0 & 0 & 0 & 0 & 0 & 1 & 0 & 1 & 1 \\
X_{45} & 0 & 0 & 1 & 0 & 0 & 1 & 1 & 1 & 0 \\
X_{42} & 0 & 0 & 1 & 1 & 0 & 0 & 1 & 0 & 0 \\
X_{51} & 0 & 0 & 0 & 1 & 0 & 0 & 0 & 0 & 1 \\
\end{array}
\right) .
\label{P_matrix_non-planar_2loops_4legs}
\eeq}

\noindent Gauging closed paths 1 and 2, a 4d moduli space is obtained with toric diagram given by

{\footnotesize
\beq
G=\left(
\begin{array}{C{0.4cm}C{0.4cm}C{0.4cm}C{0.4cm}C{0.4cm}C{0.4cm}}
 0 & 1 & 0 & 0 & -1 & 0 \\
 0 & 0 & 0 & 1 & 1 & 1 \\
 0 & 0 & 1 & -1 & 0 & 0 \\
 1 & 0 & 0 & 1 & 1 & 0 \\
\hline
  \bf{2} &  \bf{2}   &   \bf{2}   &  \bf{1}  &  \bf{1}  &  \bf{1}  \\
\hline
\end{array}\right).
\label{G_matrix_non-planar_2loops_4legs}
\eeq}

The additional gauging of loop 1 is crucial for the reducibility of this theory. The toric diagram associated to \eref{G_matrix_non-planar_2loops_4legs}, is indeed the one for the moduli space of the BFT associated to the single square box with four legs, as investigated in \cite{Franco:2012mm}. This fact already implies that this theory is reducible. Let us illustrate the reduction in more detail.

\begin{figure}[h]
 \centering
 \begin{tabular}[c]{ccccc}
 \epsfig{file=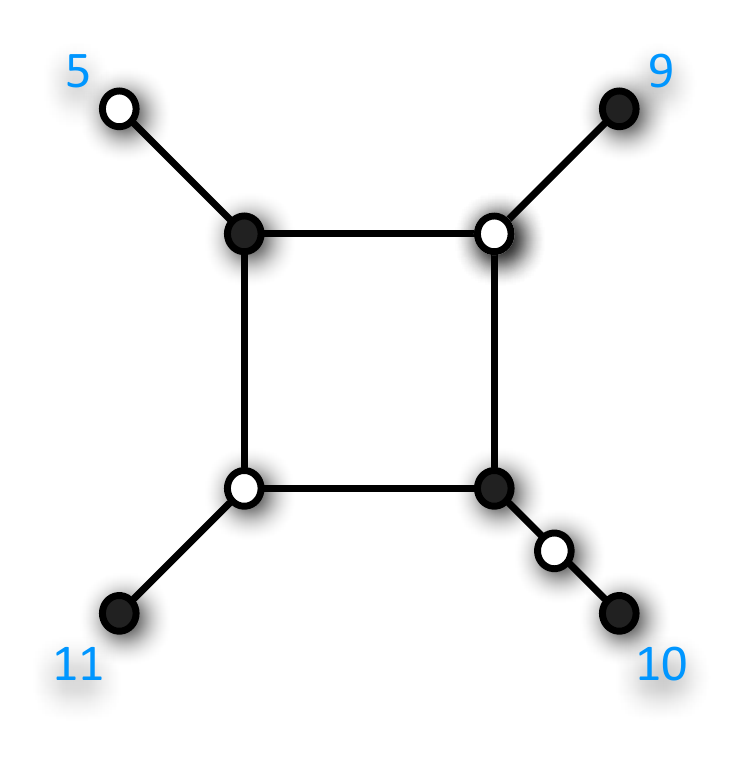,width=0.22\linewidth,clip=} & \ \ \ \ \ &
\epsfig{file=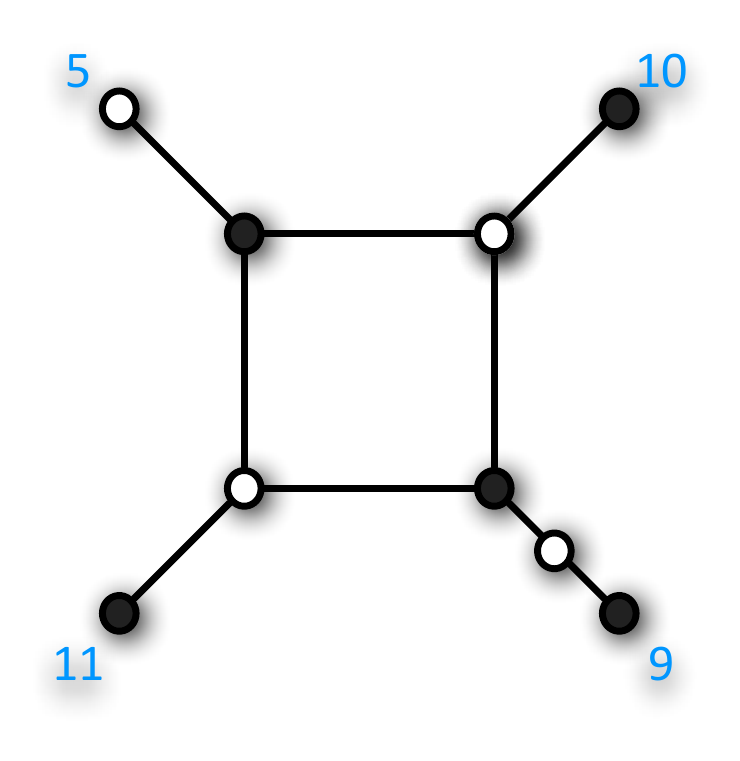,width=0.22\linewidth,clip=}  & \ \ \ \ \ &
\epsfig{file=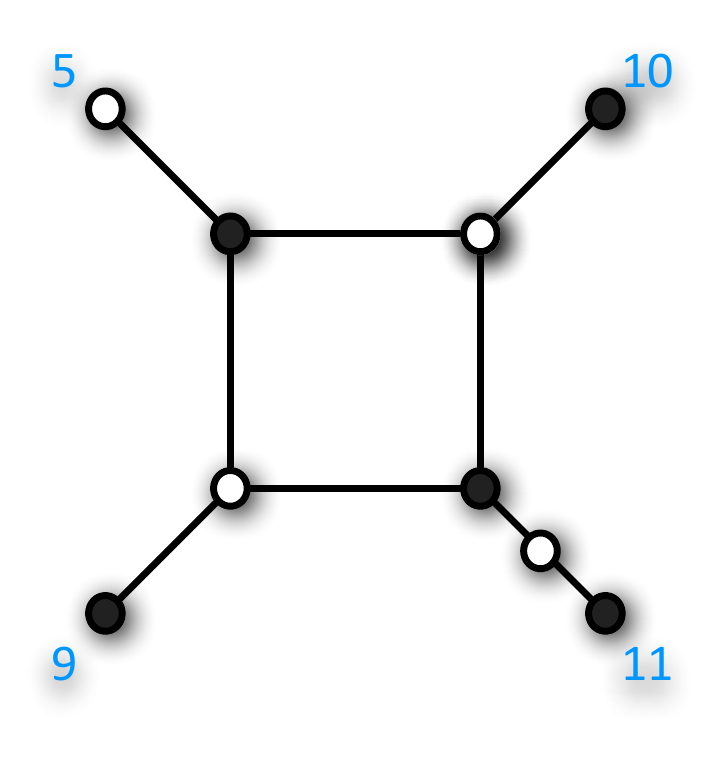,width=0.22\linewidth,clip=} \\ 
\mbox{$\langle X_{23} \rangle\neq 0$} & & \mbox{$\langle X_{12} \rangle\neq 0$} & & \mbox{$\langle X_{51} \rangle\neq 0$}
 \end{tabular}
\caption{\textit{Moduli Space preserving higgsings.} These are the result of the three possible moduli space preserving higgsings of the BFT in \fref{non-planar_2loops_4legs}.}
\label{tiling_2loops_4legs_non-planar_reduction} 
\end{figure}

Perfect matchings group themselves as follows over the 6 points in the toric diagram
\beq
\begin{array}{c}
\{p_{1}, p_{4}\} \ , \ \{p_{5}, p_{9}\} \, \ \{p_{6}, p_{7}\} \\
\{p_{2}\} \ , \ \{p_{3}\} \ , \ \{p_{8}\}
\end{array}
\eeq
It is straightforward to see that one can at most turn on a single non-zero vev while preserving the moduli space. There are three options for doing so by giving vevs to $X_{23}$, $X_{12}$ or $X_{51}$. \fref{tiling_2loops_4legs_non-planar_reduction} shows the resulting graphs after the corresponding higgsings. For a fixed cyclic ordering of the external nodes, the graphs associated to the higgsings by $X_{12}$ and $X_{51}$ are identical, so we conclude that there are only two distinct reduced graphs. Notice that there is no sequence of moves and bubble reductions capable of achieving this reduction. Interestingly, the reduced graphs are planar, unlike the original theory. This model was originally investigated in \cite{Nima} using the leading singularity approach. Remarkably, their results are in agreement with ours. 

\bigskip

\subsubsection{Reducing $C_n$ Theories}

Interestingly, $C_n$ theories become reducible when gauging 2 is considered. Below, it is shown that the $C_n'$ theories of \sref{section_Cn'} are indeed reductions of them, since they have the same moduli space and a smaller number of loops. The moduli space of $C_n$ models for gauging 2 has been discussed in \sref{section_untwisting_gauging_2}. The moduli space for $C_n'$ theories is computed in this gauging and one finds full agreement. The perfect matching multiplicities of points in the toric diagram however decrease. As discussed above, the perfect matchings that disappear are associated to the fields acquiring non-zero vevs when higgsing from $C_n$ to $C_n'$.

\bigskip

\paragraph{$C'_1$ Model.} 

The toric diagram for the moduli space of this theory is given by

{\footnotesize
\beq
G_{C'_1}=\left(
\begin{array}{C{0.4cm}C{0.4cm}}
 0 & 1 \\
 1 & 0 \\
\hline
 \bf{4}  &  \bf{1} \\
\hline
\end{array}\right),
\eeq}

\smallskip

\noindent which is indeed equal to \eref{G_C1_gauging1} up to multiplicities.

\bigskip

\paragraph{$C'_2$ Model.} 

The toric diagram for the moduli space corresponds to

{\footnotesize
\beq
G_{C'_2}=\left(
\begin{array}{C{0.4cm}C{0.4cm}C{0.4cm}C{0.4cm}C{0.4cm}C{0.4cm}}
 0 & 0 & -1 & 1 & 0 & 0 \\
 0 & 1 & 0 & 0 & -1 & 0 \\
 1 & 0 & 1 & 0 & 1 & 0 \\
 0 & 0 & 1 & 0 & 1 & 1 \\ \hline
 \bf{4}  &  \bf{2}  &  \bf{2}  &  \bf{2}  &  \bf{2}  &  \bf{1} \\ 
\hline
\end{array}
\right),
\eeq}

\smallskip

\noindent i.e. it agrees with \eref{G_C1_gauging2} up to multiplicities.

\bigskip

\paragraph{$C'_3$ Model.} 

The toric diagram of the moduli space corresponds to

{\footnotesize
\beq
G_{C'_3}=\left(
\begin{array}{C{0.4cm}C{0.4cm}C{0.4cm}C{0.4cm}C{0.4cm}C{0.4cm}C{0.4cm}C{0.4cm}C{0.4cm}C{0.4cm}C{0.4cm}C{0.4cm}C{0.4cm}C{0.4cm}C{0.4cm}C{0.4cm}C{0.4cm}C{0.4cm}C{0.4cm}C{0.4cm}}
 -1 & -2 & -1 & 0 & -1 & 0 & 0 & -1 & 0 & -1 & -1 & -1 & -1 & -1 & 0 & 0 & 1 & 0 & 0 & 0 \\
 1 & 1 & 0 & 1 & 1 & 0 & 1 & 1 & 0 & 1 & 0 & 1 & 0 & 1 & 0 & 0 & 0 & 0 & 1 & 0 \\
 2 & 1 & 1 & 1 & 0 & 0 & 1 & 0 & 0 & 1 & 0 & 1 & 0 & 1 & 1 & 1 & 0 & -1 & 0 & 0 \\
 -1 & 0 & 0 & -1 & 0 & 0 & -1 & 0 & 0 & 0 & 1 & 0 & 1 & -1 & 0 & 0 & 0 & 1 & 0 & 1 \\
 -1 & 0 & 0 & -1 & 0 & 0 & 0 & 1 & 1 & -1 & 0 & 0 & 1 & 0 & -1 & 0 & 0 & 1 & 0 & 0 \\
 1 & 1 & 1 & 1 & 1 & 1 & 0 & 0 & 0 & 1 & 1 & 0 & 0 & 1 & 1 & 0 & 0 & 0 & 0 & 0 \\
\hline
 \bf{3}  &  \bf{3}  &  \bf{3}  &  \bf{3}  &  \bf{3}  &  \bf{3}  &  \bf{3}  &  \bf{3}  &  \bf{3}  &  \bf{1}  &  \bf{1}  &  \bf{2}  &  \bf{2}  &  \bf{4}  &  \bf{1}  &  \bf{1}  &  \bf{2}  &  \bf{1}  &  \bf{1}  & \bf{1} \\ 
\hline
\end{array}
\right).
\eeq}

\smallskip

\noindent Once again, the toric diagram in \eref{G_C1_gauging3} is preserved up to reduced multiplicities.

\bigskip

\subsubsection{Full Reduction of $C_3$}

The models above can be further reduced. Let us focus on $C_3$ and show how our methods allow us to identify its maximal reductions. This theory has 24 chiral fields (6 of which correspond to external legs) and 96 perfect matchings that distribute over 20 different points of the moduli space toric diagram, as summarized by \eref{G_C1_gauging3}. 

\begin{figure}[h]
\begin{center}
\includegraphics[height=4.5cm]{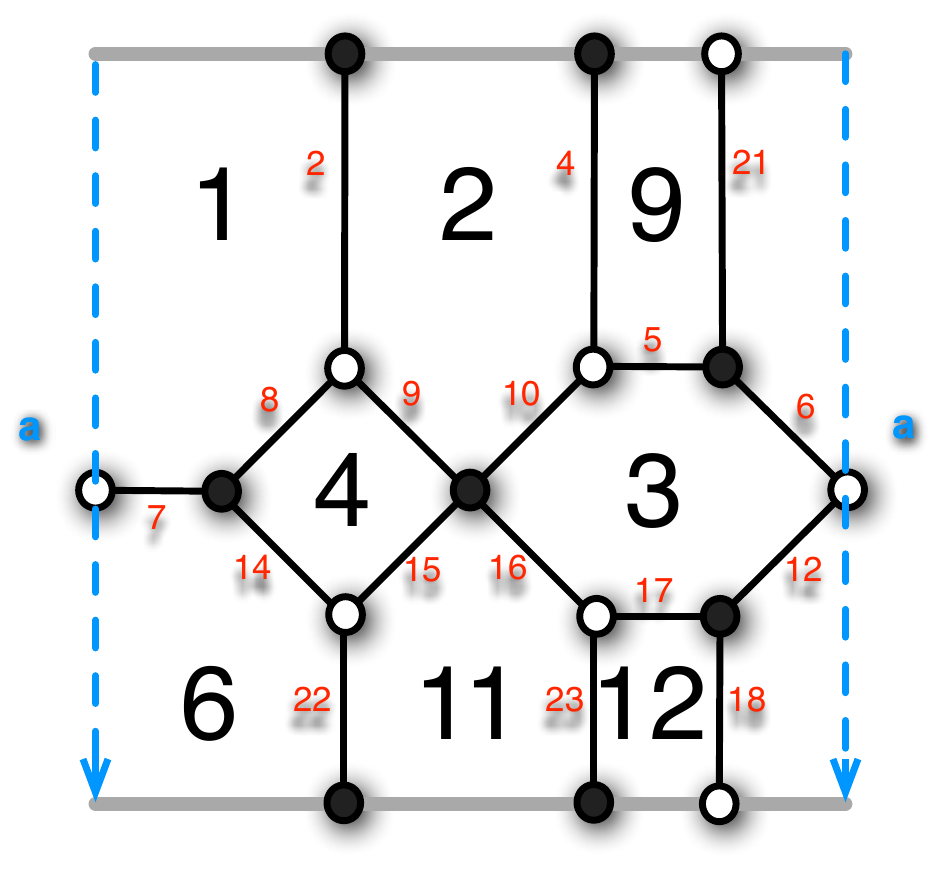}
\caption{\textit{Reduced graph from the $C_3$ theory by turning on vevs for $X_{1}$, $X_{3}$, $X_{11}$ and  $X_{13}$.}
}
\label{m3ah2}
\end{center}
\end{figure}

Using our methods, one can determine that there are 24 different combinations of four non-zero simultaneous vevs that produce reduced graphs. They are given by:

{\footnotesize\beq
\begin{array}{ccccccc}
\{X_{1}, X_{3}, X_{11}, X_{13}\} & \ \ \ \ \ & \{X_{1}, X_{5}, X_{9}, X_{17}\} & \ \ \ \ \ &  \{X_{1}, X_{8}, X_{12}, X_{13}\} & \ \ \ \ \ &  \{X_{1}, X_{9}, X_{11}, X_{13}\} \\
\{X_{1}, X_{9}, X_{13}, X_{17}\} & \ \ \ \ \ &  \{X_{2}, X_{4}, X_{12}, X_{16}\} & \ \ \ \ \ &  \{X_{2}, X_{6}, X_{10}, X_{14}\} & \ \ \ \ \ & \{X_{2}, X_{7}, X_{9}, X_{14}\} \\
\{X_{2}, X_{10}, X_{12}, X_{14}\} & \ \ \ \ \ &  \{X_{2}, X_{12}, X_{14}, X_{16}\} & \ \ \ \ \ &  \{X_{3}, X_{5}, X_{7}, X_{15}\} & \ \ \ \ \ &  \{X_{3}, X_{7}, X_{11}, X_{15}\} \\
\{X_{3}, X_{8}, X_{10}, X_{15}\} & \ \ \ \ \ &  \{X_{3}, X_{11}, X_{13}, X_{15}\} & \ \ \ \ \ &  \{X_{4}, X_{6}, X_{8}, X_{18}\} & \ \ \ \ \ & \{X_{4}, X_{8}, X_{12}, X_{16}\} \\
\{X_{4}, X_{8}, X_{16}, X_{18}\} & \ \ \ \ \ &  \{X_{4}, X_{9}, X_{11}, X_{16}\} & \ \ \ \ \ &  \{X_{5}, X_{7}, X_{9}, X_{17}\} & \ \ \ \ \ &  \{X_{5}, X_{7}, X_{15}, X_{17}\} \\
\{X_{5}, X_{10}, X_{12}, X_{17}\} & \ \ \ \ \ &  \{X_{6}, X_{7}, X_{11}, X_{18}\} & \ \ \ \ \ &  \{X_{6}, X_{8}, X_{10}, X_{18}\} & \ \ \ \ \ &  \{X_{6}, X_{10}, X_{14}, X_{18}\}
\end{array}
\eeq}
Turning on any additional vev would lead to a different moduli space and hence an inequivalent theory. \fref{m3ah2} shows the reduced graph resulting from turning on vevs for $X_{1}$, $X_{3}$, $X_{11}$ and  $X_{13}$. It is possible to check whether different combinations of vevs lead to the same reduced graph, although we do not pursue this question any further.

\bigskip

\section{Conclusions and Future Directions \label{conclusions}} 

This paper carried out a comprehensive study of BFTs, considerably extending the understanding of these theories in various directions. We recognized that there are two natural ways of assigning gauge symmetries to BFTs, which implies that in fact there are two classes of gauge theories that can be associated to bipartite graphs. BFT$_1$'s require specifying an embedding of the graph into a Riemann surface for their definition, while BFT$_2$'s do not need a Riemann surface at all. The two classes of theories are interesting in their own right and find applications in different contexts. For example, a subclass of BFT$_1$'s arise on D3-branes over toric CY 3-folds, while BFT$_2$'s are related to scattering amplitudes.

An alternative approach for connecting BFTs and the toric CYs that correspond to their master and moduli spaces was discussed. For planar BFTs, this perspective allowed us to identify the toric diagram of the moduli space with the matroid polytope arising in studies of cells in the positive Grassmannian \cite{Postnikov_toric}.

We next investigated a vast array of graph transformations and used them to generate new BFTs, including some infinite families. We also studied Seiberg dualities, reductions by higgsings, and explored some of the main differences between the two possible gaugings of BFTs.

Our work suggests several directions for future investigation of BFTs, regarding their properties, string theory realization and applications. It is also desirable to understand the physical origin of the connections between the different contexts in which BFTs appear. Below we collect some interesting open questions and thoughts on how to address them.

\bigskip

\paragraph{$\bullet$ Stringy Implementation:} We already know that BFT$_1$'s on $T^2$ arise on the worldvolume of D3-branes probing toric CY 3-folds \cite{Hanany:2005ve,Franco:2005rj,Franco:2005sm}. It would be extremely interesting to find a stringy embedding for other classes of BFTs. It is natural to expect that planar BFT$_1$'s with boundaries can be engineered by fractional D3-branes and flavor D7-branes on toric CY 3-folds. The chiral fields associated to external legs, which are singlets of all gauge symmetries and transform as bifundamentals of global symmetry groups, would correspond to higher dimensional fields living at the intersections between pairs of non-compact D7-branes. This would provide additional motivation for the special treatment we give to these fields. Orbifolds of $\mathbb{C}^3$ provide a simple setup to test these ideas, in which the D3-D3, D3-D7 and D7-D7 spectrum can be determined using standard techniques \cite{Douglas:1996sw}. After finding the spectrum of a generic brane configuration in these geometries, the BFTs would correspond to combinations of branes leading to theories in which gauge symmetries are anomaly free. Partial resolution could then be exploited to generate BFTs associated to other CY 3-folds.

\medskip

\paragraph{$\bullet$ General Graph Reducibility:} It would be interesting to determine what the most general operation leading to graph reduction is. Can all reductions be implemented by either a) moves and bubble reductions or b) edge deletion, or are there more exotic examples that do not fit into any of these two categories? Many reductions obtained by means of (a) can also be achieved by using (b), it would also be desirable to determine whether this is always true.

\medskip

\paragraph{$\bullet$ Superconformal Invariance:} It is interesting to investigate whether and under which conditions, BFTs give rise to superconformal fixed points. In the case of BFT$_1$'s, it is indeed possible to map the R-charges of fields to angles in the isoradial or rhombus embeddings of the graph \cite{Kennaway:2007tq,Hanany:2005ss,Broomhead:2008an}.\footnote{BFT$_2$'s are independent of a Riemann surface embedding and hence there is no simple graphical translation of R-charges and beta functions.} In such embeddings, the vanishing of individual beta functions for gauge and superpotential couplings translate into zero local curvature. For BFT$_1$'s it thus become natural to expect CFTs whenever the embedded graph has vanishing curvature everywhere.

It is natural to speculate that, even in cases in which a superconformal fixed point described by the full graph does not naively exist due to the curvature considerations in the previous paragraph, the graph is still useful for identifying possible fixed points. The non-vanishing curvature might be accommodated by `breaking the graph apart' at some places. The field theoretic interpretation of this operation would be that the corresponding couplings disappear from the gauge theory. We can understand this phenomenon as a graphical indication that it is possible to find a fixed point if these couplings flow to zero and the vanishing of the corresponding beta functions disappears as a constraint.

It would also be interesting to revisit the question of conformal invariance while allowing different ranks for gauge and global symmetry groups.

\medskip

\paragraph{$\bullet$ Gauge Theory and Reducibility:} From a BFT viewpoint, the reduction of degrees of freedom associated to graph reductions is strongly reminiscent of an RG flow. It would be interesting to determine whether this connection is indeed true. If so, it would provide an alternative perspective on reduced graphs, which would be mapped to fixed points of the RG flow. 

On a related front, it would be interesting to establish whether there is a simple field theoretic diagnostic for identifying reducible graphs. The graphical representation of superconformal R-charges in terms of the isoradial or rhombus embeddings discussed above provides a possible way of addressing this question. In fact, extending existing results for BFTs on $T^2$ \cite{Broomhead:2008an}, in the planar case it is possible to connect the existence of multiple intersecting zig-zag paths to non-positive values of R-charges. One limitation of this approach is that the reducibility criterion based on zig-zag paths does not seem to capture all possible reductions, particularly those that cannot be implemented in terms of moves or bubble reductions but  that require higgsing, as the one discussed in \sref{section_nonplanar_to_planar_reduction}. This phenomenon becomes important for non-planar graphs. 

\medskip

\paragraph{$\bullet$ Detailed Investigation of Non-Abelian BFTs:} The Abelian version of BFTs is sufficient for certain applications, such as scattering amplitudes, and also captures some features, like the connection by moves and bubble reductions, that are also present for non-Abelian theories. Having said that, it is extremely interesting to perform a more detailed study of non-Abelian BFTs. We envision powerful tools such as Hilbert series \cite{Hanany:2007zz,Butti:2007jv,Feng:2007ur,Benvenuti:2006qr} and the superconformal index \cite{Romelsberger:2005eg,Kinney:2005ej,Terashima:2012cx} can provide an interesting window into the dynamics of the general theories.

\medskip

\paragraph{$\bullet$ Non-Simply Connected Graphs:} Our tools apply without changes to non-simply connected graphs. Our discussion can certainly be extended to a more general class of theories that includes this possibility.
\\

The above open questions and extensions of our enquiry illustrate the vast richness of the subject of BFTs. We expect to report on new results on the subject in the near future.

\bigskip

\section*{Acknowledgments}

We would like to thank F. Cachazo, S. Cremonesi, A. Hanany, V. Jejjala, J. Maldacena, S. Ramgoolam, C. Wen and L. Williams for useful conversations. We are particularly thankful to N. Arkani-Hamed and J. Trnka for very useful discussions. S. F. is grateful to the Institute for Advanced Study for kind hospitality during the completion of this work. R.-K. S. thanks the Simons Center for Geometry and Physics at Stony Brook University, the Korea Institute for Advanced Study, the Yukawa Institute for Theoretical Physics at Kyoto University, Nagoya University, the Kavli Institute for the Physics and Mathematics of the Universe, Universit\'e libre de Bruxelles, the Center of Theoretical Science at Princeton University, and the Perimeter Institute for Theoretical Physics for hospitality during various stages of this work. The work of S.F and D. G. is supported by the U.K. Science and Technology Facilities Council (STFC).

\bigskip

\appendix

\section{$C_n$ Theories} \label{appendix_Cn}

This section summarizes as a reference the perfect matching matrices for the $C_1$ and $C_2$ models that have been in introduced in \sref{section_infinite_families}. These are
\\

{\tiny
\beq
P_{C_1}=
\left(
\begin{array}{c|cccccc}
 \; & \ p_1 \ & \ p_2 \ & \ p_3 \ & \ p_4 \ & \ p_5 \ & \ p_6 \ \\
 \hline
 X_{1} & 1 & 0 & 1 & 0 & 0 & 0 \\
 X_{2} & 0 & 1 & 0 & 1 & 0 & 0 \\
 X_{3} & 0 & 0 & 0 & 0 & 1 & 0 \\
 X_{4} & 0 & 0 & 0 & 0 & 0 & 1 \\
 X_{5} & 1 & 1 & 0 & 0 & 0 & 0 \\
 X_{6} & 0 & 0 & 1 & 1 & 0 & 0 \\
 X_{7} & 0 & 0 & 0 & 0 & 1 & 1 \\
 X_{8} & 0 & 0 & 0 & 0 & 1 & 1
\end{array}
\right)~~,~~\nn
\eeq}

\noindent and

{\tiny
\beq
P_{C_2}=
\left(
\begin{array}{c|cccccccccccccccccccccc}
 \; & \ p_1\ &\ p_2\ &\ p_3\ &\ p_4\ &\ p_5\ &\ p_6\ &\ p_7\ &\ p_8\ &\ p_9\ & p_{10} & p_{11} & p_{12} & p_{13} & p_{14} & p_{15} & p_{16} & p_{17} & p_{18} & p_{19} & p_{20} & p_{21} & p_{22}\\
 \hline
X_{1} & 1 & 0 & 0 & 0 & 0 & 0 & 1 & 1 & 0 & 0 & 1 & 0 & 0 & 0 & 0 & 0 & 1 & 0 & 1 & 0 & 0 & 0
   \\
X_{2} & 0 & 1 & 1 & 1 & 0 & 0 & 0 & 0 & 0 & 0 & 0 & 0 & 0 & 1 & 0 & 0 & 0 & 1 & 0 & 0 & 1 & 0
   \\
X_{3} & 1 & 0 & 0 & 0 & 1 & 1 & 0 & 0 & 0 & 0 & 0 & 0 & 0 & 0 & 1 & 0 & 1 & 0 & 0 & 0 & 0 & 1
   \\
X_{4} & 0 & 1 & 0 & 0 & 0 & 0 & 0 & 0 & 1 & 1 & 0 & 1 & 0 & 0 & 0 & 0 & 0 & 1 & 0 & 1 & 0 & 0
   \\
X_{5} & 0 & 0 & 1 & 1 & 1 & 1 & 0 & 0 & 0 & 0 & 0 & 0 & 1 & 0 & 0 & 0 & 0 & 0 & 0 & 0 & 0 & 0
   \\
X_{6} & 0 & 0 & 0 & 0 & 0 & 0 & 1 & 1 & 1 & 1 & 0 & 0 & 0 & 0 & 0 & 1 & 0 & 0 & 0 & 0 & 0 & 0
   \\
X_{7} & 0 & 0 & 0 & 0 & 0 & 0 & 0 & 0 & 0 & 0 & 1 & 1 & 1 & 0 & 0 & 0 & 0 & 0 & 1 & 1 & 0 & 0
   \\
X_{8} & 0 & 0 & 0 & 0 & 0 & 0 & 0 & 0 & 0 & 0 & 0 & 0 & 0 & 1 & 1 & 1 & 0 & 0 & 0 & 0 & 1 & 1
   \\
X_{9} & 0 & 0 & 0 & 0 & 0 & 0 & 0 & 0 & 0 & 0 & 0 & 0 & 0 & 0 & 0 & 0 & 1 & 1 & 1 & 1 & 1 & 1
   \\
X_{10} & 1 & 1 & 0 & 0 & 0 & 0 & 0 & 0 & 0 & 0 & 1 & 1 & 0 & 1 & 1 & 0 & 0 & 0 & 0 & 0 & 0 & 0
   \\
X_{11} & 0 & 0 & 0 & 1 & 0 & 1 & 0 & 1 & 0 & 1 & 0 & 0 & 0 & 0 & 0 & 0 & 1 & 1 & 0 & 0 & 0 & 0
   \\
X_{12} & 1 & 1 & 1 & 0 & 1 & 0 & 1 & 0 & 1 & 0 & 0 & 0 & 0 & 0 & 0 & 0 & 0 & 0 & 0 & 0 & 0 & 0
   \\
X_{13} & 0 & 0 & 0 & 0 & 1 & 1 & 0 & 0 & 1 & 1 & 0 & 1 & 1 & 0 & 1 & 1 & 0 & 0 & 0 & 1 & 0 & 1
   \\
X_{14} & 0 & 0 & 1 & 1 & 0 & 0 & 1 & 1 & 0 & 0 & 1 & 0 & 1 & 1 & 0 & 1 & 0 & 0 & 1 & 0 & 1 & 0
   \\
X_{15} & 0 & 0 & 1 & 0 & 1 & 0 & 1 & 0 & 1 & 0 & 0 & 0 & 1 & 0 & 0 & 1 & 0 & 0 & 1 & 1 & 1 & 1
   \\
X_{16} & 0 & 0 & 0 & 1 & 0 & 1 & 0 & 1 & 0 & 1 & 1 & 1 & 1 & 1 & 1 & 1 & 0 & 0 & 0 & 0 & 0 & 0
\end{array}
\right)~~.\nn
\eeq}

\noindent The perfect matching matrix for the $C_3$ and any consecutive model are too large to be included.

\newpage

\section{$C'_n$ Theories} \label{appendix_Cn'}

This section summarizes the perfect matching matrices for the $C'_1$, $C'_2$ and $C'_3$ models which are discussed in \sref{section_Cn'}. The $C'_n$ models correspond to higgsed $C_n$ models. The perfect matching matrices are

\noindent\makebox[\textwidth]{%
\tiny
$
P_{C'_{1}}=
\left(
\begin{array}{c|ccccc}
\; & \ p_1 \ & \ p_2 \ & \ p_3 \ & \ p_4 \ & \ p_5 \ \\ \hline
X_1 & 1 & 0 & 1 & 0 & 0 \\
X_2 & 0 & 1 & 0 & 1 & 0 \\
X_3 & 0 & 0 & 0 & 0 & 1 \\
X_4 & 1 & 1 & 0 & 0 & 0 \\
X_5 & 0 & 0 & 1 & 1 & 0 \\
X_6 & 0 & 0 & 0 & 0 & 1 \\
X_7 & 0 & 0 & 0 & 0 & 1
\end{array}
\right)~~,~~\nn
$
}
\vspace{0.2cm}

\noindent\makebox[\textwidth]{%
\tiny
$
P_{C'_{2}}=
\left(
\begin{array}{c|cccccccccccccccc}
\; & \ p_1 \ & \ p_2 \ & \ p_3 \ & \ p_4 \ & \ p_5 \ & \ p_6 \ & \ p_7 \ & \ p_8 \ & \ p_9 \ & \ p_{10} \ & \ p_{11} \ & \ p_{12} \ & \ p_{13} \ & \ p_{14} \ & \ p_{15} \ & \ p_{16} \ \\ \hline
X_1 & 1 & 1 & 1 & 1 & 0 & 0 & 0 & 0 & 0 & 0 & 0 & 0 & 1 & 0 & 0 & 0 \\
X_2 & 0 & 0 & 0 & 0 & 0 & 0 & 0 & 0 & 1 & 0 & 0 & 0 & 0 & 1 & 1 & 0 \\
X_3 & 0 & 0 & 0 & 0 & 1 & 1 & 1 & 1 & 0 & 2 & 1 & 1 & 1 & 0 & 0 & 2 \\
X_4 & 0 & 0 & 0 & 0 & 1 & 1 & 1 & 1 & 0 & 0 & 0 & 0 & 0 & 1 & 0 & 0 \\
X_5 & 1 & 1 & 1 & 1 & 1 & 1 & 1 & 1 & 0 & 0 & 1 & 1 & 0 & 0 & 0 & 0 \\
X_6 & 0 & 0 & 0 & 0 & 0 & 0 & 0 & 0 & 1 & 1 & 1 & 1 & 0 & 0 & 1 & 1 \\
X_7 & 0 & 0 & 0 & 0 & 0 & 0 & 0 & 0 & 0 & 0 & 0 & 0 & 1 & 1 & 1 & 1 \\
X_8 & 0 & 0 & 0 & 0 & 0 & 0 & 0 & 0 & 1 & 1 & 0 & 0 & 0 & 0 & 0 & 0 \\
X_9 & 0 & 0 & 1 & 0 & 0 & 0 & 1 & 0 & 0 & 0 & 0 & 0 & 0 & 0 & 0 & 0 \\
X_{10} & 1 & 0 & 0 & 0 & 1 & 0 & 0 & 0 & 0 & 0 & 0 & 0 & 0 & 0 & 0 & 0 \\
X_{11} & 0 & 0 & 0 & 0 & 0 & 0 & 0 & 0 & 0 & 0 & 0 & 0 & 0 & 0 & 0 & 0 \\
X_{12} & 1 & 1 & 1 & 1 & 0 & 0 & 0 & 0 & 1 & 0 & 1 & 1 & 0 & 0 & 1 & 0 \\
X_{13} & 1 & 0 & 0 & 1 & 1 & 0 & 0 & 1 & 0 & 0 & 0 & 1 & 0 & 0 & 1 & 1 \\
X_{14} & 0 & 1 & 1 & 1 & 0 & 1 & 1 & 1 & 1 & 1 & 1 & 1 & 0 & 0 & 0 & 0
\end{array}
\right)~~,~~\nn
$
}
\vspace{0.2cm}

\noindent\makebox[\textwidth]{%
\tiny
$
P_{C'_{3}}=
\left(
\begin{array}{c|ccccccccccccccccccccccccc}
\; & \ p_1 \ & \ p_2 \ & \ p_3 \ & \ p_4 \ & \ p_5 \ & \ p_6 \ & \ p_7 \ & \ p_8 \ & \ p_9 \ & \ p_{10} \ & \ p_{11} \ & \ p_{12} \ & \ p_{13} \ & \ p_{14} \ & \ p_{15} \ & \ p_{16} \ & \ p_{17} \ & \ p_{18} \ & \ p_{19} \ & \ p_{20} \
& \ p_{21} \ & \ p_{22} \ 
\\
\hline
X_1 & 0 & 0 & 1 & 1 & 0 & 1 & 0 & 0 & 0 & 0 & 0 & 0 & 0 & 1 & 1 & 0 & 0 & 0 & 0 & 0 & 0 &
   0  \\
X_2 & 1 & 1 & 0 & 0 & 0 & 0 & 0 & 1 & 1 & 1 & 1 & 1 & 1 & 0 & 0 & 0 & 0 & 0 & 1 & 1 & 1 &
   0 \\
X_3 & 0 & 0 & 1 & 1 & 1 & 1 & 1 & 0 & 0 & 0 & 0 & 0 & 0 & 1 & 1 & 1 & 1 & 1 & 0 & 0 & 0 &
   1  \\
X_4 & 0 & 0 & 0 & 0 & 0 & 0 & 0 & 1 & 1 & 1 & 0 & 0 & 0 & 0 & 0 & 0 & 0 & 0 & 0 & 0 & 0 &
   0  \\
X_5 & 0 & 0 & 0 & 0 & 0 & 0 & 0 & 0 & 0 & 0 & 1 & 1 & 1 & 1 & 1 & 1 & 1 & 1 & 0 & 0 & 0 &
   0 \\
X_6 & 0 & 0 & 0 & 0 & 0 & 0 & 0 & 0 & 0 & 0 & 0 & 0 & 0 & 0 & 0 & 0 & 0 & 0 & 1 & 1 & 1 &
   1  \\
X_7 & 0 & 0 & 0 & 0 & 0 & 0 & 0 & 0 & 0 & 0 & 0 & 0 & 0 & 0 & 0 & 0 & 0 & 0 & 0 & 0 & 0 &
   0  \\
X_8 & 1 & 1 & 1 & 1 & 1 & 1 & 1 & 0 & 0 & 0 & 0 & 0 & 0 & 0 & 0 & 0 & 0 & 0 & 1 & 1 & 1 &
   1  \\
X_9 & 1 & 1 & 0 & 0 & 1 & 0 & 1 & 1 & 1 & 1 & 1 & 1 & 1 & 0 & 0 & 1 & 1 & 1 & 0 & 0 & 0 &
   0  \\
X_{10} & 1 & 0 & 1 & 0 & 1 & 0 & 0 & 1 & 1 & 0 & 1 & 1 & 0 & 1 & 0 & 1 & 1 & 0 & 1 & 0 & 0 &
   1  \\
X_{11} & 0 & 1 & 0 & 1 & 0 & 1 & 1 & 0 & 0 & 1 & 0 & 0 & 1 & 0 & 1 & 0 & 0 & 1 & 0 & 1 & 1 &
   0  \\
X_{12} & 0 & 0 & 0 & 0 & 0 & 0 & 0 & 0 & 1 & 0 & 0 & 1 & 0 & 1 & 0 & 0 & 1 & 0 & 0 & 0 & 0 &
   0  \\
X_{13} & 0 & 0 & 0 & 0 & 0 & 0 & 0 & 1 & 0 & 1 & 1 & 0 & 1 & 0 & 1 & 1 & 0 & 1 & 0 & 0 & 0 &
   0  \\
X_{14} & 0 & 0 & 1 & 0 & 0 & 1 & 0 & 0 & 0 & 0 & 0 & 0 & 0 & 1 & 0 & 0 & 0 & 0 & 1 & 0 & 1 &
   1  \\
X_{15} & 0 & 0 & 0 & 1 & 0 & 0 & 0 & 0 & 0 & 0 & 0 & 0 & 0 & 0 & 1 & 0 & 0 & 0 & 0 & 1 & 0 &
   0  \\
X_{16} & 0 & 0 & 0 & 0 & 1 & 0 & 1 & 0 & 0 & 0 & 0 & 0 & 0 & 0 & 0 & 1 & 1 & 1 & 0 & 0 & 0 &
   1 \\
X_{17} & 1 & 1 & 0 & 0 & 0 & 0 & 0 & 0 & 0 & 0 & 1 & 1 & 1 & 0 & 0 & 0 & 0 & 0 & 1 & 1 & 1 &
   0 \\
X_{18} & 1 & 1 & 1 & 1 & 1 & 1 & 1 & 1 & 1 & 1 & 0 & 0 & 0 & 0 & 0 & 0 & 0 & 0 & 0 & 0 & 0 &
   0 \\
X_{19} & 1 & 0 & 1 & 0 & 1 & 0 & 0 & 1 & 0 & 0 & 1 & 0 & 0 & 0 & 0 & 1 & 0 & 0 & 1 & 0 & 0 &
   1 \\
X_{20} & 1 & 1 & 0 & 1 & 1 & 0 & 1 & 0 & 1 & 0 & 0 & 1 & 0 & 0 & 0 & 0 & 1 & 0 & 0 & 1 & 0 &
   0  \\
X_{21} & 0 & 1 & 0 & 0 & 0 & 1 & 1 & 0 & 0 & 1 & 0 & 0 & 1 & 0 & 0 & 0 & 0 & 1 & 0 & 0 & 1 &
   0 
\end{array}
\right. \dots
$
}
\noindent\makebox[\textwidth]{%
\tiny
$
\hspace{1cm}
\dots
\left.
\begin{array}{ccccccccccccccccccccccccc}
\ p_{23} \ & \ p_{24} \ & \ p_{25} \ & \ p_{26} \ & \ p_{27} \ & \ p_{28} \ & \ p_{29} \ & \ p_{30} \ & \ p_{31} \ & \ p_{32} \ & \ p_{33} \ & \ p_{34} \ & \ p_{35} \ & \ p_{36} \ & \ p_{37} \ & \ p_{38} \ & \ p_{39} \ & \ p_{40} \ & \ p_{41} \ & \ p_{42} \ & \ p_{43} \ & \ p_{44} \ \\
\hline
 0 & 0 & 0 & 0 & 1 & 1 & 0 & 1 & 1 & 1 & 0 & 0 & 1 & 1 & 1 & 0 & 0 & 0 & 0 & 0 & 0 &
   0 \\
 0 & 0 & 1 & 1 & 0 & 0 & 0 & 0 & 0 & 0 & 0 & 0 & 0 & 0 & 0 & 0 & 0 & 0 & 0 & 0 & 0 &
   0 \\
 1 & 1 & 0 & 0 & 0 & 0 & 0 & 0 & 0 & 0 & 0 & 0 & 0 & 0 & 0 & 0 & 0 & 0 & 0 & 0 & 0 &
   0 \\
 0 & 0 & 1 & 1 & 0 & 0 & 0 & 1 & 1 & 1 & 1 & 1 & 0 & 0 & 0 & 0 & 0 & 0 & 0 & 1 & 1 &
   1 \\
 0 & 0 & 0 & 0 & 0 & 0 & 0 & 0 & 0 & 0 & 0 & 0 & 1 & 1 & 1 & 1 & 1 & 0 & 0 & 0 & 0 &
   0 \\
 1 & 1 & 1 & 1 & 0 & 0 & 0 & 0 & 0 & 0 & 0 & 0 & 0 & 0 & 0 & 0 & 0 & 1 & 1 & 1 & 1 &
   1 \\
 0 & 0 & 0 & 0 & 1 & 1 & 1 & 1 & 1 & 1 & 1 & 1 & 1 & 1 & 1 & 1 & 1 & 1 & 1 & 1 & 1 &
   1 \\
 1 & 1 & 0 & 0 & 1 & 1 & 1 & 0 & 0 & 0 & 0 & 0 & 0 & 0 & 0 & 0 & 0 & 1 & 1 & 0 & 0 &
   0 \\
 0 & 0 & 0 & 0 & 0 & 0 & 1 & 0 & 0 & 0 & 1 & 1 & 0 & 0 & 0 & 1 & 1 & 0 & 0 & 0 & 0 &
   0 \\
 0 & 0 & 1 & 0 & 0 & 0 & 0 & 0 & 0 & 0 & 0 & 0 & 0 & 0 & 0 & 0 & 0 & 0 & 0 & 0 & 0 &
   0 \\
 1 & 1 & 0 & 1 & 0 & 0 & 0 & 0 & 0 & 0 & 0 & 0 & 0 & 0 & 0 & 0 & 0 & 0 & 0 & 0 & 0 &
   0 \\
 0 & 0 & 1 & 0 & 0 & 0 & 0 & 0 & 1 & 1 & 0 & 1 & 0 & 1 & 1 & 0 & 1 & 0 & 0 & 0 & 1 &
   1 \\
 0 & 0 & 0 & 1 & 0 & 0 & 0 & 1 & 0 & 0 & 1 & 0 & 1 & 0 & 0 & 1 & 0 & 0 & 0 & 1 & 0 &
   0 \\
 0 & 1 & 1 & 0 & 0 & 1 & 0 & 0 & 0 & 1 & 0 & 0 & 0 & 0 & 1 & 0 & 0 & 0 & 1 & 0 & 0 &
   1 \\
 1 & 0 & 0 & 1 & 1 & 0 & 0 & 1 & 1 & 0 & 0 & 0 & 1 & 1 & 0 & 0 & 0 & 1 & 0 & 1 & 1 &
   0 \\
 1 & 1 & 0 & 0 & 0 & 0 & 1 & 0 & 0 & 0 & 1 & 1 & 0 & 0 & 0 & 1 & 1 & 1 & 1 & 1 & 1 &
   1 \\
 0 & 0 & 0 & 0 & 1 & 1 & 1 & 0 & 0 & 0 & 0 & 0 & 1 & 1 & 1 & 1 & 1 & 1 & 1 & 0 & 0 &
   0 \\
 0 & 0 & 0 & 0 & 1 & 1 & 1 & 1 & 1 & 1 & 1 & 1 & 0 & 0 & 0 & 0 & 0 & 0 & 0 & 0 & 0 &
   0 \\
 0 & 0 & 0 & 0 & 1 & 1 & 1 & 1 & 0 & 0 & 1 & 0 & 1 & 0 & 0 & 1 & 0 & 1 & 1 & 1 & 0 &
   0 \\
 1 & 0 & 0 & 0 & 1 & 0 & 1 & 0 & 1 & 0 & 0 & 1 & 0 & 1 & 0 & 0 & 1 & 1 & 0 & 0 & 1 &
   0 \\
 0 & 1 & 0 & 0 & 0 & 1 & 1 & 0 & 0 & 1 & 1 & 1 & 0 & 0 & 1 & 1 & 1 & 0 & 1 & 0 & 0 &
   1
 \end{array}
\right)~~.
$
}

\newpage

\section{Sewed Models} \label{appendix_sewed_models}

This section summarizes the perfect matching matrices for the sewed models $\sigma_{0}(C_1)$, $\sigma_{0}(C_2)$ and $\sigma_{0}(C_3)$. These models are discussed in \sref{ssew}. The perfect matching matrices are

\noindent\makebox[\textwidth]{%
\tiny
$
P_{\sigma_{0}(C_1)}=
\left(
\begin{array}{c|cccccc}
\; & p_1 & p_2 & p_3 & p_4 & p_5 & p_6 \\
\hline
X_1 & 1 & 0 & 1 & 0 & 0 & 0 \\
X_2 & 0 & 1 & 0 & 1 & 0 & 0 \\
X_3 & 0 & 0 & 0 & 0 & 1 & 0 \\
X_4 & 0 & 0 & 0 & 0 & 0 & 1 \\
X_5 & 1 & 1 & 0 & 0 & 0 & 0 \\
X_6 & 0 & 0 & 1 & 1 & 0 & 0 \\
X_7 & 0 & 0 & 0 & 0 & 1 & 1
\end{array}
\right)~~,~~
$
}
\vspace{0.2cm}

\noindent\makebox[\textwidth]{%
\tiny
$
P_{\sigma_{0}(C_2)}=
\left(
\begin{array}{c|cccccccccccccc}
\; & \ p_1 \ & \ p_2 \ & \ p_3 \ &  \ p_4 \ & \ p_5 \ &  \ p_6 \ & \ p_7 \ & \ p_8 \  & \ p_9 \ & p_{10}
& p_{11} & p_{12} & p_{13} & p_{14}\\
\hline
X_{1} & 1 & 0 & 0 & 0 & 1 & 0 & 0 & 0 & 0 & 0 & 1 & 0 & 1 & 0 \\
X_{2} & 0 & 1 & 1 & 0 & 0 & 0 & 0 & 0 & 0 & 0 & 0 & 1 & 0 & 1 \\
X_{3} & 1 & 0 & 0 & 1 & 0 & 0 & 0 & 0 & 1 & 0 & 1 & 0 & 0 & 0 \\
X_{4} & 0 & 1 & 0 & 0 & 0 & 1 & 1 & 0 & 0 & 0 & 0 & 1 & 0 & 0 \\
X_{5} & 0 & 0 & 1 & 1 & 0 & 0 & 0 & 1 & 0 & 0 & 0 & 0 & 0 & 0 \\
X_{6} & 0 & 0 & 0 & 0 & 1 & 1 & 0 & 0 & 0 & 1 & 0 & 0 & 0 & 0 \\
X_{7} & 0 & 0 & 0 & 0 & 0 & 0 & 1 & 1 & 0 & 0 & 0 & 0 & 1 & 0 \\
X_{8} & 0 & 0 & 0 & 0 & 0 & 0 & 0 & 0 & 1 & 1 & 0 & 0 & 0 & 1 \\
X_{9} & 0 & 0 & 0 & 0 & 0 & 0 & 0 & 0 & 0 & 0 & 1 & 1 & 1 & 1 \\
X_{10} & 1 & 1 & 0 & 0 & 0 & 0 & 1 & 0 & 1 & 0 & 0 & 0 & 0 & 0 \\
X_{11} & 0 & 0 & 0 & 1 & 0 & 1 & 0 & 0 & 0 & 0 & 1 & 1 & 0 & 0 \\
X_{12} & 1 & 1 & 1 & 0 & 1 & 0 & 0 & 0 & 0 & 0 & 0 & 0 & 0 & 0 \\
X_{13} & 0 & 0 & 0 & 1 & 0 & 1 & 1 & 1 & 1 & 1 & 0 & 0 & 0 & 0 \\
X_{14} & 0 & 0 & 1 & 0 & 1 & 0 & 0 & 1 & 0 & 1 & 0 & 0 & 1 & 1
\end{array}
\right)
~~,~~
$}
\vspace{0.2cm}

\noindent\makebox[\textwidth]{%
\tiny
$
P_{\sigma_{0}(C_3)}=
\left(
\begin{array}{c|cccccccccccccccccc}
\; &\ p_1 \ & \ p_2 \ & \ p_3 \ & \ p_4 \ & \ p_5 \ & \ p_6 \ & \ p_7 \ & \ p_8 \ & \ p_9 \ & p_{10}
& p_{11} & p_{12} & p_{13} & p_{14} & p_{15} & p_{16} & p_{17} & p_{18}\\
\hline
X_{1} & 0 & 1 & 1 & 0 & 0 & 0 & 1 & 1 & 0 & 0 & 0 & 0 & 0 & 0 & 0 & 0 & 0 & 0\\
X_{2} & 1 & 0 & 0 & 0 & 1 & 1 & 0 & 0 & 0 & 1 & 1 & 0 & 0 & 1 & 1 & 1 & 0 & 1\\
X_{3} & 0 & 1 & 1 & 1 & 0 & 0 & 1 & 1 & 1 & 0 & 0 & 1 & 1 & 0 & 0 & 0 & 1 & 0\\
X_{4} & 0 & 0 & 0 & 0 & 1 & 0 & 0 & 0 & 0 & 0 & 0 & 0 & 0 & 1 & 1 & 0 & 0 & 1\\
X_{5} & 0 & 0 & 0 & 0 & 0 & 1 & 1 & 1 & 1 & 0 & 0 & 0 & 0 & 0 & 0 & 0 & 0 & 0\\
X_{6} & 0 & 0 & 0 & 0 & 0 & 0 & 0 & 0 & 0 & 1 & 1 & 1 & 1 & 1 & 1 & 0 & 0 & 0\\
X_{7} & 0 & 0 & 0 & 0 & 0 & 0 & 0 & 0 & 0 & 0 & 0 & 0 & 0 & 0 & 0 & 1 & 1 & 1\\
X_{8} & 0 & 0 & 0 & 0 & 0 & 0 & 0 & 0 & 0 & 0 & 0 & 0 & 0 & 0 & 0 & 0 & 0 & 0\\
X_{9} & 0 & 0 & 0 & 0 & 0 & 0 & 0 & 0 & 0 & 0 & 0 & 0 & 0 & 0 & 0 & 0 & 0 & 0\\
X_{10} & 1 & 0 & 1 & 1 & 0 & 0 & 0 & 0 & 0 & 0 & 1 & 0 & 1 & 0 & 0 & 1 & 0 & 0\\
X_{11} & 0 & 1 & 0 & 0 & 0 & 0 & 0 & 0 & 0 & 1 & 0 & 1 & 0 & 0 & 0 & 0 & 1 & 0\\
X_{12} & 1 & 0 & 0 & 1 & 1 & 1 & 0 & 0 & 1 & 0 & 0 & 0 & 0 & 0 & 0 & 0 & 0 & 0\\
X_{13} & 1 & 1 & 0 & 0 & 1 & 1 & 1 & 0 & 0 & 1 & 1 & 0 & 0 & 1 & 0 & 0 & 0 & 0\\
X_{14} & 0 & 0 & 1 & 1 & 0 & 0 & 0 & 1 & 1 & 0 & 0 & 1 & 1 & 0 & 1 & 0 & 0 & 0\\
X_{15} & 0 & 1 & 0 & 0 & 1 & 0 & 1 & 0 & 0 & 0 & 0 & 0 & 0 & 1 & 0 & 0 & 1 & 1\\
X_{16} & 0 & 0 & 0 & 0 & 0 & 1 & 0 & 1 & 1 & 1 & 0 & 1 & 0 & 0 & 1 & 0 & 0 & 0\\
X_{17} & 0 & 0 & 0 & 0 & 0 & 0 & 1 & 0 & 0 & 0 & 1 & 0 & 1 & 1 & 0 & 0 & 0 & 0\\
X_{18} & 0 & 0 & 1 & 0 & 0 & 0 & 0 & 1 & 0 & 0 & 0 & 0 & 0 & 0 & 1 & 1 & 0 & 1\\
X_{19} & 0 & 0 & 0 & 1 & 0 & 0 & 0 & 0 & 1 & 0 & 0 & 1 & 1 & 0 & 0 & 0 & 1 & 0\\
X_{20} & 1 & 0 & 0 & 0 & 0 & 1 & 0 & 0 & 0 & 1 & 1 & 0 & 0 & 0 & 0 & 1 & 0 & 0\\
X_{21} & 1 & 1 & 1 & 1 & 1 & 0 & 0 & 0 & 0 & 0 & 0 & 0 & 0 & 0 & 0 & 1 & 1 & 1
\end{array}
\right.
\dots
$}

\noindent\makebox[\textwidth]{%
\tiny
$
\hspace{1cm}
\dots
\left.
\begin{array}{cccccccccccccccccc}
 p_{19} & p_{20}
& p_{21} & p_{22} &
p_{23} & p_{24} & p_{25} & p_{26} & p_{27} & p_{28} & p_{29} & p_{30} & p_{31} & p_{32}
& p_{33} & p_{34} & p_{35} & p_{36}\\
\hline
 0 & 0 & 1 & 0 & 1 & 0 & 1 & 0 & 0 & 0 & 0 & 1 & 1 & 0 & 1 & 0 & 0 & 0 \\
 1 & 0 & 0 & 0 & 0 & 0 & 0 & 0 & 0 & 0 & 0 & 0 & 0 & 0 & 0 & 0 & 0 & 0 \\
 0 & 1 & 0 & 0 & 0 & 0 & 0 & 0 & 0 & 0 & 0 & 0 & 0 & 0 & 0 & 0 & 0 & 0 \\
 0 & 0 & 0 & 0 & 1 & 1 & 0 & 0 & 0 & 0 & 1 & 0 & 1 & 1 & 0 & 1 & 0 & 0 \\
 1 & 1 & 0 & 0 & 0 & 0 & 1 & 1 & 0 & 0 & 0 & 0 & 0 & 0 & 1 & 0 & 0 & 1 \\
 0 & 0 & 0 & 0 & 0 & 0 & 0 & 0 & 1 & 1 & 1 & 0 & 0 & 0 & 0 & 1 & 0 & 0 \\
 1 & 1 & 0 & 0 & 0 & 0 & 0 & 0 & 0 & 0 & 0 & 0 & 0 & 0 & 0 & 0 & 1 & 1 \\
 0 & 0 & 1 & 1 & 1 & 1 & 1 & 1 & 1 & 1 & 1 & 0 & 0 & 0 & 0 & 0 & 0 & 0 \\
 0 & 0 & 0 & 0 & 0 & 0 & 0 & 0 & 0 & 0 & 0 & 1 & 1 & 1 & 1 & 1 & 1 & 1 \\
 0 & 0 & 1 & 1 & 0 & 0 & 0 & 0 & 0 & 1 & 0 & 0 & 0 & 0 & 0 & 0 & 0 & 0 \\
 0 & 0 & 0 & 0 & 0 & 0 & 0 & 0 & 1 & 0 & 0 & 1 & 0 & 0 & 0 & 0 & 1 & 0 \\
 0 & 0 & 0 & 1 & 0 & 1 & 0 & 1 & 0 & 0 & 0 & 0 & 0 & 1 & 0 & 0 & 0 & 0 \\
 0 & 0 & 0 & 0 & 0 & 0 & 0 & 0 & 0 & 0 & 0 & 1 & 0 & 0 & 1 & 0 & 0 & 0 \\
 0 & 0 & 0 & 0 & 0 & 0 & 0 & 0 & 0 & 0 & 0 & 0 & 1 & 1 & 0 & 1 & 0 & 0 \\
 0 & 1 & 0 & 0 & 1 & 1 & 0 & 0 & 0 & 0 & 1 & 0 & 0 & 0 & 0 & 0 & 0 & 0 \\
 1 & 0 & 0 & 0 & 0 & 0 & 1 & 1 & 1 & 0 & 0 & 0 & 0 & 0 & 0 & 0 & 0 & 0 \\
 0 & 1 & 0 & 0 & 0 & 0 & 0 & 0 & 0 & 1 & 1 & 0 & 0 & 0 & 1 & 1 & 0 & 1 \\
 1 & 0 & 1 & 0 & 1 & 0 & 1 & 0 & 0 & 0 & 0 & 0 & 1 & 0 & 0 & 0 & 0 & 0 \\
 0 & 1 & 0 & 1 & 0 & 1 & 0 & 1 & 1 & 1 & 1 & 0 & 0 & 1 & 0 & 1 & 1 & 1 \\
 1 & 0 & 1 & 1 & 0 & 0 & 1 & 1 & 1 & 1 & 0 & 1 & 0 & 0 & 1 & 0 & 1 & 1 \\
 0 & 0 & 1 & 1 & 1 & 1 & 0 & 0 & 0 & 0 & 0 & 1 & 1 & 1 & 0 & 0 & 1 & 0
\end{array}
\right)
~~.
$}




\end{document}